# Adding Photonic Entanglement to Superradiance by Using Multilevel Atoms

**Amir Sivan[1,2] and Meir Orenstein[1,2]**

[1]*Andrew and Erna Viterbi department of Electrical & Computer Engineering, Technion – Israel Institute of Technology, Haifa 32000, Israel*
[2]*Helen Diller Quantum Center, Technion – Israel Institute of Technology, Haifa 32000, Israel*
Author e-mail address: *amirsi@campus.technion.ac.il*

**Abstract:** We show here that the photonic states emitted by ensembles of multilevel atoms via a superradiance process exhibit entanglement in the modal (frequency) degree of freedom, making this collective emission process a favorable candidate for a fast, bright and deterministic source of entangled photons. This entanglement is driven by two mechanisms: (i) selective excitation of the atomic ensemble to a superposition state and (ii) degeneracies of the optical transitions due to internal structure of the emitting atoms. The latter induces intricate non-radiative virtual transitions in the ensemble, which create interatomic correlations that are imprinted onto the emitted photons. One of the important outcomes of this complexity is the generation of mode-independent entangled multiphoton states. In addition, we study the dynamics of the correlations of the superradiating multilevel atom ensembles, and demonstrate a case where they exhibit beating in steady-state due to the aforementioned virtual transitions.

## I.  INTRODUCTION

The area of quantum optics, combining quantum mechanics and optical physics, has undergone tremendous development in the last few decades and today encapsulates many theoretical advancements and practical applications related to quantum information processing, communications, metrology, creation of photonic qubits and continuous variable computation to name just a few [1-8]. One central photonic quantum resource is quantum entanglement, and in particular multiphoton entanglement is crucial for cluster states required for quantum computing [9-12] and for quantum sensing [13-17]. Generation of multiphoton entanglement can be achieved in various ways; for example, entangled photon pairs can be generated deterministically from quantum dots (QDs) and wells [18-28]. Generation of multiphoton states consisting of more than two photons is highly challenging, and can be achieved using intricate QD schemes systems requiring intervention at precise timing to create photonic cluster states [29-31] or exact energy-level engineering for generation of exotic frequency-entangled multipartite states [32]. Although reliable, these systems are strictly limited by electronic decoherence times, imposing complicated engineering and operational challenges on their usability. Another approach is generation of photon states via nonlinear processes [33-50], however – this method is based on low-probability, low-efficiency interactions which are not deterministic nor easily scalable to a large number of photons. A different method entailing the utilization of linear optics in order to create the entangled states of interest [51-59] has the disadvantage of requiring post-selection.

A deterministic multiphoton source can be realized by exploiting a collective spontaneous emission effect, known as Dicke superradiance [60], where an excited dense ensemble of identical two-level atoms spontaneously emits an extremely short and strong pulse of light driven by the inter-atomic entanglement, thereby acting as a deterministic high-yield source of as many photons as there are excited atoms. This collective effect has been intensively studied and generalized to include the effects of various geometries and the introduction of materials and structures on the superradiant emission process [61-74]. However, although the emitting atoms are highly entangled [75-77], the emitted photonic state is not entangled due to the absence of a multitude of quantum evolution paths for the photonic degrees of freedom. Furthermore, since all photons are emitted in a single frequency, there is no degree of freedom for entanglement.

Here we combine the two quantum-optics subfields of superradiance and photon entanglement to realize a superradiant photon entanglement source. This is achieved by generating a cooperative emission from an ensemble of indistinguishable *multilevel* atoms. Although generalizations of the Dicke model that also include the dynamics of multilevel atoms exist [78-90], with recent works describing atomic dynamics due to atomic entanglement invoked by such multilevel structures [91-95] – to the best of our knowledge, the entanglement in the photonic degrees of freedom was not yet considered in the context of superradiance.

Furthermore, it is important to note that the common modeling of the phenomenon of superradiance is obtained by tracing-out the photonic states and solving the master equation for the atoms under the Born-Markov approximation [96-99]. This method is constructive for purposes of describing the temporal evolution of the system, however – in doing so we lose the detailed information on the photonic state, and the photonic fields are obtained only in terms of ensemble averages deduced from the atomic operators. Although this may not be a disadvantage in the case of Dicke superradiance from two-level atoms in which the emitted photon states are not entangled – it renders the standard approach irrelevant for the case of spontaneous emission from multilevel atoms, when complex photonic states are expected. Therefore, a new modeling approach is employed.

In this paper, we propose superradiant light sources that exhibit photon entanglement (as well as dynamics and features related to superradiance) which are based on ensembles of multilevel atoms. Although in general photons may be entangled in several degrees of freedom, such as polarization, time-binning, spatial modes, etc., we focus here on entanglement between photonic Fock number states that differ in frequency. In order to correctly analyze the generation of the photonic entangled state we develop a method that is based on the Weisskopf-Wigner description of spontaneous emission [100], generalized to the case of indistinguishable (symmetrized) ensembles of multilevel atoms. We demonstrate the concept by focusing on two configurations: ensembles of three-level V-atoms and ensembles of four-level atoms (FLAs). We study the photonic state emanated from both ensemble types, and use our method to quantify the entanglement negativity [10,101-103] and conditional entanglement entropy [104,105] of the photonic fields. Additionally, we demonstrate the intricate dynamics of the atomic entanglement properties of FLA ensembles. Furthermore, we dwell on the subject of mode-independent entanglement, which has been a subject of active discussions in the quantum physics community recently [1,106-109,111]. This type of entanglement is a desired quantum resource, since it does not depend on the choice of a measurement basis and cannot be disentangled under any unitary transformation, and is thus considered to be a more fundamental quantum-mechanical property. We show that while the photon entanglement in the V-atom ensemble is mode-dependent, the photons emitted from the FLA ensemble exhibits mode-independent entanglement, meaning that the photonic Fock numbers are entangled in a way that no measurement-base exists in which the entanglement disappears.

## II. SUPERRADIANT ENSEMBLE OF V-ATOMS

### A. The non-degenerate case

We study the generation of photonic entangled states emanated from a superradiance process by considering an ensemble of three-level atoms in a V-shaped configuration ("V-atoms") (Figure 1a). We first consider the non-degenerate case for which the emitted photons are discernable in their frequency degree of freedom (either $\omega_1$ or $\omega_2$), and show that the two photon frequency-modes emitted from an ensemble of indistinguishable non-degenerate V-atoms are entangled in the frequency-mode Fock-number basis. We assume that no direct dipole-dipole interactions occur between the atoms in the ensemble.

The Hamiltonian for $N$ non-degenerate V-atoms is given by

$$\hat{H} = \hbar\omega_1 \sum_{i=1}^{N} \hat{n}_{e_1}^i + \hbar\omega_2 \sum_{i=1}^{N} \hat{n}_{e_2}^i + \hbar \sum_{\mathbf{k}} \omega_{\mathbf{k}} \left( \hat{a}_{1,\mathbf{k}}^\dagger \hat{a}_{1,\mathbf{k}} + \hat{a}_{2,\mathbf{k}}^\dagger \hat{a}_{2,\mathbf{k}} \right)$$
$$+ \hbar \sum_{i=1}^{N} \sum_{\mathbf{k}} \left\{ g_{1,\mathbf{k}}^i(\mathbf{r}) \hat{a}_{1,\mathbf{k}}^\dagger \hat{\sigma}_{ge_1}^i e^{i(\omega_{\mathbf{k}}-\omega_1)t} + g_{2,\mathbf{k}}^i(\mathbf{r}) \hat{a}_{2,\mathbf{k}}^\dagger \hat{\sigma}_{ge_2}^i e^{i(\omega_{\mathbf{k}}-\omega_2)t} + h.c. \right\} \quad (1)$$

Here $\hbar\omega_1$ and $\hbar\omega_2$ are the energies of the first and second excited states $|e_1\rangle$, $|e_2\rangle$ with respect to the ground state $|g\rangle$, the atomic operators for the $i$'th atom are $\hat{\sigma}_{ge_j}^i = |g\rangle_{ii}\langle e_j|$ and $\hat{n}_{e_j}^i = |e_j\rangle_{ii}\langle e_j|$ with the level index $j \in \{1,2\}$. The photonic annihilation operator $\hat{a}_{j,\mathbf{k}}$ annihilates a single photon in the mode characterized by the wave vector $\mathbf{k}$ centered around the angular frequency $\omega_j$, and obeys the commutation relations $[\hat{a}_{j,\mathbf{k}}, \hat{a}_{j',\mathbf{k}'}^\dagger] = \delta_{j,j'}\delta_{\mathbf{k},\mathbf{k}'}$. We assume that the photon linewidths around $\omega_1$ and $\omega_2$ are narrow with negligible overlapping. Throughout the paper, these will be referred to as frequency modes.

The atom-field coupling coefficients are given by $g_{j,\mathbf{k}}^i(\mathbf{r}_i) = -i\sqrt{\frac{2\pi c k}{V}} \mathbf{d}_j^i \cdot \mathbf{e}_{\mathbf{k}} e^{i\mathbf{k}\cdot\mathbf{r}_i}$ where $\mathbf{d}_j^i$ is the dipole moment vector of the $j$'th transition in the $i$'th atom, $\mathbf{r}_i$ is the position of the atom, $\mathbf{e}_{\mathbf{k}}$ is the unit vector in the direction of the polarization of the photon in the mode $\mathbf{k}$, $c$ is the speed of light in vacuum and $V$ is the modal volume, which will be taken to infinity when considering emission in free space.

We start by describing the dynamics of the superradiance process in the standard manner, i.e., by tracing out the photonic degrees of freedom and obtaining the atomic master equation in the Born-Markov approximation for the atomic density matrix $\hat{\Theta}$ in the Schrödinger picture [99]. We consider indistinguishable atoms; expressly, we assume that we are not able to assign the atoms indices to differentiate between them, and we will only be able to count the total number of atomic levels occupations across the ensemble. This amounts to symmetrizing the atomic system as was done for example in refs. [81-84,110]. For indistinguishable V-atoms this results in a master equation for the six-dimensional tensor describing the evolution of the atomic system, with each element obeying the coupled first-order linear differential equation (APPENDIX A)

$$\frac{\partial}{\partial t}\Theta_{n_1,n_2,n_g}^{m_1,m_2,m_g} = -i\left[(m_1-n_1)\omega_1 + (m_2-n_2)\omega_2\right]\Theta_{n_1,n_2,n_g}^{m_1,m_2,m_g}$$
$$-\frac{\Gamma_1}{2}\left\{\left[m_1(m_g+1)+n_1(n_g+1)\right]\Theta_{n_1,n_2,n_g}^{m_1,m_2,m_g} - 2\sqrt{m_g(m_1+1)n_g(n_1+1)}\Theta_{n_1+1,n_2,n_g-1}^{m_1+1,m_2,m_g-1}\right\}$$
$$-\frac{\Gamma_2}{2}\left\{\left[m_2(m_g+1)+n_2(n_g+1)\right]\Theta_{n_1,n_2,n_g}^{m_1,m_2,m_g} - 2\sqrt{m_g(m_2+1)n_g(n_2+1)}\Theta_{n_1,n_2+1,n_g-1}^{m_1,m_2+1,m_g-1}\right\} \quad (2)$$

with $\Theta_{n_1,n_2,n_g}^{m_1,m_2,m_g} = \langle m_1,m_2,m_g|\hat{\Theta}|n_1,n_2,n_g\rangle$, where $m_1,n_1$ denote the number of atoms in the first excited level $|e_1\rangle$, $m_2,n_2$ the number of atoms in the second excited level $|e_2\rangle$ and $m_g,n_g$ the number of atoms in the ground state $|g\rangle$, satisfying the conditions $m_1,n_1,m_2,n_2,m_g,n_g \geq 0$, $m_1+m_2+m_g=N$ and $n_1+n_2+n_g=N$. The energy ladder describing such a symmetrized ensemble is illustrated in Figure 1b for the case of three indistinguishable V-atoms where the atomic state is described by the atomic occupation numbers $|n_{e_1},n_{e_2},n_g\rangle$. The intrinsic decay rate of the spontaneous transition $j \in \{1,2\}$ is

$$\Gamma_j = \frac{|\mathbf{d}_j|^2 \omega_j^3}{3\pi\varepsilon_0 \hbar c^3}. \quad (3)$$

Here $\varepsilon_0$ is the vacuum permittivity. The initial photonic state is the vacuum state, and we choose here the initial atomic state as a single electron excitation state in a combination of the excited levels for each atom with equal amplitudes. The density matrix can be written by considering symmetric excitations as $\rho_{V,i} = \hat{\Theta} \otimes |0,0\rangle_{ff}\langle 0,0| = |\psi_{V,i}^S\rangle\langle\psi_{V,i}^S| \otimes |0,0\rangle_{ff}\langle 0,0|$ where $|\psi_{V,i}^S\rangle$ is the state vector of the atomic part given by $|\psi_{V,i}^S\rangle = 2^{-N/2}\hat{S}\{\prod_{j=1}^{N}(|e_1\rangle_j + |e_2\rangle_j)\}$ (with $\hat{S}$ denoting the symmetrization operator defined in APPENDIX A) and $|0,0\rangle_f$ is the photonic vacuum state for the two frequency modes. In Figure 1d, we show by solving master equation (2) that the collective spontaneous decay process consists of two competing Dicke-superradiance-like processes, as detailed in APPENDIX A. We have hereby demonstrated that the temporal dynamics of the atomic degree of freedom describe concurrent superradiance processes, by deriving the ensemble-averaged emission intensity from the expectation value of the atomic operators. The maximal intensities of both emission processes are fitted to curve of the form $\beta N^\alpha + c$ to an excellent agreement, as shown in Figure 1c. Different scaling laws for different transitions in collectively emitting ensembles are evident, in agreement with literature (e.g. [90]).

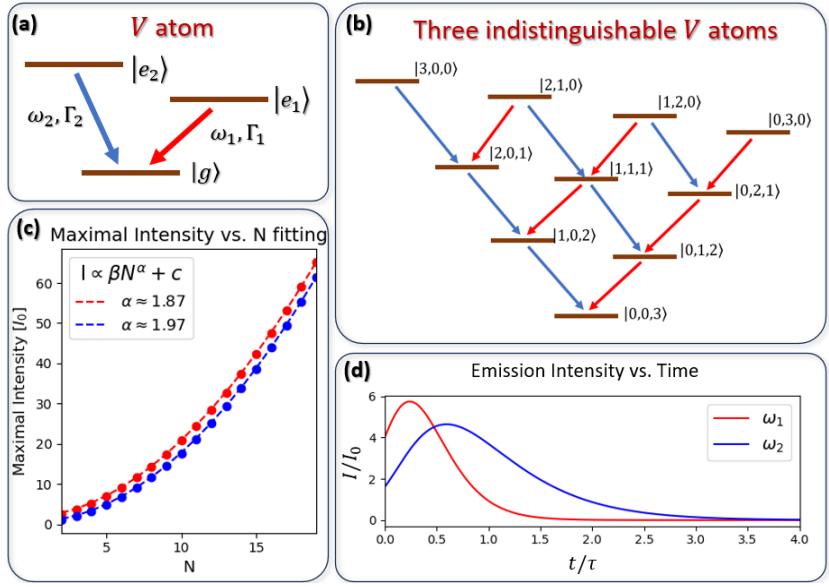

Figure 1 – (a) Energy level diagram of a single V-atom. Red (blue) arrow denotes a radiative transition of angular frequency $\omega_1$ ($\omega_2$) at a rate $\Gamma_1$ ($\Gamma_2$) from the excited state $|e_1\rangle$ ($|e_2\rangle$) to the ground state $|g\rangle$. (b) Energy level diagram of three indistinguishable V-atoms. The atomic states are represented in the atomic occupation numbers basis $|n_{e_2}, n_{e_1}, n_g\rangle$. (c) Scaling laws for the intensity peaks of each of the three photonic modes with respect to the number of indistinguishable V-atoms, color coded as in subfigure (a), with $\omega_2 = 5\omega_1/8$ and $\Gamma_1 = 4\Gamma_2$. Dots: intensity peaks values. Dashed lines: fitted curves of the form $\beta N^\alpha + c$. The parameter $\alpha$ denotes the exponential scaling parameter. (d) Intensity vs. time for $N = 8$ indistinguishable, non-degenerate V-atoms, initialized in the fully excited symmetric state. Time is given in units of $\tau = 1/\Gamma_1$ and intensity is given in units of $I_0$ - intensity of a two-level atom with angular frequency and decay rate $\omega_1$ and $\Gamma_1$.

The main result of this section is the accomplishment of our goal, namely – generating entanglement between the two photonic frequency-modes emitted from the V-atom ensemble. As mentioned above, the ensemble-average approach that we used to calculate the total intensity erases the information on the quantum properties of the photons. Therefore, to quantify the photonic entanglement at the end of the collective spontaneous emission process we take a different approach in which the photonic degrees of freedom are not traced out. To do so, we consider the states of the entire atoms and photons system, and employ the Weisskopf-Wigner method with the usual Markov approximation [99,112]. The full state of the indistinguishable (symmetrized) atomic ensemble and the photons can be described as a single state vector consisting of a superposition of all possible combinations of atomic excitations and emitted photons satisfying energy conservation, weighed by infinite sums of time-dependent coefficients describing the continuum of modes into which each photon can be spontaneously emitted. For example, for the single V-atom, we have

$$|\psi_V\rangle(t) = a_1(t)|1,0,0\rangle_A |0,0\rangle_f + a_2(t)|0,1,0\rangle_A |0,0\rangle_f + |0,0,1\rangle_A \sum_{\mathbf{k}} \left( b_{1,\mathbf{k}}(t)|1_\mathbf{k},0\rangle_f + b_{2,\mathbf{k}}(t)|0,1_\mathbf{k}\rangle_f \right) \quad (4)$$

where in each term the left ket (with subscript $A$) denotes the atomic occupation numbers $|n_{e_1}, n_{e_2}, n_g\rangle_A$ and the right ket (with subscript $f$) denotes the modal photonic occupation numbers $|n_{\omega_1,\mathbf{k}}, n_{\omega_2,\mathbf{k}}\rangle_f$, with coefficients $b_{1,\mathbf{k}}(t)$ and $b_{2,\mathbf{k}}(t)$. In this representation, the two distinct photonic Fock numbers describe emitted photons with a Lorentzian distribution with angular frequencies (resulting from the Weisskopf-Wigner approximation, as shown in APPENDIX B) centered around $\omega_1$ or $\omega_2$. The subscript denoting the wavevector $\mathbf{k}$ encapsulates information on the amplitude and direction of a single mode from this distribution. At time $t \to \infty$,

$$|\psi_V\rangle(\infty) = |0,0,1\rangle_A \left( \sum_{\mathbf{k}} b_{1,\mathbf{k}}(\infty)|1_\mathbf{k},0\rangle_f + \sum_{\mathbf{q}} b_{2,\mathbf{q}}(\infty)|0,1_\mathbf{q}\rangle_f \right) \quad (5)$$

and in the general $N$ indistinguishable V-atom case at $t \to \infty$,

$$|\psi_{N-V}\rangle(\infty) = |0,0,N\rangle_A \sum_{n=0}^{N} \left[ \sum_{\mathbf{k}_1} \cdots \sum_{\mathbf{k}_N} c_{n,\mathbf{k}_1 \ldots \mathbf{k}_N}(\infty) \Big|(N-n)_{\{N-n\}}, n_{\{n\}}\Big\rangle_f \right]. \quad (6)$$

The subscripts $\{i\}$ of a photonic modal Fock numbers vector consisting of $N$ total photons denote a tuple of $i$ unique wave vector indices from the set $\{\mathbf{k}_1, \ldots, \mathbf{k}_N\}$, to account for the fact that the order of summations is interchangeable. The sets of vector indices for the two frequency mode photon numbers are disjoint. The coefficients $c_{n,k_1 \ldots k_N}(\infty)$ are solved using Schrödinger's equation under the Markov approximation as per the Weisskopf-Wigner approach, resulting in an intricate system of coupled first-order ordinary differential equations (ODE). As thoroughly outlined in APPENDIX B, this system of ODE's can be solved by Laplace transformation, replacing a laborious calculation of an excessively large number of coupled differential equations with simple recursion rules that amount to solving polynomial fractions. The procedure outlined in APPENDIX B yields the exact atomic and photonic states rather than ensemble averages, which is necessary in order to obtain the information on the photonic state.

To prove that the two emitted photonic fields are entangled after the atoms have reached their ground state, we consider the density matrix of the state at $t \to \infty$ and trace out the atomic degree of freedom. Because at $t \to \infty$ the atomic state is the common ground state $|G\rangle = \hat{S} \prod_{j=1}^{N} |g_j\rangle = |0, 0, N\rangle_A$, there is no entanglement between the atoms and the photons and therefore the photonic state after tracing out the atomic degree of freedom remains a pure state. In APPENDIX C we show that the photonic state is bipartite and non-separable and thus the two photonic frequency modes are entangled.

Interestingly, for an ensemble of indistinguishable V-atoms the final photonic superposition state is an imprint of the initial atomic excitation state. In other words, the coefficient of the photonic state vector with energy $\hbar(n\omega_1 + (N-n)\omega_2)$ in the final photonic superposition state (namely, $n$ photons of angular frequency $\omega_1$ and $N-n$ photons of angular frequency $\omega_2$) is equal to the excitation amplitude of the atomic state with the same energy (with respect to the ground level), as shown in APPENDIX C. Consequently, the initial entanglement between the symmetrized initial atomic states is imprinted onto the emitted photonic state, and the measure of entanglement negativity [103] between the two photonic modal Fock numbers does not depend on the parameters of the atoms (e.g. decay rates and transition frequencies), but only on their initial excitation amplitudes. In Figure 2a we illustrate the dependence of the entanglement negativity for general symmetrized initial combination states,

$$|\psi_{N-V,i}\rangle = N\hat{S}\left\{\prod_{j=1}^{N}\left(\alpha|e_1\rangle_j + \beta|e_2\rangle_j\right)\right\}, \qquad (7)$$

for $\alpha$ and $\beta$ fulfilling $|\alpha|^2 + |\beta|^2 = 1$ where $N = \left(\sum_{j=0}^{N}\left|\binom{N}{j}\alpha^j \beta^{N-j}\right|^2\right)^{-1/2}$ is a normalization constant. It is evident that the largest entropy negativity for a given number of atoms $N$ occurs when $|\alpha|^2 = 1/2$, and is equal to zero when $|\alpha|$ is either 0 or 1, corresponding to a deterministic excitation to either excited state in all atoms which is equivalent to the two-level atoms case. Figure 2b shows the maximal value of the entropy negativity as a function of the number of atoms in the ensemble. Therefore, the entropy negativity is shown to be a monotonically increasing, unbounded function with respect to $N$; meaning that the larger the ensemble, the larger the measure of entanglement of the emitted photonic state is.

The meaning of this result is that the entanglement ingredient of the emitted photons originates from, and is controlled by, the electronic excitation of the atomic ensemble. Namely, by exciting a sample of identical V-atoms into preselected identical superposition states, one can realize a deterministic source of entangled photons.

In Figure 2c-e we show the two-mode Wigner distribution (spanning a 4D phase-space) of the photonic state resulting from (7) for different excitation amplitudes $\alpha$ and $\beta$, by plotting the 2D projections on the hypersurfaces corresponding to the six pairs of phase-space coordinates of the two modes, where in each plot the 4D phase-space is sliced at the origin with respect to the two phase-space coordinates that are not visualized. The derivation of the two-mode Wigner distribution is detailed and discussed in APPENDIX D. In Figure 2c,d, we illustrate four indistinguishable V-atoms each excited to the superposition states $\alpha = \beta = 2^{-1/2}$ and $\alpha = 10^{-1/2}, \beta = -i(9/10)^{-1/2}$, respectively. We see that pairs of phase-space coordinates corresponding to different modes (e.g. $(X_1, P_2)$) are correlated, indicating that the two photonic modes are entangled (see discussion in APPENDIX D). In Figure 2e we show the case of a deterministic excitation, $\beta = 0$; in this case, the Wigner distributions of the first and second modes correspond to those of a Fock number state of four and the vacuum state, respectively, and the four cross-mode phase-space coordinate pairs are uncorrelated – proving that in this case the photonic modes are separable.

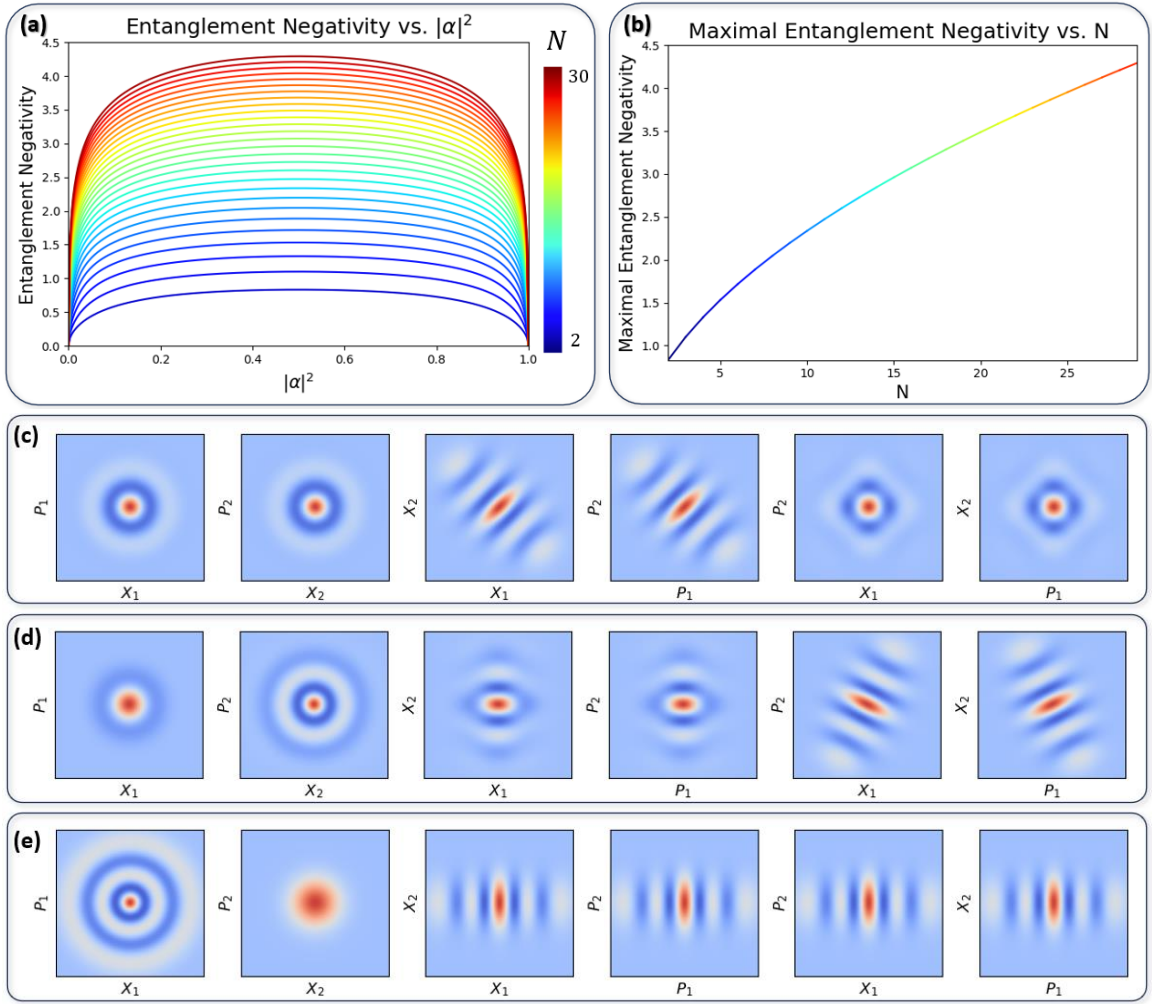

Figure 2 – (a) The entanglement negativity of the photonic state vs. $|\alpha|^2$ for different numbers $N$ of indistinguishable V-atoms. (b) The entanglement negativity vs. $N$ for $|\alpha|^2 = 1/2$, for which the maximal value of the negativity of the photonic state is attained. (c-e) Bi-modal Wigner distributions $W^{2D}(X_1, P_1, X_2, P_2)$ for four indistinguishable non-degenerate V-atoms. In each illustration the 4D bi-modal phase-space is sliced at the origin of the pair of phase-space coordinates not shown. Entanglement is shown for the superposition excitations $\alpha = \beta = 2^{-1/2}$ (subfigure (c)) and $\alpha = 10^{-1/2}, \beta = -i(9/10)^{-1/2}$ (subfigure (d)). For the deterministic excitation $\beta = 0$ (subfigure (e)), the phase-space coordinate pairs of the different modes are not cross-correlated and therefore no entanglement between the photonic modes occurs.

Finally, we remark that entanglement between emitted photons does not occur for a Λ-type atom [113] since the entanglement in the V-atom case is copied from the entanglement between the initial atomic states, which does not exist in the Λ-atom which has a single fully excited state.

### B. Atomic dynamics in the degenerate case

We briefly mention the case of degenerate V atoms, i.e. – when the two radiative transitions have the same frequency. Although here the emitted photons occupy a single frequency mode, and thus frequency entanglement is absent, the aim here is to introduce another effect of indistinguishability of a multilevel atom ensemble (beyond superradiance), pertaining to the overall emission dynamics introduced by the degeneracy of transitions within the constituent multilevel atoms of the ensemble. This will motivate the next section, where we examine ensembles of indistinguishable atoms with partially degenerate transitions and analyze how such intra-atomic degeneracy influences photon entanglement.

To elucidate this effect we consider a single V-atom, where the degeneracy results in the existence of finite probability (even at $t \to \infty$) for the atom to remain excited in either the $|e_1\rangle$ or $|e_2\rangle$ states, as well as nonvanishing correlations [82,99]. Only for a symmetric combination state initialization $|\psi^S\rangle = 2^{-1/2}(|e_1\rangle + |e_2\rangle)$, the atom completely decays at $t \to \infty$, whereas for an antisymmetric excitation $|\psi^{AS}\rangle = 2^{-1/2}(|e_1\rangle - |e_2\rangle)$ the decay is completely suppressed (dark state) and the system does not evolve. This will also be the case for an ensemble of *N distinguishable* V atoms (that are described as a tensor product of *N* single atoms). However, in an ensemble of *indistinguishable* V atoms an excitation to the symmetric combination state $|\psi_i^V\rangle$ in each atom results in a final state wherein the initial populations do not decay completely to the ground state. This means that the indistinguishability of the superradiant ensemble supports partial dark states that do not exist for an ensemble of distinguishable V atoms. This indistinguishability-related effect is repeated also for a "dark state" initialization, that non-intuitively undergoes a partial decay to the ground state only in the indistinguishable ensembles.

This phenomenon arises from coupling between the degenerate intra- and interatomic $e_1 \to g$ and $g \to e_2$ transitions (or the complementary pair). Such coupling describes energy-conserving virtual photon transitions which link the atomic coherences and the atomic population terms in the master equation of the ensemble of indistinguishable atoms. Consequently, the evolution of the atomic populations is affected, resulting in a further accelerated superradiant emission and different steady-state atomic populations compared to the scenario of distinguishable V-atoms. A full derivation and discussion can be found in APPENDIX E.

## III. SUPERRADIANCE OF AN ENSEMBLE OF FOUR-LEVEL ATOMS WITH PARTIAL DEGENERACY

We saw that internal degeneracy in each atom energy ladder introduces additional dynamical effects of the superradiance in an indistinguishable ensemble of such atoms. However, for degenerate V atoms, the photonic entanglement is absent due to emission into a single frequency mode. To combine the intricate effects of superradiance and photon entanglement, we consider here the collective emission from an ensemble of the four-level atoms (FLA) illustrated in Figure 3. This structure contains both ingredients: a degeneracy in two of the four transitions which couples between all four radiative transitions, as well as entangled multi-frequency mode emission. In a different setup, several FLAs have been shown to produce entangled photonic states when cascaded, due to the multitude of evolution paths [32]. Here, we analyze an indistinguishable ensemble of such atoms that results in a superradiant process.

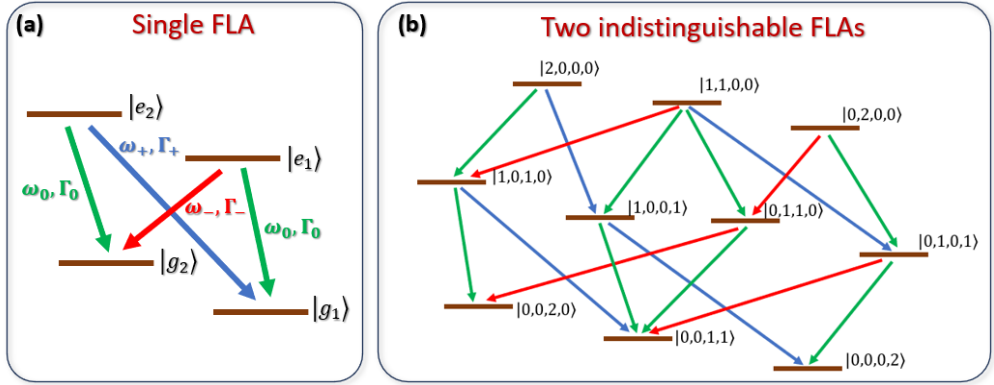

Figure 3 – (a) Energy level diagram of a single FLA. Red, green and blue arrows denote radiative transitions of angular frequencies $\omega_-, \omega_0$ and $\omega_+$ at rates $\Gamma_-, \Gamma_0$ and $\Gamma_+$, respectively, From the excited states $|e_1\rangle$ and $|e_2\rangle$ to the ground states $|g_1\rangle$ and $|g_2\rangle$. (b) Energy level diagram of two indistinguishable FLAs. The atomic states are represented in the atomic occupation numbers basis $|n_{e_2}, n_{e_1}, n_{g_2}, n_{g_1}\rangle$.

### A. Evolution of the atomic part

We first examine the dynamics of the atomic part in order to obtain the superradiance dynamics and study the effect of the intra-atomic degeneracy. We develop the atomic master equation governing the evolution of an ensemble of indistinguishable FLAs producing the superradiant emission. We show that the collective emission process of such atoms drives novel entanglement properties of the atomic system. The main result of this subsection is the intricate entanglement between the atoms resulting from virtual transitions (due to the intra-atomic degeneracy), in addition to the conventional Dicke superradiance atomic entanglement conferred by the indistinguishability of the atoms. Since the focus of this work is the creation of photonic entanglement from superradiance, we leave the full details on the atomic dynamics and entanglement to APPENDIX G.

For each FLA there are four atomic states – two excited states $|e_2\rangle$, $|e_1\rangle$ and two ground states $|g_2\rangle$, $|g_1\rangle$, and a single photon will be spontaneously emitted from a singly-excited FLA in one of three modes with angular frequency $\omega_0$, $\omega_+$ or $\omega_-$, as

shown in Figure 3a, such that $\omega_+ - \omega_0 = \omega_0 - \omega_- \equiv \Delta$. The photon of angular frequency $\omega_0$ will correspond to the degenerate transitions $e_2 \to g_2$ and $e_1 \to g_1$, while the $\omega_+$ and $\omega_-$ photons will correspond to the $e_2 \to g_1$ and $e_1 \to g_2$ transitions respectively.

The internal degeneracy invokes entanglement within the atomic system, beyond what stems from the atomic ensemble indistinguishability. This is because the degenerate transitions introduce cross-terms to the master equation of the atomic density matrix which couple between the atomic operators $\hat{\sigma}_{g_2 e_2}$ and $\hat{\sigma}_{g_1 e_1}$ that govern the transitions between the excited and ground states, as is detailed in APPENDIX F. These cross-terms describe virtual interactions within the FLA ensemble between the degenerate $e_2 \to g_2$ and $g_1 \to e_1$ transitions (and the complementary pair), analogous to the degenerate V-atom case described in Section II.B. These virtual transitions play a crucial role in the intricate entanglement of the atomic ensemble. Since the two radiative transitions emitting the $\omega_-$ and $\omega_+$ photons share either a ground or an excited state with one of the $\omega_0$-photon emitting transitions, this degeneracy results in a correlation between the $\omega_+$ and $\omega_-$ transitions. We will see in the next section that this degeneracy also entangles the emitted photonic modes.

To develop the master equation for an ensemble of FLAs we define the atomic and photonic states in the occupation number basis, $|n_{e_2}, n_{e_1}, n_{g_2}, n_{g_1}\rangle_A$ and $|n_{\omega_-}, n_{\omega_0}, n_{\omega_+}\rangle_f$ respectively. An illustration of the energy level structure of two indistinguishable FLA is shown in Figure 3b. The evolution of the atomic system of such an ensemble is described in the Schrodinger picture by the master equation (APPENDIX F)

$$\frac{d}{dt}\hat{\Theta}(t) = -\frac{i}{\hbar}\left[\hat{H}_0^{FLA}, \hat{\Theta}(t)\right] - \hat{L}\{\hat{\Theta}\} \tag{8}$$

$$\hat{L}\{\hat{\Theta}\} = \frac{1}{2} \underbrace{\sum_{l,m=\{1,2\}} \Gamma_{lm} \left\{ \hat{\Sigma}^\dagger_{g_m e_m} \hat{\Sigma}_{g_l e_l} \hat{\Theta}(t) + \hat{\Theta}(t) \hat{\Sigma}^\dagger_{g_m e_m} \hat{\Sigma}_{g_l e_l} - 2\hat{\Sigma}_{g_l e_l} \hat{\Theta}(t) \hat{\Sigma}^\dagger_{g_m e_m} \right\}}_{A}$$

$$+ \underbrace{\frac{\Gamma_+}{2} \left\{ \hat{\Sigma}^\dagger_{g_1 e_2} \hat{\Sigma}_{g_1 e_2} \hat{\Theta}(t) + \hat{\Theta}(t) \hat{\Sigma}^\dagger_{g_1 e_2} \hat{\Sigma}_{g_1 e_2} - 2\hat{\Sigma}_{g_1 e_2} \hat{\Theta}(t) \hat{\Sigma}^\dagger_{g e_{g_1 e_2}} \right\}}_{B} + \underbrace{\frac{\Gamma_-}{2} \left\{ \hat{\Sigma}^\dagger_{g_2 e_1} \hat{\Sigma}_{g_2 e_1} \hat{\Theta}(t) + \hat{\Theta}(t) \hat{\Sigma}^\dagger_{g_2 e_1} \hat{\Sigma}_{g_2 e_1} - 2\hat{\Sigma}_{g_2 e_1} \hat{\Theta}(t) \hat{\Sigma}^\dagger_{g_2 e_1} \right\}}_{C} \tag{9}$$

for the free Hamiltonian $\hat{H}_0^{FLA} = \hat{H}_A^{FLA} + \hat{H}_f^{FLA}$ defined in (F1) and the symmetrized atomic operators $\hat{\Sigma}_{g_i e_j}$ defined in (F6). The decay rates commensurate with the transitions are

$$\Gamma_{lm} = \frac{\mathbf{d}_{g_l e_l} \cdot \mathbf{d}_{g_m e_m} \omega_0^3}{3\pi \varepsilon_0 \hbar c^3}, \quad \Gamma_+ = \frac{|\mathbf{d}_{g_1 e_2}|^2 \omega_+^3}{3\pi \varepsilon_0 \hbar c^3}, \quad \Gamma_- = \frac{|\mathbf{d}_{g_2 e_1}|^2 \omega_-^3}{3\pi \varepsilon_0 \hbar c^3}, \tag{10}$$

where $\mathbf{d}_{g_l e_m}$ is the dipole moment of the $|e_m\rangle \to |g_l\rangle$ transitions and $l, m \in \{1, 2\}$. The first term of the Lindblad operator (9) denoted by $A$ describes the coupling to the $\omega_0$ photonic field through the $e_2 \to g_2$ and $g_1 \to e_1$ transitions (and the complementary pair), and contains the aforementioned cross-terms involving the two atomic operators. The $B$ and $C$ terms correspond to the $\omega_+$ and $\omega_-$ transitions, respectively, each describing a Dicke-like process with no common atomic states, so that the two processes are seemingly decoupled. However, a key property of (8) that stems from the degeneracy, is that the term $A$ (describing the virtual transitions) also couples the terms $B$ and $C$, since the transition operators in $A$ have common excited and ground states with $B$ and $C$. Therefore, the field of angular frequency $\omega_0$ couples between all four intra-atomic levels, thereby also between the $\omega_+$ and $\omega_-$ intra-atomic radiative transitions. This is the manifestation of the dynamics introduced by the degeneracy, that results in the intricate photonic entanglement scheme as we will show in the next subsection.

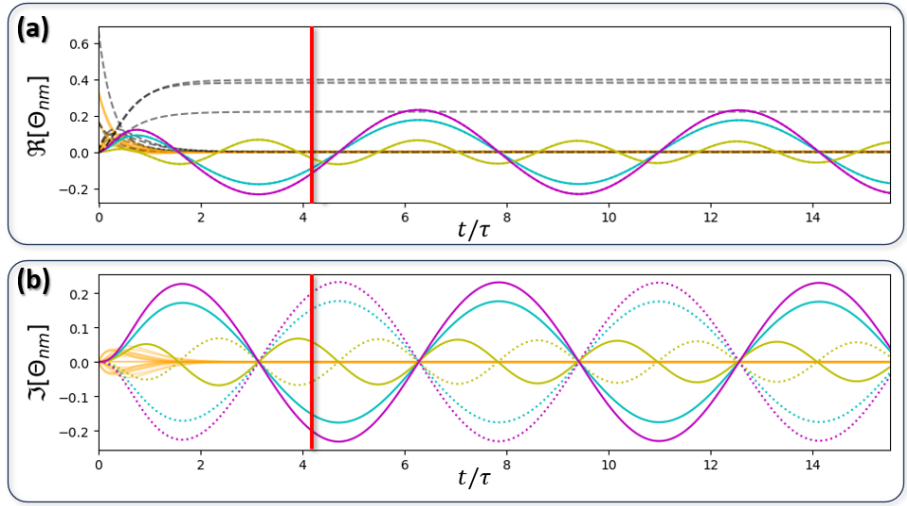

Figure 4 – Atomic density matrix entries $\hat{\theta}$ vs. time, for $N = 2$ and $\forall l, m$: $\Gamma_{lm} = \Gamma_- = 2\Gamma_+$. Time is given in units of $\tau = 1/\Gamma_-$. (a) Real part. (b) Imaginary part. Black dashed lines denote the population (diagonal) terms of $\hat{\theta}$. The vertical red line marks the time after which all terms containing an excited atom are smaller than $10^{-4}$ in their absolute value, to approximate steady-state. Full cyan, magenta and yellow lines denote the symmetrized ground-states coherences $\theta^{0011}_{0002}$, $\theta^{0020}_{0011}$ and $\theta^{0020}_{0002}$ respectively. Dotted cyan, magenta and yellow lines denote the Hermitian conjugates of the coherences. Orange lines denote all other terms of $\hat{\theta}$.

As mentioned, this coupling due to the inter- and intra-atom degeneracy in the $\omega_0$ transitions introduces atomic entanglement, driven by virtual transitions of quanta between the atoms in the ensemble. Consequentially, the FLA ensemble exhibits periodic beating of the atomic coherences in steady state (after the termination of the radiative process) due to the nonclassical correlations between the atomic ground state Fock numbers (Figure 4). By modifying the ratios of the decay rates of the different radiative transitions given in (10), we can control the probabilities of the evolution paths of the atomic system and their overlaps, and therefore the atomic entanglement. More details on the atomic dynamics and entanglement are given in APPENDIX G.

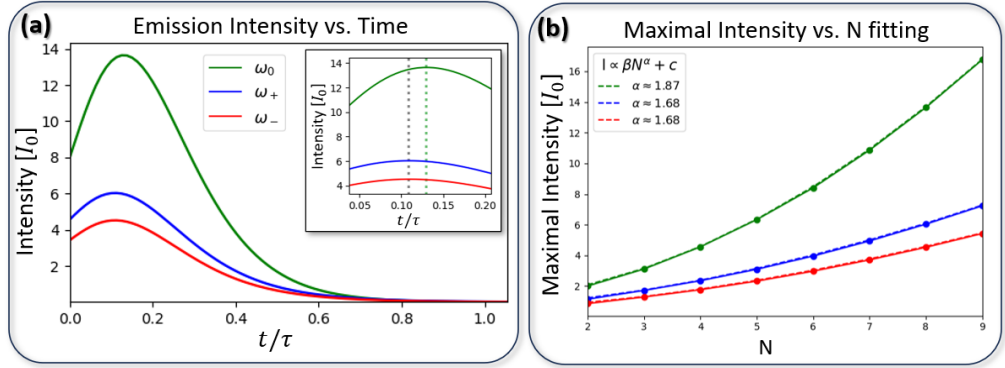

Figure 5 – (a) Intensity vs. time plots for $N = 8$ indistinguishable FLAs initialized in the fully excited symmetric state, for $\forall l, m$: $\Gamma_{lm} = \Gamma_+ = \Gamma_- = \Gamma$ and $\omega_+ = 4\omega_-/3$. Time is given in units of $\tau = 1/\Gamma$. Green, blue and red solid lines correspond to intensities of the $\omega_0$, $\omega_+$ and $\omega_-$ photonic modes, respectively. Inset: green vertical line marks the intensity peak time of the $\omega_0$ emission, gray vertical line marks the intensity peak times of the $\omega_\pm$ emissions. (b) Scaling laws for the intensity peaks of each of the three photonic modes with respect to the number of indistinguishable FLAs, color coded as in subfigure (a). Intensity is given in units of $I_0$ - intensity of a two-level atom with angular frequency and decay rate $\omega_0$ and $\Gamma$. Dots: intensity peaks values. Dashed lines: fitted curves of the form $\beta N^\alpha + c$. The parameter $\alpha$ denotes the exponential scaling parameter, and is found to be $\alpha \approx 1.68$ for the $\omega_+, \omega_-$ photonic modes and $\alpha \approx 1.87$ for the $\omega_0$ photonic mode.

We calculate the dynamics of FLA superradiance using (8). In Figure 5(a), we show the emission intensity from an ensemble of $N = 8$ indistinguishable FLA, where we choose for simplicity the parameters $\Gamma_{lm} = \Gamma_\pm = \Gamma$ for any $l, m$ - meaning identical intrinsic decay rates for all transitions. The three competing emission processes exhibit Dicke superradiance behavior, namely, a sharp peak in intensity followed by a rapid decay. It is interesting to note that the highest peak corresponding to the emission of the $\omega_0$ photons has the longest delay time (Figure 5a inset); this is due to the aforementioned virtual interactions between the two degenerate intra-atomic paths that interfere and affect the population evolution. Figure 5(b) shows the scaling law of the maximal peak intensity as a function of the number of atoms. The graphs show the result of the numerical calculation of the respective peaks for each of the three photon modes (in full lines), compared to functions of the form $\beta N^\alpha + c$ (in dashed lines) that are fitted to an excellent agreement. We see that the parameter $\alpha$ is identical for the $\omega_+$ and $\omega_-$ transitions, but larger for the $\omega_0$ transition. This is expected since all intrinsic atomic decay rates are identical, and the $\omega_0$ transition has more possible paths than the other two transitions.

### B. Photon entanglement from the superradiant FLA ensemble

We study the entanglement of the multiphoton state emanated from a superradiating FLA ensemble and elucidate how the internal atomic degeneracies affect it. The entanglement between the three photonic frequency modes emitted from an FLA ensemble can be quantified in a manner similar to what was done for the V-atom case (Section II.A) using the Weisskopf-Wigner approximation. The increased complexity of a single FLA with respect to a single V-atom is reflected in the complexity of the state which we will denote $|\psi_{FLA}\rangle$,

$$|\psi_{FLA}\rangle(t) = a_2(t)|1,0,0,0\rangle_A|0,0,0\rangle_f + a_1(t)|0,1,0,0\rangle_A|0,0,0\rangle_f$$
$$+ \sum_{\mathbf{k}}\left[|0,0,1,0\rangle_A\left(b_{22,\mathbf{k}}(t)|0,1_\mathbf{k},0\rangle_f + b_{21,\mathbf{k}}(t)|1_\mathbf{k},0,0\rangle_f\right) + |0,0,0,1\rangle_A\left(b_{12,\mathbf{k}}(t)|0,0,1_\mathbf{k}\rangle_f + b_{11,\mathbf{k}}(t)|0,1_\mathbf{k},0\rangle_f\right)\right] \quad (11)$$

Since we are interested in the photonic state after the termination of the emission process, i.e. $t \to \infty$, we have

$$|\psi_{FLA}\rangle(\infty) = \sum_{\mathbf{k}}\left[|0,0,1,0\rangle_A\left(b_{22,\mathbf{k}}(\infty)|0,1_\mathbf{k},0\rangle_f + b_{21,\mathbf{k}}(\infty)|1_\mathbf{k},0,0\rangle_f\right)\right.$$
$$\left. + |0,0,0,1\rangle_A\left(b_{12,\mathbf{k}}(\infty)|0,0,1_\mathbf{k}\rangle_f + b_{11,\mathbf{k}}(\infty)|0,1_\mathbf{k},0\rangle_f\right)\right] \quad (12)$$

where, as before, the infinite time notation $(\infty)$ shall be henceforth omitted. The coefficients $b_{ij,k}(t)$ are calculated from Schrödinger's equations and are detailed in APPENDIX H. In contrast to the V-atom ensemble case, here there is cross-entanglement between the atomic and photonic states, thus the final photonic density matrix obtained from (12) after tracing out the atomic degrees of freedom is not pure but rather a mixture of pure states. This originates from the existence of several atomic ground states into which the system can evolve, similar to the case of a $\Lambda$-type atom. However, unlike in the $\Lambda$-atom case, each photonic state in the mixture is entangled due to the internal degeneracy of the FLA, as is evident from the non-diagonal photonic density matrix at long time $t \to \infty$ derived from (12),

$$\rho_f^{FLA} = \sum_{\mathbf{k},\mathbf{k}'}\left[\left(b_{22,\mathbf{k}}|0,1_\mathbf{k},0\rangle + b_{21,\mathbf{k}}|1_\mathbf{k},0,0\rangle\right)\left(b_{22,\mathbf{k}'}^*\langle 0,1_{\mathbf{k}'},0| + b_{21,\mathbf{k}'}^*\langle 1_{\mathbf{k}'},0,0|\right)\right.$$
$$\left. + \left(b_{12,\mathbf{k}}|0,0,1_\mathbf{k}\rangle + b_{11,\mathbf{k}}|0,1_\mathbf{k},0\rangle\right)\left(b_{12,\mathbf{k}'}^*\langle 0,0,1_{\mathbf{k}'}| + b_{11,\mathbf{k}'}^*\langle 0,1_{\mathbf{k}'},0|\right)\right]. \quad (13)$$

We focus on the simple setup where all decay rates are equal. In this case, we can define the basis of orthonormal vectors $|010\rangle\rangle = \sqrt{2}\sum_{\mathbf{k}}\tilde{b}_{\mathbf{k}}|0,1_\mathbf{k},0\rangle$, $|100\rangle\rangle = \sqrt{2}\sum_{\mathbf{k}}\tilde{b}_{\mathbf{k}}|1_\mathbf{k},0,0\rangle$, $|001\rangle\rangle = \sqrt{2}\sum_{\mathbf{k}}\tilde{b}_{\mathbf{k}}|0,0,1_\mathbf{k}\rangle$ with the coefficients $\tilde{b}_{\mathbf{k}}\alpha \equiv b_{11,\mathbf{k}} = b_{21,\mathbf{k}}$ and $\tilde{b}_{\mathbf{k}}\beta \equiv b_{22,\mathbf{k}} = b_{12,\mathbf{k}}$ and write the photonic density matrix (13) as

$$\rho_f^{FLA} = \frac{1}{2}|\varphi_{f,(1)}^{2-FLA}\rangle\langle\varphi_{f,(1)}^{2-FLA}| + \frac{1}{2}|\varphi_{f,(2)}^{2-FLA}\rangle\langle\varphi_{f,(2)}^{2-FLA}| \quad (14)$$

With

$$|\varphi_{f,(1)}^{FLA}\rangle = \alpha|100\rangle\rangle + \beta|010\rangle\rangle$$
$$|\varphi_{f,(2)}^{FLA}\rangle = \alpha|010\rangle\rangle + \beta|001\rangle\rangle \quad (15)$$

The coefficients $\alpha$ and $\beta$ are the initial atomic excitation coefficients defined in (11), such that $\alpha = a_1(0)$ and $\beta = a_2(0)$ and $|\alpha|^2 + |\beta|^2 = 1$. The density matrix (14) describes a mixture of two pure states (15), each describing the photonic state emitted from a non-degenerate V-atom with angular frequencies $\omega_-, \omega_0$ and $\omega_0, \omega_+$ respectively.

We are also interested in the photonic density matrix $\rho_f^{2-FLA}$ for two indistinguishable FLAs (composing the smallest possible ensemble), each initially excited to the atomic state $\alpha|e_1\rangle + \beta|e_2\rangle$. Analogously to the single FLA case, we define a basis consisting of the six orthonormal state vectors describing the combinations of two photons in three modes, namely $|200\rangle\rangle$, $|110\rangle\rangle$, $|020\rangle\rangle$, $|101\rangle\rangle$, $|011\rangle\rangle$ and $|022\rangle\rangle$ – and obtain (APPENDIX H)

$$\rho_f^{2-FLA} = \frac{1}{3}|\varphi_{f,(0)}^{2-FLA}\rangle\langle\varphi_{f,(0)}^{2-FLA}| + \frac{1}{3}|\varphi_{f,(1)}^{2-FLA}\rangle\langle\varphi_{f,(1)}^{2-FLA}| + \frac{1}{3}|\varphi_{f,(2)}^{2-FLA}\rangle\langle\varphi_{f,(2)}^{2-FLA}|, \tag{16}$$

which similarly to (14) also describes a uniform mixture of the pure photonic states

$$|\varphi_{f,(0)}^{2-FLA}\rangle = a_0|200\rangle\rangle + a_1|110\rangle\rangle + a_2|020\rangle\rangle$$
$$|\varphi_{f,(1)}^{2-FLA}\rangle = a_0|110\rangle\rangle + 2^{-1/2}a_1(|020\rangle\rangle + |101\rangle\rangle) + a_2|011\rangle\rangle \tag{17}$$
$$|\varphi_{f,(2)}^{2-FLA}\rangle = a_0|020\rangle\rangle + a_1|011\rangle\rangle + a_2|002\rangle\rangle$$

Here, $a_0 = N\alpha^2$, $a_1 = 2N\alpha\beta$ and $a_2 = N\beta^2$ with the normalization constant $N = (|\alpha|^4 + 4|\alpha\beta|^2 + |\beta|^4)^{-1/2}$. The first and third states correspond to the photonic states emanated from a pair of indistinguishable V-atoms either with angular frequencies $\omega_-, \omega_0$ or $\omega_0, \omega_+$, respectively. The second state contains a superposition of photonic states in all three angular frequencies.

As mentioned before, the internal degeneracy in the FLA energy ladder has a profound contribution to the photonic entanglement. We first note that in the single FLA case, the $\omega_+, \omega_-$ modes are entangled to one another only through the $\omega_0$ mode. This can be seen from the reduced photonic density matrix obtained by tracing out the $\omega_0$ mode from (14), which is diagonal,

$$\rho_{f,0}^{FLA} = \frac{1}{2}\begin{pmatrix} 1 & 0 & 0 \\ 0 & |\alpha|^2 & 0 \\ 0 & 0 & |\beta|^2 \end{pmatrix} \tag{18}$$

and therefore, there are no coherences between the $\omega_+$ and $\omega_-$ modes. In contrast, tracing out any one of the other two modes from (14) results in a non-diagonal density matrix.

Contrary to the single FLA case, tracing out the $\omega_0$ mode from the two FLAs photonic density matrix (16) yields a density matrix which is not diagonal, namely

$$\rho_{f,0}^{2-FLA} = \frac{1}{3}\begin{pmatrix} |a_0|^2 & 0 & 0 & 0 & 0 & 0 \\ 0 & |a_0|^2 + |a_1|^2 & 0 & 0 & a_0 a_2^* & 0 \\ 0 & 0 & |a_0|^2 + 2^{-1}|a_1|^2 + |a_2|^2 & 0 & 0 & 0 \\ 0 & 0 & 0 & 2^{-1}|a_1|^2 & 0 & 0 \\ 0 & a_2 a_0^* & 0 & 0 & |a_1|^2 + |a_2|^2 & 0 \\ 0 & 0 & 0 & 0 & 0 & |a_2|^2 \end{pmatrix}. \tag{19}$$

The coherence terms originate from interference of the two paths consisting of either one $\omega_+$ and one $\omega_0$ photon or one $\omega_-$ and one $\omega_0$ photon; in other words, the entanglement between the $\omega_+$ and $\omega_-$ modes in the two FLAs case is conferred due to the multiplicity of paths by which a single $\omega_0$ photon emission occurs.

In order to quantify this entanglement, we calculate the three conditional entanglement entropies [104] for the three photonic frequency modes $S_f^{FLA}(\mathbf{0},+|-) = S_f^{FLA} - S_{f,-}^{FLA}$, $S_f^{FLA}(-,+|\mathbf{0}) = S_f^{FLA} - S_{f,\mathbf{0}}^{FLA}$ and $S_f^{FLA}(-,\mathbf{0}|+) = S_f^{FLA} - S_{f,+}^{FLA}$. The calculations for the one and two indistinguishable FLAs cases are outlined in APPENDIX H. In Figure 6 we illustrate the conditional entanglement entropies for a single FLA and for two indistinguishable FLAs as a function of the modulus squared of their atomic excitation coefficient $|\alpha|^2$.

All conditional entropies are negative; thus, it is demonstrated that each photonic frequency mode is entangled to the other two. We note that in the limiting cases $|\alpha|^2 = 0$ or $|\alpha|^2 = 1$, each corresponding to a Λ-atom configuration, all three conditional entropies are zero – meaning no evidence of entanglement – in agreement with literature [113]. This calculation can be extended to any number of FLAs.

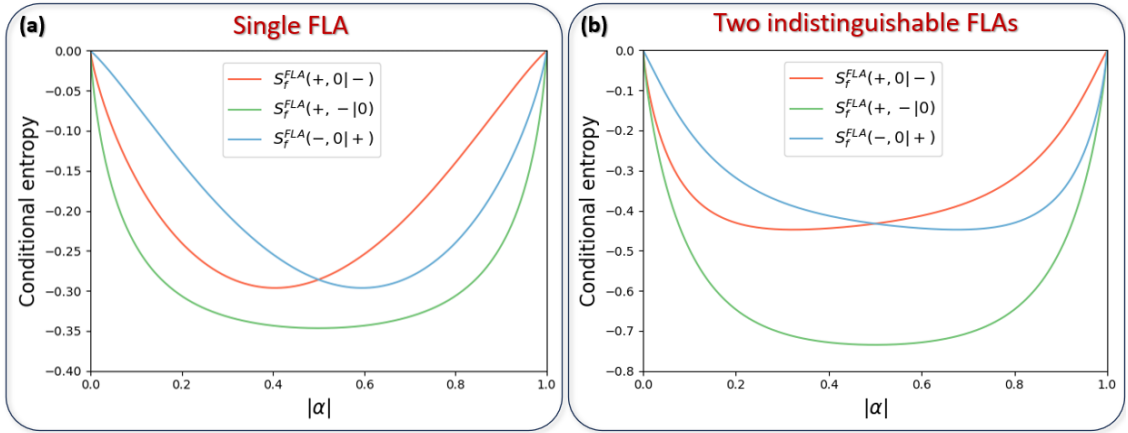

Figure 6 – Conditional entanglement entropies for each of the three emitted photonic modes vs. the absolute value of the atomic excitation coefficient, $|\alpha|$. (a) A single FLA. (b). Two indistinguishable FLAs.

We now analyze the three-mode Wigner distribution of the tripartite photonic state. Analogously to Section II.A, for each Wigner distribution we show fifteen 2D maps representing all phase-space coordinate pairs spanning the 6D space, where in each map the remaining four phase-space coordinates are set to zero. In Figure 7 we show the Wigner distribution of two indistinguishable FLAs with equal internal radiative transition rates, for the case $\alpha = \beta = 2^{-1/2}$. Each of the 2D projections representing a single-mode is symmetric with respect to the origin, with two concentric ring-like features indicating Fock number occupancy of up to two for each mode. It is evident that the phase-space coordinates corresponding to the $\omega_-$ and $\omega_+$ modes are correlated to the phase-space coordinates of the $\omega_0$ mode, indicating non-separability of the Wigner distribution and therefore entanglement between those mode pairs. Correlations also exist directly between the $\omega_-$ and $\omega_+$, however they are less visible as is evident from the more axial-symmetric 2D projections of the Wigner distribution. This therefore indicates that all three modes are entangled to one another, as discussed above and as expected from the internal atomic structure. Derivation of the three-modal Wigner distribution and further discussion is presented in APPENDIX D.

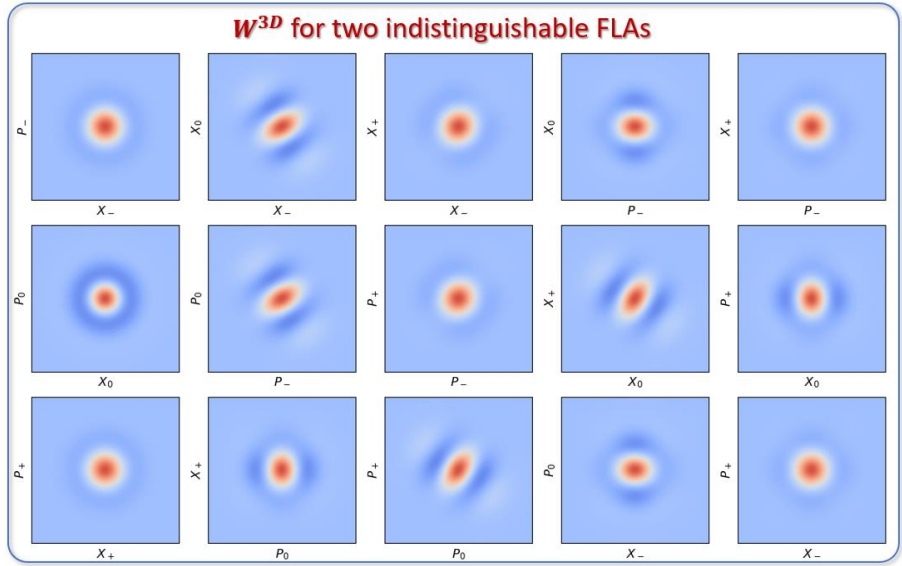

Figure 7 – Three-modal Wigner distribution $W^{3D}(X_-, P_-, X_0, P_0, X_+, P_+)$ for two indistinguishable FLAs with equal radiative decay rates. Both atoms are excited to a superposition state $\alpha = \beta = \sqrt{1/2}$. Intermodal correlations between all three modes are evident from correlations between phase-space coordinates corresponding to different modes, manifesting as 2D Wigner distribution projections which are not axially-symmetrical.

In summary, we have shown that superradiance from an ensemble of FLAs emits a tri-partite multiphoton state that exhibits entanglement between each mode to the two other modes, originating from virtual transitions stemming from the internal degeneracy of the atoms that is absent in the nondegenerate V-atoms ensemble. Moreover, similarly to the nondegenerate V-atoms case, this photon entanglement is also highly dependent on the coherent excitation of the FLAs. We have hereby demonstrated that the ingredient of internal degeneracy adds yet another mechanism of generation of multiphoton entanglement.

## IV. MODE-INDEPENDENT ENTANGLEMENT

We remark here on another important property of the multilevel atom ensembles, pertaining to the dependence of the photonic entanglement on the chosen measurement basis. The significance of this property is that some entangled quantum states can be disentangled by a proper choice of basis, whereas other entangled quantum states remain entangled under any unitary transformation. The latter states possess an intrinsic form of entanglement that cannot be nullified by passive optical elements, making it an important quantum resource for various communication protocols [106,111].

For the nondegenerate V-atom ensemble, the entanglement or separability of final photonic state depends on the choice of basis of modes – i.e., one can apply a unitary transformation to the photonic state to make it separable (e.g., defining a dressed photonic creation operator that creates a dressed mode which is an addition or subtraction of both modes). In contrast, for the FLA ensemble case there is no basis in which the photonic state is separable. The difference stems from the fact that in the V-atom case each creation of a photon in a mode is associated with annihilation of a different excited atomic level, whereas in the FLA case annihilation of either excited atomic level contributes to the creation of photons in the $\omega_0$ mode (superimposed with either the $\omega_+$ or $\omega_-$ modes). Therefore, since both photon creation operators in the FLA ensemble will involve creating three photon modes in total, with both operators creating $\omega_0$-mode photons, they are not orthogonal and the photonic state is not separable under any change of basis.

To see that, we write the final photonic state in the most general way as a series of creation operators $\hat{a}^\dagger_{\omega_1}$ and $\hat{a}^\dagger_{\omega_2}$, each creating a photon packet with Lorentzian distribution around $\omega_1$ and $\omega_2$ respectively, acting on the initial vacuum state. This will result in the photonic state $|\varphi_{N-V}\rangle(\infty) = \prod_{i=1}^{N}(\alpha \hat{a}^\dagger_{\omega_1} + \beta \hat{a}^\dagger_{\omega_2})|0,0\rangle_f$. If we define the symmetric and antisymmetric dressed photonic creation operators by $\hat{a}^\dagger_s \equiv \alpha \hat{a}^\dagger_{\omega_1} + \beta \hat{a}^\dagger_{\omega_2}$ and $\hat{a}^\dagger_a \equiv \alpha \hat{a}^\dagger_{\omega_1} - \beta \hat{a}^\dagger_{\omega_2}$, respectively, then by substituting this and choosing a basis of symmetric/antisymmetric modes $|0,0\rangle_{S/AS}$, we obtain the bipartite state $|\varphi_{N-V}\rangle_{S/AS}(\infty) = (\hat{a}^\dagger_s)^N |0,0\rangle_{S/AS}$ which is separable and thus not entangled.

In contrast, for an ensemble of indistinguishable FLA, we saw in Section III.B that the final photonic state is a mixture of $N + 1$ states commensurate with different final atomic states with different energies, each corresponding to a different combination of Fock states with $N$ emitted photons in total. Considering the final photonic state,

$$\rho_f^{N-FLA} = \sum_{q=0}^{N} p_q \left| \varphi_{f,(q)}^{N-FLA} \right\rangle \left\langle \varphi_{f,(q)}^{N-FLA} \right| \tag{20}$$

with $\sum_{q=0}^{N} p_q = 1$ classical probabilities and the pure states

$$\left| \varphi_{f,(q)}^{N-FLA} \right\rangle = \mathcal{N}_q \hat{P}_q \left\{ \prod_{v \in \{\mathbf{q}\}} \left( f_v^{\mathbf{q}} \hat{a}_{\omega_0}^\dagger + g_v^{\mathbf{q}} \hat{a}_{\omega_+}^\dagger \right) \prod_{v \in \{\bar{\mathbf{q}}\}} \left( h_v^{\mathbf{q}} \hat{a}_{\omega_-}^\dagger + w_v^{\mathbf{q}} \hat{a}_{\omega_0}^\dagger \right) \right\} \left| 0,0,0 \right\rangle \tag{21}$$

with $\{\mathbf{q}\}$ a subset of q different integers from the set $\{0,...,N\}$, $\{\bar{\mathbf{q}}\}$ is the complementary subset of the remaining $N+1-q$ integers $\hat{P}_q$ the permutation operator on all possible subsets $\{\mathbf{q}\}$ containing q integers and $\mathcal{N}_q$ is a normalization constant for $\left| \varphi_{f,(q)}^{N-FLA} \right\rangle$ and the constants $f_v^{\mathbf{q}}, g_v^{\mathbf{q}}, h_v^{\mathbf{q}}$ and $w_v^{\mathbf{q}}$ are derived from the Weisskopf-Wigner method, the index $v$ encapsulating combinatoric factors corresponding to the atomic occupation numbers pertaining to the photon emissions as outlined in APPENDIX H. In this case, by defining $\hat{a}_{H,v}^{\mathbf{q}\dagger} = f_v^{\mathbf{q}} \hat{a}_{\omega_0}^\dagger + g_v^{\mathbf{q}} \hat{a}_{\omega_+}^\dagger$ and $\hat{a}_{L,v}^{\mathbf{q}\dagger} = h_v^{\mathbf{q}} \hat{a}_{\omega_-}^\dagger + w_v^{\mathbf{q}} \hat{a}_{\omega_0}^\dagger$, it is clear that any two of these operators are neither orthogonal nor parallel and so no unitary transformation exists under which the photonic state (20) is separable [106]. Therefore, the intra-atomic degeneracy present in the case of indistinguishable FLAs introduces entanglement between the photonic Fock states that is mode-independent.

This is elucidated by considering the examples shown in Section III.B. For the single FLA case, each of the pure states (15) consists of a single creation operator $\hat{a}_{H,v}^\dagger = \alpha \hat{a}_{\omega_0}^\dagger + \beta \hat{a}_{\omega_+}^\dagger$ or $\hat{a}_{L,v}^\dagger = \alpha \hat{a}_{\omega_-}^\dagger + \beta \hat{a}_{\omega_0}^\dagger$. Each is therefore separable under a unitary transformation, however no unitary transformation exists under which both pure states become simultaneously separable. For two indistinguishable FLAs the pure states (17) are obtained by operating $\left( \hat{a}_{H,v}^\dagger \right)^2$, $\hat{a}_{H,v}^\dagger \hat{a}_{L,v}^\dagger$ and $\left( \hat{a}_{L,v}^\dagger \right)^2$, respectively, on the vacuum. Although the first and third pure states can each be independently transformed to a separable state, the second pure state cannot, and therefore the state (16) is also entangled under any unitary transformation.

## V. CONCLUSIONS AND OUTLOOK

The main goal of this paper was to develop schemes for generation of superradiant emission of photons that are entangled in their modal degree of freedom, thereby unifying two distinct subfields of quantum optics. In order to invoke photonic entanglement in a superradiance process, which does not occur in Dicke superradiance of two-level atoms – we proposed and analyzed ensembles of identical, noninteracting multilevel atoms. We studied in detail two types of ensembles of multilevel atoms – V-atoms and FLAs. Superradiance dynamics and photonic entanglement were demonstrated in both types of ensembles.

We found that the photonic states emitted from those superradiating samples are entangled. The photon entanglement originates from two distinct mechanisms: (i) the indistinguishability of the atoms – namely, the entanglement between the symmetrized combination atomic states is imprinted onto the photonic states; and (ii) the internal energetic structure of each individual atom, e.g., degenerate transitions involving different atomic levels. The degenerate intra-atomic transitions induce virtual interactions that invoke correlations between the atoms as well as the emitted photons.

The nondegenerate V-atom ensemble emits photonic states that mimic the initial atomic excitation. Therefore, if the ensemble of indistinguishable atoms is excited to a (symmetrized) combination state, the final photonic state will have the same entanglement negativity as the initial atomic excitation. This result means that a source of photons with controlled entanglement can be realized from a dense sample of identical emitters with a V-shaped energy ladder by exciting each of its constituent atoms to a prescribed superposition state.

The effect of virtual transitions on the entanglement properties manifests in an ensemble of FLAs. We showed that the virtual transitions stemming from the internal atomic degeneracy introduces correlations between the atomic states that exhibit beating long after the ensemble has emitted all photons. These correlations are manifested as fringes in the Wigner distributions describing the atomic state of the decayed system. Furthermore, the entanglement of the photonic state is affected from both the virtual transitions acting between atoms in the ensemble, and the initial excitation scheme of the ensemble. We emphasize that an entangled photon source based on FLAs can be also realized and controlled by its initial excitation.

Another important aspect of the entanglement of photons emitted from the FLA ensemble, is that it exists regardless of the choice of measurement basis. This notion is significant as mode-independent entanglement is another type of entanglement resource that cannot be lost under any linear transformation.

The work presented here demonstrates the possibility to utilize the superradiance phenomena as a source of entangled photons by engineering specialized multilevel atoms. It is important to note that although the superradiant emission from the FLAs ensemble produces a mixture of entangled photonic states – its conditional entropy is negative. This indicates that the entanglement persists despite the mixing, allowing the photonic states to be distilled further and used in applications that require entanglement as a resource, such as quantum communication and sensing. In light of our findings, many other appealing research possibilities are merited, such as more complicated atoms; effects of different atomic excitation schemes on the properties of the emitted photons; driven superradiance in multilevel atom ensembles; and introduction of geometry and dimensionality.

**ACKNOWLEDGEMENTS**

A.S. acknowledges fellowship by the Helen Diller Quantum Center at the Technion, Israel Institute of Technology.

**APPENDIX A – MASTER EQUATION FOR NON-DEGENERATE V-ATOMS**

Here we describe the temporal evolution of a system composed of $N$ indistinguishable V-atoms, each consisting of two excited levels $|e_1\rangle$ and $|e_2\rangle$ and a ground level $|g\rangle$. For this part, we follow the method of evolving the atomic system's density matrix (after tracing out the photonic degrees of freedom) under the Born-Markov approximation as was done in refs. [96-99] and others resulting in a system of coupled first-order differential equations for the density matrix entries in the Schrödinger picture

$$\frac{d}{dt}\hat{\Theta}(t) = -\frac{i}{\hbar}\left[\hat{H}_0, \hat{\Theta}(t)\right] - \hat{L}\left\{\hat{\Theta}(t)\right\} \tag{A1}$$

where $\hat{\Theta}(t) = Tr_f(\hat{\rho}(t))$ is the atomic part of density matrix obtained by tracing out the fields degrees of freedom of the total system density matrix $\hat{\rho}(t)$. Here, $\hat{H}_0$ is the free Hamiltonian and $\hat{L}\{\hat{X}\}$ is the Lindblad operator acting on some density matrix $\hat{X}$. We start by writing the Hamiltonian

$$\hat{H} = \hat{H}_0 + \hat{H}_I \tag{A2}$$

consisting of the free Hamiltonian $\hat{H}_0 = \hat{H}_A + \hat{H}_f$ and the interaction Hamiltonian $\hat{H}_I$, where

$$\hat{H}_A = \hbar\omega_1 \sum_{i=1}^{N} \hat{n}_{e_1}^i + \hbar\omega_2 \sum_{i=1}^{N} \hat{n}_{e_2}^i \tag{A3}$$

$$\hat{H}_f = \hbar \sum_{\mathbf{k}} \omega_{\mathbf{k}} \left( \hat{a}_{1,\mathbf{k}}^{\dagger} \hat{a}_{1,\mathbf{k}} + \hat{a}_{2,\mathbf{k}}^{\dagger} \hat{a}_{2,\mathbf{k}} \right) \tag{A4}$$

are the atomic and field Hamiltonians, respectively. Here, $\hbar\omega_1$ ($\hbar\omega_2$) is the energy gap between the excited state $e_1$ ($e_2$) and the ground state $g$. The atomic operators $\hat{\sigma}_{ge_j}^i = |g\rangle_{ii}\langle e_j|$ and $\hat{n}_j^i = |e_j\rangle_{ii}\langle e_j|$ acts on the $i$'th atom for the transition $j \in \{1,2\}$. The photonic annihilation operator $\hat{a}_{j,\mathbf{k}}$ annihilates a single photon of angular frequency $\omega_j$ in the mode characterized by the wave vector $\mathbf{k}$ and obeys the commutation relations

$$\left[\hat{a}_{j,\mathbf{k}}, \hat{a}_{j',\mathbf{k}'}^{\dagger}\right] = \delta_{j,j'} \delta_{\mathbf{k},\mathbf{k}'}. \tag{A5}$$

We have implicitly assumed in the above that the modal distributions of the photon packets corresponding to the photon emission associated with the $e_1 \to g$ and $e_2 \to g$ transitions do not overlap. The interaction Hamiltonian in the interaction picture takes the form

$$\hat{H}_I = \hbar \sum_{i=1}^{N} \sum_{\mathbf{k}} \left\{ g_{1,\mathbf{k}}^{i}(\mathbf{r}) \hat{a}_{1,\mathbf{k}}^{\dagger} \hat{\sigma}_{ge_1}^{i} e^{i(\omega_\mathbf{k} - \omega_1)t} + g_{2,\mathbf{k}}^{i}(\mathbf{r}) \hat{a}_{2,\mathbf{k}}^{\dagger} \hat{\sigma}_{ge_2}^{i} e^{i(\omega_\mathbf{k} - \omega_2)t} + h.c. \right\} \tag{A6}$$

with the coupling coefficients for the $j$'th transition of the $i$'th, atom

$$g_{j,\mathbf{k}}^{i}(\mathbf{r}_i) = -i\sqrt{\frac{2\pi c k}{V}} \mathbf{d}_j^{i} \cdot \mathbf{e}_\mathbf{k} e^{i\mathbf{k}\cdot\mathbf{r}_i} \tag{A7}$$

where $\mathbf{d}_j^{i}$ is the dipole moment vector of the $j$'th transition in the $i$'th atom, $\mathbf{r}_i$ is the position of the atom, $\mathbf{e}_\mathbf{k}$ is the unit vector in the direction of the polarization of the photon in the mode $\mathbf{k}$, $c$ is the speed of light in vacuum and $V$ is the modal volume, which will be taken to infinity when considering emission in free space.

We apply second-order perturbation theory on the atomic part of the density matrix in order to develop the atomic master equation, from which the evolution of the emission process can be obtained. As in previous treatments of the problem, we consider the Born-Markov approximation [96-99], meaning that during the emission process, the photons do not act back on the atom so that the atomic system at any time can be considered to be in vacuum, and that the atom-field correlation time is negligibly short. Under these assumptions we have the integrodifferential equation for the atomic density matrix in the interaction picture (denoted by the "$I$" superscript),

$$\frac{d}{dt}\hat{\Theta}^{I}(t) = -\frac{1}{\hbar^2} Tr_f \left\{ \int_0^\infty d\tau \left[ \hat{H}_I(t), \left[ \hat{H}_I(t-\tau), \hat{\Theta}^{I}(t-\tau) \otimes \hat{\Phi}(0) \right] \right] \right\} \tag{A8}$$

with the photonic density matrix $\hat{\Phi}(0) = |0,0\rangle\langle 0,0|$. Substituting (A6) in (A8) in yields

$$\frac{d}{dt}\hat{\Theta}^{I}(t) = -Tr_f \left\{ \int_0^\infty d\tau \left[ \sum_{i=1}^{N} \sum_{\mathbf{k}} \left\{ g_{1,\mathbf{k}}^{i}(\mathbf{r}) \hat{a}_{1,\mathbf{k}}^{\dagger} \hat{\sigma}_{ge_1}^{i} e^{i(\omega_\mathbf{k} - \omega_1)t} + g_{2,\mathbf{k}}^{i}(\mathbf{r}) \hat{a}_{2,\mathbf{k}}^{\dagger} \hat{\sigma}_{ge_2}^{i} e^{i(\omega_\mathbf{k} - \omega_2)t} \right. \right.$$
$$+ g_{1,\mathbf{k}}^{i*}(\mathbf{r}) \hat{a}_{1,\mathbf{k}} \hat{\sigma}_{ge_1}^{i\dagger} e^{-i(\omega_\mathbf{k} - \omega_1)t} + g_{2,\mathbf{k}}^{i*}(\mathbf{r}) \hat{a}_{2,\mathbf{k}} \hat{\sigma}_{ge_2}^{i\dagger} e^{-i(\omega_\mathbf{k} - \omega_2)t} \right\}$$
$$, \left[ \sum_{i=1}^{N} \sum_{\mathbf{q}} \left\{ g_{1,\mathbf{q}}^{i}(\mathbf{r}) \hat{a}_{1,\mathbf{q}}^{\dagger} \hat{\sigma}_{ge_1}^{i} e^{i(\omega_\mathbf{q} - \omega_1)(t-\tau)} + g_{2,\mathbf{q}}^{i}(\mathbf{r}) \hat{a}_{2,\mathbf{q}}^{\dagger} \hat{\sigma}_{ge_2}^{i} e^{i(\omega_\mathbf{q} - \omega_2)(t-\tau)} \right. \right.$$
$$\left. \left. + g_{1,\mathbf{q}}^{i*}(\mathbf{r}) \hat{a}_{1,\mathbf{q}} \hat{\sigma}_{ge_1}^{i\dagger} e^{-i(\omega_\mathbf{q} - \omega_1)(t-\tau)} + g_{2,\mathbf{q}}^{i*}(\mathbf{r}) \hat{a}_{2,\mathbf{q}} \hat{\sigma}_{ge_2}^{i\dagger} e^{-i(\omega_\mathbf{q} - \omega_2)(t-\tau)} \right\}, \hat{\Theta}^{I}(t-\tau) \otimes \hat{\Phi}(0) \right] \right] \tag{A9}$$

Using the property

$$Tr(\Phi(0)\hat{a}_{j,\mathbf{k}}^{\dagger}\hat{a}_{j',\mathbf{k}'}) = 0, \qquad Tr(\Phi(0)\hat{a}_{j,\mathbf{k}}\hat{a}_{j',\mathbf{k}'}) = 0,$$
$$Tr(\Phi(0)\hat{a}_{j,\mathbf{k}}^{\dagger}\hat{a}_{j',\mathbf{k}'}^{\dagger}) = 0, \quad Tr(\Phi(0)\hat{a}_{j,\mathbf{k}}\hat{a}_{j',\mathbf{k}'}^{\dagger}) = \delta_{j,j'}\delta_{\mathbf{k},\mathbf{k}'} \tag{A10}$$

derived from (A5), and emphasizing that the photon packets centered around the angular frequency $\omega_1$ do not overlap with those centered around $\omega_2$, a straightforward calculation gives

$$\frac{d}{dt}\hat{\Theta}^{I}(t) = -\int_0^\infty d\tau \sum_{j=\{1,2\}} \sum_{\mathbf{k}} \sum_{i,i'=1}^{N} \left\{ g_{j,\mathbf{k}}^{i}(\mathbf{r}_i) g_{j,\mathbf{k}}^{i'*}(\mathbf{r}_{i'}) e^{-i\omega_\mathbf{k}\tau} \left\{ e^{i\omega_j\tau} \left[ \hat{\sigma}_{ge_j}^{i\dagger}, \hat{\sigma}_{ge_j}^{i'} \hat{\Theta}^{I}(t-\tau) \right] + e^{-i\omega_j\tau} \left[ \hat{\sigma}_{ge_j}^{i}, \hat{\sigma}_{ge_j}^{i'\dagger} \hat{\Theta}^{I}(t-\tau) \right] \right\} \right.$$
$$\left. + g_{j,\mathbf{k}}^{i*}(\mathbf{r}_i) g_{j,\mathbf{k}}^{i'}(\mathbf{r}_{i'}) e^{i\omega_\mathbf{k}\tau} \left\{ e^{-i\omega_j\tau} \left[ \hat{\sigma}_{ge_j}^{i}, \hat{\sigma}_{ge_j}^{i'\dagger} \hat{\Theta}^{I}(t-\tau) \right] + e^{i\omega_j\tau} \left[ \hat{\sigma}_{ge_j}^{i\dagger}, \hat{\sigma}_{ge_j}^{i'} \hat{\Theta}^{I}(t-\tau) \right] \right\} \right\} \tag{A11}$$

We now consider the V-atoms to be indistinguishable. This notion entails treatment of all atoms as being located in a single point in space with identical dipole moments, such that $g_{j,\mathbf{k}}^{i}(\mathbf{r}_i) = g_{j,\mathbf{k}}^{i}(0) \equiv g_{j,\mathbf{k}}$ for $\forall i,j$ (although various symmetric structures can in principle be considered [61], here we will treat only the simplified case of atoms at a single point in space), as well as re-definition of the atomic operators in the following manner. Because the atoms are indistinguishable, one cannot assign an atomic index to the state of any particular atom; one can, however, count the total number of levels occupied across the entire ensemble. Consequently we move to work in the Fock number basis of atomic levels in which the atomic state vector is defined as $|n_{e_1}, n_{e_2}, n_g\rangle$ for the number of excitations of levels $e_1$, $e_2$ and $g$, respectively. This amounts to a symmetrization operation in the basis of discernible atoms,

$$\hat{S} = \frac{1}{\sqrt{N!}} \sum_i^N \hat{P}_i \tag{A12}$$

where $\hat{P}_i$ is the permutation operator, ordering a vector of $N$ identical particles in the $i$'th permutation. The sums of the indexed atomic operators in (A11) are therefore replaced with operators that annihilate a quantum from one Fock number and create a quantum in another one [81-84,110], or explicitly,

$$\begin{aligned}
\hat{\Sigma}_{ge_2} |n_{e_2}, n_{e_1}, n_g\rangle &= \sqrt{n_{e_2}(n_g+1)} |n_{e_2}-1, n_{e_1}, n_g+1\rangle & \hat{\Sigma}^\dagger_{ge_2} |n_{e_2}, n_{e_1}, n_g\rangle &= \sqrt{n_g(n_{e_2}+1)} |n_{e_2}+1, n_{e_1}, n_g-1\rangle \\
\hat{\Sigma}_{ge_1} |n_{e_2}, n_{e_1}, n_g\rangle &= \sqrt{n_{e_1}(n_g+1)} |n_{e_2}, n_{e_1}-1, n_g+1\rangle & \hat{\Sigma}^\dagger_{ge_1} |n_{e_2}, n_{e_1}, n_g\rangle &= \sqrt{n_g(n_{e_1}+1)} |n_{e_2}, n_{e_1}+1, n_g-1\rangle
\end{aligned} \tag{A13}$$

and the number operators from the free atomic Hamiltonian are

$$\hat{n}_{e_2} |n_{e_2}, n_{e_1}, n_g\rangle = n_{e_2} |n_{e_2}, n_{e_1}, n_g\rangle \quad \hat{n}_{e_1} |n_{e_2}, n_{e_1}, n_g\rangle = n_{e_1} |n_{e_2}, n_{e_1}, n_g\rangle \tag{A14}$$

so we may write (A11) for the indistinguishable ensemble as

$$\begin{aligned}
\frac{d}{dt}\hat{\Theta}^I(t) = -\int_0^\infty d\tau \sum_{j=\{1,2\}} \sum_{\mathbf{k}} |g_{j,\mathbf{k}}|^2 &\left\{ e^{-i\omega_\mathbf{k}\tau} \left\{ e^{i\omega_j\tau} \left[\hat{\Sigma}^\dagger_{ge_j}, \hat{\Sigma}_{ge_j} \hat{\Theta}^I(t-\tau)\right] + e^{-i\omega_j\tau} \left[\hat{\Sigma}_{ge_j}, \hat{\Sigma}^\dagger_{ge_j} \hat{\Theta}^I(t-\tau)\right] \right\} \right. \\
&\left. + e^{i\omega_\mathbf{k}\tau} \left\{ e^{-i\omega_j\tau} \left[\hat{\Sigma}_{ge_j}, \hat{\Sigma}^\dagger_{ge_j} \hat{\Theta}^I(t-\tau)\right] + e^{i\omega_j\tau} \left[\hat{\Sigma}^\dagger_{ge_j}, \hat{\Sigma}_{ge_j} \hat{\Theta}^I(t-\tau)\right] \right\} \right\}
\end{aligned} \tag{A15}$$

Following ref. [99] and others, assuming no dipole-dipole interactions, we take the Laplace transform of (A15),

$$s\hat{\tilde{\Theta}}^I(s) - \hat{\Theta}^I(0) = -\sum_{j=\{1,2\}} \Gamma_j \left\{ \hat{\Sigma}^\dagger_{ge_j} \hat{\Sigma}_{ge_j} \hat{\tilde{\Theta}}^I(s) + \hat{\tilde{\Theta}}^I(s) \hat{\Sigma}^\dagger_{ge_j} \hat{\Sigma}_{ge_j} - 2\hat{\Sigma}_{ge_j} \hat{\tilde{\Theta}}^I(s) \hat{\Sigma}^\dagger_{ge_j} \right\} \tag{A16}$$

with the decay rate $\Gamma_j$ obtained from Fermi's golden rule in the continuum limit of the mode (by taking the modal volume to infinity, $\sum_\mathbf{k} \to \frac{2V}{(2\pi)^3} \int d^3k$) in the Markov approximation,

$$\Gamma_j = 2\pi \sum_\mathbf{k} |g_{j,\mathbf{k}}|^2 \delta(\omega_j - \omega_\mathbf{k}) \to \frac{|\mathbf{d}_j|^2 \omega_j^3}{3\pi\varepsilon_0 \hbar c^3}. \tag{A17}$$

Applying the inverse Laplace transformation and transforming the result to Schrödinger picture finally yields

$$\frac{d}{dt}\hat{\Theta}(t) = -\frac{i}{\hbar} \sum_{j=\{1,2\}} \omega_j \left[\hat{n}_{e_j}, \hat{\Theta}(t)\right] - \hat{L}\{\hat{\Theta}(t)\} \tag{A18}$$

for the Lindblad operator $\hat{L}\{\hat{\Theta}(t)\} = \sum_{j=\{1,2\}} \frac{\Gamma_j}{2} \left\{ \hat{\Sigma}^\dagger_{ge_j} \hat{\Sigma}_{ge_j} \hat{\Theta}(t) + \hat{\Theta}(t) \hat{\Sigma}^\dagger_{ge_j} \hat{\Sigma}_{ge_j} - 2\hat{\Sigma}_{ge_j} \hat{\Theta}(t) \hat{\Sigma}^\dagger_{ge_j} \right\}$, and for the atomic density matrix elements we obtain the ODE system (2) from the main text,

$$\frac{\partial}{\partial t}\Theta_{n_{e_1},n_{e_2},n_g}^{m_{e_1},m_{e_2},m_g} = -i\left[(m_1-n_1)\omega_1 + (m_2-n_2)\omega_2\right]\Theta_{n_{e_1},n_{e_2},n_g}^{m_{e_1},m_{e_2},m_g}$$

$$-\frac{\Gamma_1}{2}\left\{\left[m_1(m_g+1)+n_1(n_g+1)\right]\Theta_{n_{e_1},n_{e_2},n_g}^{m_{e_1},m_{e_2},m_g} - 2\sqrt{m_g(m_1+1)n_g(n_1+1)}\Theta_{n_1+1,n_{e_2},n_g-1}^{m_{e_1}+1,m_{e_2},m_g-1}\right\}$$

$$-\frac{\Gamma_2}{2}\left\{\left[m_2(m_g+1)+n_2(n_g+1)\right]\Theta_{n_{e_1},n_{e_2},n_g}^{m_{e_1},m_{e_2},m_g} - 2\sqrt{m_g(m_2+1)n_g(n_2+1)}\Theta_{n_{e_1},n_{e_2}+1,n_g-1}^{m_{e_1},m_{e_2}+1,m_g-1}\right\} \quad \text{(A19)}$$

To calculate the emitted intensities, we use for the emission intensity of photons with angular frequency $\omega_j$

$$I_{\omega_j}(t) = 2\omega_j \Gamma_j \left\langle \hat{\Sigma}_{ge_j}^\dagger \hat{\Sigma}_{ge_j} \right\rangle, \quad \text{(A20)}$$

where $\left\langle \hat{\Sigma}_{ge_j}^\dagger \hat{\Sigma}_{ge_j} \right\rangle = Tr\left\{\hat{\Theta}(t)\hat{\Sigma}_{ge_j}^\dagger \hat{\Sigma}_{ge_j}\right\}$ [99].

## APPENDIX B – WEISSKOPF-WIGNER TREARMENT OF A V-ATOM ENSEMBLE

To describe the photonic state emitted from the atomic system another approach must be taken, since the "standard" method of evolving the density matrix taken in APPENDIX A inevitably involves tracing out the photonic degrees of freedom. We therefore develop in the following a framework involving the Weisskopf-Wigner description of spontaneous emission [100], generalized to fit our system of an ensemble of indistinguishable V-atoms. We describe the general state vector of our system as a product of an atomic part described in the atomic Fock number basis $\left|n_{e_1},n_{e_2},n_g\right\rangle_A$ and a photonic part described in the photonic modal Fock number basis $\left|n_{\omega_1},n_{\omega_2}\right\rangle_f$. As a simple example, we can see that a system consisting of a single V-atom can be described at any time as

$$\left|\psi_V\right\rangle(t) = a_1(t)\left|1,0,0\right\rangle_A\left|0,0\right\rangle_f + a_2(t)\left|0,1,0\right\rangle_A\left|0,0\right\rangle_f + \left|0,0,1\right\rangle_A\sum_{\mathbf{k}}\left(b_1(t)\left|1_{\mathbf{k}},0\right\rangle_f + b_2(t)\left|0,1_{\mathbf{k}}\right\rangle_f\right). \quad \text{(B1)}$$

The subscript of the photonic Fock numbers denotes the wavevector characterizing the photon mode. The interaction Hamiltonian

$$\hat{H}_I = \hbar\sum_{\mathbf{k}}\left(g_{1,\mathbf{k}}\hat{\sigma}_{ge_1}\hat{a}_{1,\mathbf{k}}^\dagger e^{i(\omega_{\mathbf{k}}-\omega_1)t} + g_{2,\mathbf{k}}\hat{\sigma}_{ge_2}\hat{a}_{2,\mathbf{k}}^\dagger e^{i(\omega_{\mathbf{k}}-\omega_2)t}\right) + h.c \quad \text{(B2)}$$

does not contain direct interaction terms between the two transitions (only through the ground level) and so the coefficients can be solved in a straightforward fashion from the system of ODE's

$$\frac{d}{dt}a_1(t) = -i\sum_{\mathbf{k}}g_{1,\mathbf{k}}^* b_{1,\mathbf{k}}(t)e^{i(\omega_1-\omega_{\mathbf{k}})t}, \quad \frac{d}{dt}b_{1,\mathbf{k}}(t) = -ig_{1,\mathbf{k}}a_1(t)e^{-i(\omega_{\mathbf{k}}-\omega_2)t},$$

$$\frac{d}{dt}a_2(t) = -i\sum_{\mathbf{k}}g_{2,\mathbf{k}}^* b_{2,\mathbf{k}}(t)e^{i(\omega_2-\omega_{\mathbf{k}})t}, \quad \frac{d}{dt}b_{2,\mathbf{k}}(t) = -ig_{2,\mathbf{k}}a_2(t)e^{-i(\omega_{\mathbf{k}}-\omega_1)t} \quad \text{(B3)}$$

By solving this system using the Weisskopf-Wigner method, within the validity of Markovian assumption [112], one obtains

$$\left|\psi_V\right\rangle(t) = a_1(0)e^{-\Gamma_1 t/2}\left|1,0,0\right\rangle_A\left|0,0\right\rangle_f + a_2(0)e^{-\Gamma_2 t/2}\left|0,1,0\right\rangle_A\left|0,0\right\rangle_f$$

$$+\left|0,0,1\right\rangle_A\sum_{\mathbf{k}}\left(a_1(0)g_{1,\mathbf{k}}\frac{1-e^{i(\omega_1-\omega_{\mathbf{k}})t-\Gamma_1 t/2}}{\omega_{\mathbf{k}}-\omega_1+i\Gamma_1/2}\left|1_{\mathbf{k}},0\right\rangle_f + a_2(0)g_{2,\mathbf{k}}\frac{1-e^{i(\omega_2-\omega_{\mathbf{k}})t-\Gamma_2 t/2}}{\omega_{\mathbf{k}}-\omega_2+i\Gamma_2/2}\left|0,1_{\mathbf{k}}\right\rangle_f\right) \quad \text{(B4)}$$

with the decay rates $\Gamma_j$ obtained from the Weisskopf-Wigner ansatz, from the solution of

$$-\frac{\Gamma_j}{2}a_j(t) = -i\sum_{\mathbf{k}}g_{j,\mathbf{k}}^* b_{j,\mathbf{k}}(t)e^{i(\omega_j-\omega_{\mathbf{k}})t} \quad \text{(B5)}$$

from which

$$\Gamma_j = \frac{|\mathbf{d}_j|^2 \omega_j^3}{3\pi\varepsilon_0 \hbar c^3}, \tag{B6}$$

in agreement with (A17), that are assumed to satisfy the linewidth condition $\Gamma_j \ll min(\omega_1, \omega_2, |\omega_1 - \omega_2|)$ ensuring that the photonic transitions effectively do not overlap. The photonic population terms at $t \to \infty$, after the atom has reached its ground level, can be immediately obtained from (B4) by imposing the continuum limit on the modal sum, $p_{|1,0\rangle} = |a_1(0)|^2$, $p_{|0,1\rangle} = |a_2(0)|^2$. This means, unsurprisingly, that the probability ratio of a photon in either angular frequency $\omega_1$ or $\omega_2$ is equal to the ratio of the excitation probabilities of the two excited levels.

We can generalize (B4) to the case of $N$ indistinguishable V-atoms,

$$|\psi_{N-V}\rangle(t) = \sum_{n=0}^{N} a_n(t)|N-n,n,0\rangle_A |0,0\rangle_f + \sum_{n_g=1}^{N-1} \sum_{n_2=0}^{N-n_g} |N-n_g-n_2, n_2, n_g\rangle_A \sum_{\mathbf{k}_1} \cdots \sum_{\mathbf{k}_p} \sum_{j=0}^{n_g} b_{j,\mathbf{k}_1\ldots\mathbf{k}_{n_g}}^{n_1,n_2,n_g}(t) \Big| (n_g-j)_{\{n_g-j\}}, j_{\{j\}} \Big\rangle_f$$

$$+|0,0,N\rangle_A \sum_{\mathbf{k}_1} \cdots \sum_{\mathbf{k}_N} \sum_{n=0}^{N} c_{n,\mathbf{k}_1\ldots\mathbf{k}_N}(t) \Big| (N-n)_{\{N-n\}}, n_{\{n\}} \Big\rangle_f \tag{B7}$$

The subscripts $\{i\}$ of a photonic Fock numbers vector consisting of $n_g$ total photons denote a tuple of $i$ unique wave vector indices from the set $\{\mathbf{k}_1, \ldots, \mathbf{k}_{n_g}\}$, to account for the fact that the order of summations is interchangeable. To find the coefficients we again solve Schrödinger's equation in the interaction picture with the interaction Hamiltonian, considering again the symmetrized atomic operators of the V-atom ensemble (B2),

$$\hat{H}_I = \hbar \sum_{\mathbf{k}} g_{1,\mathbf{k}} \hat{\Sigma}_1 \hat{a}_{1,\mathbf{k}}^\dagger e^{i(\omega_\mathbf{k}-\omega_1)t} + g_{2,\mathbf{k}} \hat{\Sigma}_2 \hat{a}_{2,\mathbf{k}}^\dagger e^{i(\omega_\mathbf{k}-\omega_2)t} + h.c \tag{B8}$$

with the symmetrized atomic annihilation operators defined as in (A13). Applying the Weisskopf-Wigner ansatz yields a system of coupled differential equations of the form

$$\frac{d}{dt} b_0^{N-n,n,0}(t) \equiv \frac{d}{dt} a_n(t) = -\frac{\Gamma_n}{2} a_n(t)$$

$$\frac{d}{dt} b_{j,\mathbf{k}_1\ldots\mathbf{k}_{n_g}}^{n_1,n_2,n_g}(t) = -i g_{1,\mathbf{k}_{n_g}}^{n_1+1,n_2,n_g-1} b_{j,\mathbf{k}_1\ldots\mathbf{k}_{n_g-1}}^{n_1+1,n_2,n_g-1}(t) e^{-i(\omega_1-\omega_{\mathbf{k}_{n_g}})t} - i g_{2,\mathbf{k}_{n_g}}^{n_1,n_2+1,n_g-1} b_{j-1,\mathbf{k}_1\ldots\mathbf{k}_{n_g-1}}^{n_1,n_2+1,n_g-1}(t) e^{-i(\omega_2-\omega_{\mathbf{k}'_{n_g}})t} - \frac{\Gamma_{n_1,n_2,n_g}}{2} b_{j,\mathbf{k}_1\ldots\mathbf{k}_{n_g}}^{n_1,n_2,n_g}(t)$$

$$\frac{d}{dt} c_{n,\mathbf{k}_1\ldots\mathbf{k}_N}(t) = -i g_{1,\mathbf{k}_N}^{n_1+1,n_2,n_g-1} b_{j,\mathbf{k}_1\ldots\mathbf{k}_{N-1}}^{n_1+1,n_2,n_g-1}(t) e^{-i(\omega_1-\omega_{\mathbf{k}_N})t} - i g_{2,\mathbf{k}_N}^{n_1,n_2+1,n_g-1} b_{j-1,\mathbf{k}_1\ldots\mathbf{k}_{N-1}}^{n_1,n_2+1,n_g-1}(t) e^{-i(\omega_2-\omega_{\mathbf{k}'_N})t} \tag{B9}$$

The primed notation $\mathbf{k}'_{n_g}$ denotes wavevector indices corresponding to the variables centered around $\omega_2$, in order to emphasize that the photon packets around $\omega_1$ and around $\omega_2$ do not overlap and so independent variables are used to describe them. This index is taken from the set of indices $\{\mathbf{k}_1 \ldots \mathbf{k}_{n_g}\}$. Note that $n_1 = N - n_2 - n_g$. Here, $g_{a,\mathbf{k}_p}^{n_1,n_2,n_g}$ is defined to include the combinatoric factors stemming from the symmetrized operators in (A13),

$$g_{a,\mathbf{k}_p}^{n_1,n_2,n_g} = g_{a,\mathbf{k}_p} \sqrt{n_a(n_g+1)}, \tag{B10}$$

and the decay rates $\Gamma_{n_1,n_2,n_g}$ are calculated either from the Weisskopf-Wigner ansatz as in (B5) or directly from Fermi's golden rule for the transition between the state $|n_1, n_2, n_g\rangle_A |n_g - j, j\rangle_f$ and the two states to which it may evolve, $|n_1 - 1, n_2, n_g + 1\rangle_A |n_g - j + 1, j\rangle_f$ and $|n_1, n_2 - 1, n_g + 1\rangle_A |n_g - j, j + 1\rangle_f$ (wherein the Markov approximation – namely, no back-action of the emitted photon on the ensemble – translates into considering the photon emission to occur in vacuum, i.e., no combinatoric factor is added as a result of applying the photonic operators), finally yielding

$$\Gamma_{n_1,n_2,n_g} = \frac{1}{3\pi\varepsilon_0 \hbar c^3}\left(n_1|\mathbf{d}_1|^2 \omega_1^3 + n_2|\mathbf{d}_2|^2 \omega_2^3\right)(n_g+1). \tag{B11}$$

Equation system (B9) is highly complicated and solving it as a system of coupled ODE's is an exorbitant task. However, these equations can be stated as a relatively simple recursion rule by applying Laplace transform, $\mathcal{L}: f(t) \to F(s)$ with $s$ the Laplace-space parameter. Then, (B9) becomes

$$B_0^{N-n,n,0}(s) \equiv A_n(s) = \frac{a_n(0)}{s+\Gamma_n/2}$$

$$B_{j,\mathbf{k}_1...\mathbf{k}_{n_g}}^{n_1,n_2,n_g}(s) = -i\frac{g_{1,\mathbf{k}_{n_g}}^{n_1+1,n_2,n_g-1}B_{j,\mathbf{k}_1...\mathbf{k}_{n_g-1}}^{n_1+1,n_2,n_g-1}\left(s-i\left(\omega_{\mathbf{k}_{n_g}}-\omega_1\right)\right)+g_{2,\mathbf{k}_{n_g}}^{n_1,n_2+1,n_g-1}B_{j-1,\mathbf{k}_1...\mathbf{k}_{n_g-1}}^{n_1,n_2+1,n_g-1}\left(s-i\left(\omega_{\mathbf{k'}_{n_g}}-\omega_2\right)\right)}{s+\Gamma_{n_1,n_2,n_g}^j/2}$$

$$C_{n,\mathbf{k}_1...\mathbf{k}_N}(s) = -i\frac{g_{1,\mathbf{k}_N}^{n_1,n_2+1,N-1}B_{j,\mathbf{k}_1...\mathbf{k}_{N-1}}^{n_1+1,n_2,N-1}\left(s-i\left(\omega_{\mathbf{k}_N}-\omega_1\right)\right)+g_{2,\mathbf{k}_N}^{n_1,n_2+1,N-1}B_{j-1,\mathbf{k}_1...\mathbf{k}_{N-1}}^{n_1,n_2+1,N-1}\left(s-i\left(\omega_{\mathbf{k'}_N}-\omega_2\right)\right)}{s} \tag{B12}$$

Each function $B_{n_g,j,\mathbf{k}_1...\mathbf{k}_{n_g}}^{n_1,n_2,n_g}(s)$ in (B12) is recursively obtained by summing the two functions associated with the two preceding states, shifting their poles by either $-i(\omega_{\mathbf{k}_{n_g}}-\omega_1)$ or $-i(\omega_{\mathbf{k}_{n_g}}-\omega_2)$ (depending on the central angular frequency of the photon emitted in either process), and adding a new pole at $s = -\Gamma_{n_1,n_2,n_g}/2$. Because all poles are located on the left complex half-plane of the Laplace-domain coordinate $s$, we may use the known property $\lim_{t\to\infty} f(t) = \lim_{s\to 0^+} sF(s)$ in order to calculate the coefficients $c_{n,\mathbf{k}_1...\mathbf{k}_N}(t)$ in the long-time limit, from which the photonic state at the end of the superradiance emission process can be obtained.

As an example, consider two indistinguishable V-atoms. The total state vector can be given by

$$|\psi_{2-V}\rangle(t) = a_0(t)|2,0,0\rangle_A|0,0\rangle_f + a_1(t)|1,1,0\rangle_A|0,0\rangle_f + a_2(t)|0,2,0\rangle_A|0,0\rangle_f$$
$$+|1,0,1\rangle_A\sum_{\mathbf{k}}\left(b_{0,\mathbf{k}}^{1,0,1}(t)_A|1_\mathbf{k},0\rangle_f + b_{1,\mathbf{k}}^{1,0,1}(t)|0,1_\mathbf{k}\rangle_f\right) + |0,1,1\rangle_A\sum_{\mathbf{k}}\left(b_{0,\mathbf{k}}^{0,1,1}(t)|1_\mathbf{k},0\rangle_f + b_{0,\mathbf{k}}^{0,1,1}(t)|0,1_\mathbf{k}\rangle_f\right)$$
$$+\sum_{\mathbf{k}}\sum_{\mathbf{q}}c_{0,\mathbf{kq}}(t)|0,0,2\rangle_A|2_{\mathbf{kq}},0\rangle_f + \sum_{\mathbf{k}}\sum_{\mathbf{q}}c_{1,\mathbf{kq}}(t)|0,0,2\rangle_A|1_\mathbf{k},1_\mathbf{q}\rangle_f + \sum_{\mathbf{k}}\sum_{\mathbf{q}}c_{2,\mathbf{kq}}(t)|0,0,2\rangle_A|0,2_{\mathbf{kq}}\rangle_f \tag{B13}$$

The recursion rule (B12) yields

$$A_n(s) = \frac{a_n(0)}{s+\Gamma_n/2},$$

$$B_{0,\mathbf{k}}^{1,0,1}(s) = -i\frac{g_{1,\mathbf{k}}^{2,0,0}A_0\left(s-i(\omega_\mathbf{k}-\omega_1)\right)}{s+\Gamma_{1,0,1}/2}, \quad B_{1,\mathbf{q}}^{1,0,1}(s) = -i\frac{g_{2,\mathbf{q}}^{1,1,0}A_1\left(s-i(\omega_\mathbf{q}-\omega_2)\right)}{s+\Gamma_{1,0,1}/2},$$

$$B_{0,\mathbf{k}}^{0,1,1}(s) = -i\frac{g_{1,\mathbf{k}}^{1,1,0}A_1\left(s-i(\omega_\mathbf{k}-\omega_1)\right)}{s+\Gamma_{0,1,1}/2}, \quad B_{1,\mathbf{q}}^{0,1,1}(s) = -i\frac{g_{2,\mathbf{q}}^{0,2,0}A_2\left(s-i(\omega_\mathbf{q}-\omega_2)\right)}{s+\Gamma_{0,1,1}/2},$$

$$C_{0,\mathbf{kq}}(s) = -i\frac{g_{1,\mathbf{k}}^{1,0,1}B_{0,\mathbf{q}}^{1,0,1}\left(s-i(\omega_\mathbf{k}-\omega_1)\right)}{s}, \quad C_{2,\mathbf{kq}}(s) = -i\frac{g_{2,\mathbf{q}}^{0,1,1}B_{1,\mathbf{k}}^{0,1,1}\left(s-i(\omega_\mathbf{q}-\omega_2)\right)}{s},$$

$$C_{1,\mathbf{kq}}(s) = -i\frac{g_{1,\mathbf{k}}^{1,0,1}B_{1,\mathbf{q}}^{1,0,1}\left(s-i(\omega_\mathbf{k}-\omega_1)\right)+g_{2,\mathbf{q}}^{0,1,1}B_{0,\mathbf{k}}^{0,1,1}\left(s-i(\omega_\mathbf{q}-\omega_2)\right)}{s} \tag{B14}$$

so that the superradiance process from two indistinguishable V-atoms can be decomposed into three disjoint paths

$$|2,0,0\rangle_A |0,0\rangle_f \to \sum_{\mathbf{k}}|1,0,1\rangle_A |1_{\mathbf{k}},0\rangle_f \to \sum_{\mathbf{kq}}|0,0,2\rangle_A |2_{\mathbf{kq}},0\rangle_f$$

$$|1,1,0\rangle_A |0,0\rangle_f \to \left\{\sum_{\mathbf{k}}|1,0,1\rangle_A |0,1_{\mathbf{k}}\rangle_f, \sum_{\mathbf{k}}|0,1,1\rangle_A |1_{\mathbf{k}},0\rangle_f\right\} \to \sum_{\mathbf{kq}}|0,0,2\rangle_A |1_{\mathbf{k}},1_{\mathbf{q}}\rangle_f$$

$$|0,2,0\rangle_A |0,0\rangle_f \to \sum_{\mathbf{k}}|0,1,1\rangle_A |0,1_{\mathbf{k}}\rangle_f \to \sum_{\mathbf{kq}}|0,0,2\rangle_A |0,2_{\mathbf{kq}}\rangle_f \quad (B15)$$

In (B14) – $g_{a,\mathbf{k}}^{1,1,0} = g_{a,\mathbf{k}}$, $g_{1,\mathbf{k}}^{1,0,1} = g_{1,\mathbf{k}}^{2,0,0} = \sqrt{2}g_{1,\mathbf{k}}$ and $g_{2,\mathbf{k}}^{0,2,0} = g_{2,\mathbf{k}}^{0,1,1} = \sqrt{2}g_{2,\mathbf{k}}$, and the decay rates are calculated from (B11) are $\Gamma_{1,0,1} = \frac{2|\mathbf{d}_1|^2 \omega_1^3}{3\pi\varepsilon_0 \hbar c^3}$, $\Gamma_{0,1,1} = \frac{2|\mathbf{d}_2|^2 \omega_2^3}{3\pi\varepsilon_0 \hbar c^3}$, $\Gamma_0 = \frac{2|\mathbf{d}_1|^2 \omega_1^3}{3\pi\varepsilon_0 \hbar c^3}$, $\Gamma_2 = \frac{2|\mathbf{d}_2|^2 \omega_2^3}{3\pi\varepsilon_0 \hbar c^3}$, and $\Gamma_1 = \frac{1}{3\pi\varepsilon_0 \hbar c^3}\left(|\mathbf{d}_1|^2 \omega_1^3 + |\mathbf{d}_2|^2 \omega_2^3\right)$, which yield the final state amplitudes

$$c_{0,\mathbf{kq}}(\infty) = -\frac{2g_{1,\mathbf{q}}g_{1,\mathbf{k}}a_0(0)}{\left(i(2\omega_1 - \omega_{\mathbf{k}} - \omega_{\mathbf{q}}) + \Gamma_0/2\right)\left(-i(\omega_{\mathbf{k}} - \omega_1) + \Gamma_0/2\right)}$$

$$c_{1,\mathbf{kq}}(\infty) = -\frac{\sqrt{2}g_{1,\mathbf{k}}g_{2,\mathbf{q}}a_1(0)}{i(\omega_1 + \omega_2 - \omega_{\mathbf{k}} - \omega_{\mathbf{q}}) + \Gamma_1/2}\left(\frac{1}{-i(\omega_{\mathbf{k}} - \omega_1) + \Gamma_0/2} + \frac{1}{-i(\omega_{\mathbf{q}} - \omega_2) + \Gamma_2/2}\right)$$

$$c_{2,\mathbf{kq}}(\infty) = -\frac{2g_{2,\mathbf{q}}g_{2,\mathbf{k}}a_2(0)}{\left(i(2\omega_2 - \omega_{\mathbf{k}} - \omega_{\mathbf{q}}) + \Gamma_2/2\right)\left(-i(\omega_{\mathbf{q}} - \omega_2) + \Gamma_2/2\right)} \quad (B16)$$

Note that in the above we have denoted by $\omega_{\mathbf{k}}$ and $\omega_{\mathbf{q}}$ the modal variables corresponding to the $\omega_1$ and $\omega_2$ photons, respectively. To find the populations of the final photonic states $p_{|2,0\rangle}$, $p_{|1,1\rangle}$ and $p_{|0,2\rangle}$ we calculate the overlaps of the final state terms of the density matrix derived from the (pure) state (B13), for example

$$p_{|2,0\rangle} = \left(\sum_{\mathbf{k'q'}}c_{0,\mathbf{k'q'}}^*(\infty)_f \langle 2_{\mathbf{k'q'}},0|\right)\left(\sum_{\mathbf{kq}}c_{0,\mathbf{kq}}(\infty)_f |2_{\mathbf{kq}},0\rangle_f\right) = \sum_{\mathbf{kq}}|c_{0,\mathbf{kq}}(\infty)|^2 \to \left(\frac{2V}{(2\pi)^3}\right)^2 \int d^3k\, d^3q\, |c_{0,\mathbf{kq}}(\infty)|^2 \quad (B17)$$

where the transition from summation to integration was performed in the large modal volume limit, $V \to \infty$. Substituting (B16) into the above, applying the Weisskopf-Wigner assumption of narrow linewidth and using contour integration methods [112] yield

$$p_{|2,0\rangle} = |a_0(0)|^2, \quad p_{|1,1\rangle} = |a_1(0)|^2, \quad p_{|0,2\rangle} = |a_2(0)|^2. \quad (B18)$$

This result demonstrates that the probabilities of the populations of the final photonic states are equal to the excitation probabilities of the initial states. The correspondence agrees with energy conservation considerations; the initial symmetrized atomic states have energies $2\hbar\omega_1, \hbar(\omega_1 + \omega_2)$ and $2\hbar\omega_2$ (with respect to the symmetrized atomic ground level), and the final photonic state must be a superposition of states with the same energies. Since the interaction Hamiltonian conserves energy, a decay process from any initial atomic state into a lower atomic state with the corresponding emission of photons must conserve probability. This is well demonstrated in (B18). This of course can be generalized to the case of $N$ indistinguishable V-atoms.

*Proposition:* The population probability of the final photonic state $|n, N-n\rangle$ is identical to the probability of the $n$'th excited symmetrized atomic state, expressly given by

$$p_{|n,N-n\rangle} = \sum_{\mathbf{k}_1,...,\mathbf{k}_N}|c_{n,\mathbf{k}_1...\mathbf{k}_N}(\infty)|^2 = |a_n(0)|^2. \quad (B19)$$

*Proof*:

From the recurrence definition (B12), it is evident that any coefficient $C_{n,\mathbf{k}_1\ldots\mathbf{k}_{N-n}\mathbf{q}_1\ldots\mathbf{q}_{N-n}}(s)$ can be written as a sum of multiplications of Lorentzian distributions with different widths, peak heights and center angular frequencies, of the form $C_{n,\mathbf{k}_1\ldots\mathbf{k}_{N-n}\mathbf{q}_1\ldots\mathbf{q}_{N-n}}(s) = \sum_{j_n}^{N_{j_n}} C_{n,\mathbf{k}_1\ldots\mathbf{k}_{N-n}\mathbf{q}_1\ldots\mathbf{q}_{N-n}}^{j_n}(s)$, so that

$$c_{n,\mathbf{k}_1\ldots\mathbf{k}_{N-n}\mathbf{q}_1\ldots\mathbf{q}_{N-n}}(\infty) = \sum_{j_n}^{N_{j_n}} c_{n,\mathbf{k}_1\ldots\mathbf{k}_{N-n}\mathbf{q}_1\ldots\mathbf{q}_{N-n}}^{j_n}(\infty). \tag{B20}$$

We have denoted the $n$ modal variables corresponding to the photons with central angular frequency $\omega_1$ by $\mathbf{k}_1,\ldots,\mathbf{k}_n$ and the $N-n$ modal variables corresponding to the $\omega_2$ photons by $\mathbf{q}_1,\ldots,\mathbf{q}_{N-n}$. Each term $c_{n,\mathbf{k}_1\ldots\mathbf{k}_{N-n}\mathbf{q}_1\ldots\mathbf{q}_{N-n}}^{j_n}(\infty)$ denotes a single evolution path from the initial state $a_n(0)$, consisting of $n$ symmetrized excitations of the $|e_1\rangle$ atomic level and $N-n$ excitations of the $|e_2\rangle$ atomic level, across the ensemble of $N$ indistinguishable atoms. From energy conservation considerations, a system initially excited to $a_n(0)$ will emit in total $n$ photons into the $\omega_1$ mode and $N-n$ photons into the $\omega_2$ mode. The number of evolution paths $N_{j_n}$ appearing in (B20) via which this combination of photons is emitted in total where a single photon is emitted at a time is $N_{j_n} = N!/n!(N-n)!$.

Since $c_{n,\mathbf{k}_1\ldots\mathbf{k}_{N-n}\mathbf{q}_1\ldots\mathbf{q}_{N-n}}^{j_n}(\infty)$ consists of a product of $N$ Lorentzian distributions subdivided into $n$ distributions describing a photon in the $\omega_1$ mode and $N-n$ distributions describing a photon in the $\omega_2$ mode, using the definition (B10) we may write

$$c_{n,\mathbf{k}_1\ldots\mathbf{k}_n\mathbf{q}_1\ldots\mathbf{q}_{N-n}}^{j_n}(\infty) = a_n\sqrt{N!}\prod_{m=n-1}^{0} g_{1,\mathbf{k}_{m+1}}\sqrt{m+1}\prod_{p=N-n-1}^{0} g_{2,\mathbf{q}_{p+1}}\sqrt{p+1}\prod_{i=0}^{N}\frac{1}{\tilde{\omega}_i - i\Gamma_i^{j_n}/2}. \tag{B21}$$

The variables $\tilde{\omega}_i$ are dummy variables, defined as sums of central angular frequencies and angular frequency variables, resulting from the recursive rule (B12) when $s \to 0$. For example, if after $i$ emissions a total of $t$ ($r$) photons in the $\omega_1$ ($\omega_2$) mode were emitted, then

$$\tilde{\omega}_i = \sum_{i1=1}^{t}\omega_{\mathbf{k}_{i1}} + \sum_{i2=1}^{r}\omega_{\mathbf{q}_{i2}} - t\omega_1 - r\omega_2. \tag{B22}$$

An important notion is that by defining these dummy variables we effectively decouple all the variables $\omega_{\mathbf{k}_i}$ and $\omega_{\mathbf{q}_j}$ in the summation (in the Riemann integral limit), as we will see in (B25). The constants $\Gamma_i^j$ are the corresponding decay rates of the $i$'th emission encountered in the $j$'th evolution path, using the definition (B11). Equation (B21) can be more conveniently stated,

$$c_{n,\mathbf{k}_1\ldots\mathbf{k}_n\mathbf{q}_1\ldots\mathbf{q}_{N-n}}^{j_n}(\infty) = a_n(g_{1,\mathbf{k}})^n(g_{2,\mathbf{q}})^{N-n}\sqrt{N!}\sqrt{n!}\sqrt{(N-n)!}\prod_{i=0}^{N}\frac{1}{\tilde{\omega}_i - i\Gamma_i^{j_n}/2}. \tag{B23}$$

We will omit the infinite time notation from now on. Then,

$$\left|c_{n,\mathbf{k}_1\ldots\mathbf{k}_n\mathbf{q}_1\ldots\mathbf{q}_{N-n}}^{j_n}\right|^2 = |a_n|^2|g_{1,\mathbf{k}}|^{2n}|g_{2,\mathbf{q}}|^{2(N-n)}N!n!(N-n)!\prod_{i=0}^{N}\frac{1}{\tilde{\omega}_i^2 + \Gamma_i^{j_n\,2}/4}. \tag{B24}$$

Summing (B24) over all indices and taking the continuum limit to perform Riemann integration,

$$\sum_{\mathbf{k}_1\ldots\mathbf{k}_n}\sum_{\mathbf{q}_1\ldots\mathbf{q}_{N-n}}\left|c_{n,\mathbf{k}_1\ldots\mathbf{k}_n\mathbf{q}_1\ldots\mathbf{q}_{N-n}}^{j_n}\right|^2$$

$$\to |a_n|^2 N!n!(N-n)!\left(\frac{1}{6\varepsilon_0\hbar\pi^2 c^3}\right)^N\int_0^\infty d\omega_{\mathbf{k}_1}\omega_{\mathbf{k}_1}^3\ldots\int_0^\infty d\omega_{\mathbf{k}_n}\omega_{\mathbf{k}_n}^3\int_0^\infty d\omega_{\mathbf{q}_1}\omega_{\mathbf{q}_1}^3\ldots\int_0^\infty d\omega_{\mathbf{q}_{N-n}}\omega_{\mathbf{q}_{N-n}}^3\prod_{i=0}^{N}\frac{4}{4\tilde{\omega}_i^2 + \Gamma_i^{j_n\,2}} \tag{B25}$$

We make the assumption that due to the narrowness of the linewidths around the respective central frequencies, we may replace the $\omega_\mathbf{k}$'s and $\omega_\mathbf{q}$'s by $\omega_1$ and $\omega_2$, respectively, and take the lower integration limit to be $\to -\infty$ in the usual manner [112]. We now perform the change of variables defined by (B22) in the integrals, and integrate with respect to the dummy variables without any other modification. This is possible because under the narrow linewidth approximation, all integrations are performed on the entire frequency axis and the Jacobian matrices of the transformations (B22) are triangular matrices of 1's, as can be immediately verified, so that the Jacobians are equal to 1 for all integrals under the above changes of variables. Additionally, note that Lorentzian distributions satisfy the integral relation

$$\int_{-\infty}^{\infty} dx \frac{4}{4x^2 + a^2} = \frac{2\pi}{a}. \tag{B26}$$

From (B23), (B24), (B26) and (B10) we have in the continuum limit

$$\sum_{\mathbf{k}_1...\mathbf{k}_n} \sum_{\mathbf{q}_1...\mathbf{q}_{N-n}} \left| c_{n,\mathbf{k}_1...\mathbf{k}_n\mathbf{q}_1...\mathbf{q}_{N-n}}^{j_n} \right|^2 \to |a_n|^2 N! n! (N-n)! \frac{\left(\mathbf{d}_1^2 \omega_1^3\right)^n \left(\mathbf{d}_2^2 \omega_2^3\right)^{N-n}}{\left(6\pi\varepsilon_0 \hbar c^3\right)^N} \prod_{i=0}^{N-1} \frac{2}{\Gamma_i^{j_n}}. \tag{B27}$$

Using (B11), we write a general expression for $\Gamma_i^{j_n}$, describing the decay are of a system after the $i$'th emission in the $j$'th evolution path,

$$\Gamma_i^{j_n} = \frac{1}{3\pi\varepsilon_0 \hbar c^3} \left( |\mathbf{d}_1|^2 \omega_1^3 x_i^{j_n} + |\mathbf{d}_2|^2 \omega_2^3 y_i^{j_n} \right)(i+1). \tag{B28}$$

Here $\mathbf{x}^{j_n} = \{x_i^{j_n}\}_{i=0}^{i=N-1}$ and $\mathbf{y}^{j} = \{y_i^{j_n}\}_{i=0}^{i=N-1}$ are tuples containing the remaining excitation numbers after the $i$'th emission in the $j_n$'th evaluation path; for all $j_n$, $x_0^{j_n} = n_{e_1}$ and $y_0^{j_n} = n_{e_2}$. Plugging this into (B27), we have

$$\sum_{\mathbf{k}_1...\mathbf{k}_n} \sum_{\mathbf{q}_1...\mathbf{q}_{N-n}} \left| c_{n,\mathbf{k}_1...\mathbf{k}_n\mathbf{q}_1...\mathbf{q}_{N-n}}^{j_n} \right|^2 \to |a_n|^2 n!(N-n)! \left(\mathbf{d}_1^2 \omega_1^3\right)^n \left(\mathbf{d}_2^2 \omega_2^3\right)^{N-n} \prod_{i=0}^{N-1} \frac{1}{\mathbf{d}_1^2 \omega_1^3 x_i^{j_n} + \mathbf{d}_2^2 \omega_2^3 y_i^{j_n}} \tag{B29}$$

and in total

$$\sum_{j_n}^{\binom{N}{n}} \sum_{\mathbf{k}_1...\mathbf{k}_n} \sum_{\mathbf{q}_1...\mathbf{q}_{N-n}} \left| c_{n,\mathbf{k}_1...\mathbf{k}_n\mathbf{q}_1...\mathbf{q}_{N-n}}^{j_n} \right|^2 \to |a_n|^2 n!(N-n)! \left(\mathbf{d}_1^2 \omega_1^3\right)^n \left(\mathbf{d}_2^2 \omega_2^3\right)^{N-n} \sum_{j_n}^{\binom{N}{n}} \prod_{i=0}^{N-1} \frac{1}{\mathbf{d}_1^2 \omega_1^3 x_i^{j_n} + \mathbf{d}_2^2 \omega_2^3 y_i^{j_n}}. \tag{B30}$$

Under the definition

$$G_n^N(s,t) \equiv \sum_{j_n}^{\binom{N-s}{n-t}} \prod_{i=s}^{N-1} \frac{1}{\mathbf{d}_1^2 \omega_1^3 x_i^{j_n} + \mathbf{d}_2^2 \omega_2^3 y_i^{j_n}} \tag{B31}$$

(for $N-s \geq n-t$, $t \leq s$), we may state a recursive rule illustrating that after the $s$'th emitted photon the total number of paths is divided into a group of paths in which the next transition emits a $\omega_1$ photon and a group of paths in which the next emitted photon is a $\omega_2$ photon. Using the identity

$$\binom{N-s}{n-t} = \binom{N-(s+1)}{n-t} + \binom{N-(s+1)}{n-(t+1)}$$

we obtain

$$G_n^N(s,t) = \frac{1}{\mathbf{d}_1^2 \omega_1^3 (n-t) + \mathbf{d}_2^2 \omega_2^3 (N-n+t-s)} \left( G_n^N(s+1,t+1) + G_n^N(s+1,t) \right) \tag{B32}$$

with the conditions

$$G_n^N(s,n) = \frac{\left(\mathbf{d}_2^2\omega_2^3\right)^{s-N}}{(N-s)!}, \quad G_n^N(s,n-N+s) = \frac{\left(\mathbf{d}_1^2\omega_1^3\right)^{s-N}}{(N-s)!}, \quad G_n^N(N-1,0) = \frac{1}{\mathbf{d}_2^2\omega_2^3}, \quad G_n^N(N-1,1) = \frac{1}{\mathbf{d}_1^2\omega_1^3}. \quad \text{(B33)}$$

It can be shown that

$$G_n^N(0,0) = \sum_{j_n}^{\binom{N}{n}} \prod_{i=0}^{N-1} \frac{1}{\mathbf{d}_1^2\omega_1^3 x_i^{j_n} + \mathbf{d}_2^2\omega_2^3 y_i^{j_n}} = \frac{1}{n!(N-n)!\left(\mathbf{d}_1^2\omega_1^3\right)^n \left(\mathbf{d}_2^2\omega_2^3\right)^{N-n}} \quad \text{(B34)}$$

and plugging this in (B30) c.f. (B20) yields

$$\sum_{\mathbf{k}_1\ldots\mathbf{k}_n} \sum_{\mathbf{q}_1\ldots\mathbf{q}_{N-n}} \left|c_{n,\mathbf{k}_1\ldots\mathbf{k}_n\mathbf{q}_1\ldots\mathbf{q}_{N-n}}\right|^2 \to \left|a_n\right|^2, \quad \text{(B35)}$$

thus proving (B19). ∎

### APPENDIX C – ENTANGLEMENT MEASURE BETWEEN PHOTONIC FIELDS IN V-ATOMS

In order to quantify entanglement between the two Fock-numbers describing the two photonic modes, we will use the measure of entanglement negativity [102],

$$\mathcal{N}(X) = \frac{1}{2}\left(\left\|X_A^{T_p}\right\|_1 - 1\right) \quad \text{(C1)}$$

where $\|X\|_1 = Tr\sqrt{X^\dagger X}$ is the trace norm of the bipartite density matrix $X$ and the superscript $T_p$ denotes partial transposition with respect to either degree of freedom of the density matrix. Non-zero values of (C1) attest to non-separability of $X$ and therefore to entanglement of its subsystems. The final state of the ensemble of $N$ indistinguishable V-atoms takes the form of the last term in (B7) at $t \to \infty$, in which the atomic part that consists of only the atomic ground level is common to all photonic states. Therefore, the photonic state

$$\left|\varphi_{f,N}\right\rangle \equiv \sum_{\mathbf{k}_1}\ldots\sum_{\mathbf{k}_N}\sum_{n=0}^{N} c_{n,\mathbf{k}_1\ldots\mathbf{k}_N}(t)\left|(N-n)_{\{N-n\}}, n_{\{n\}}\right\rangle_f \quad \text{(C2)}$$

is a pure state and we may write its density matrix

$$\hat{\rho}_{f,N}(\infty) = \left(\sum_{m=0}^{N}\sum_{\mathbf{k}_1}\ldots\sum_{\mathbf{k}_N} c_{m,\mathbf{k}_1\ldots\mathbf{k}_N}(\infty)\left|m_{\{m\}},(N-m)_{\{N-m\}}\right\rangle\right)\left(\sum_{n=0}^{N}\sum_{\mathbf{k}'_1}\ldots\sum_{\mathbf{k}'_N} c^*_{n,\mathbf{k}'_1\ldots\mathbf{k}'_N}(\infty)\left\langle n_{\{n\}'},(N-n)_{\{N-n\}'}\right|\right). \quad \text{(C3)}$$

From (B12) it is evident that for each $n$, the coefficient $C_{n,\mathbf{k}'_1\ldots\mathbf{k}'_N}(s)$, and consequently also $c_{n,\mathbf{k}'_1\ldots\mathbf{k}'_N}(\infty)$, are proportional to $a_n(0)$. Denoting

$$c_{n,\mathbf{k}'_1\ldots\mathbf{k}'_N}(\infty) \equiv a_n(0)\tilde{c}_{n,\mathbf{k}'_1\ldots\mathbf{k}'_N}(\infty), \quad \text{(C4)}$$

we may now define the orthonormal vectors describing the total number of photons in each of the two modes,

$$\left|n, N-n\right\rangle\!\rangle \equiv \sum_{\mathbf{k}_1}\ldots\sum_{\mathbf{k}_N} \tilde{c}_{n,\mathbf{k}_1\ldots\mathbf{k}_N}(\infty)\left|n_{\{n\}},(N-n)_{\{N-n\}}\right\rangle, \quad \text{(C5)}$$

where the orthonormality emerges from (C4) and (B19). We emphasize that due to the narrow linewidth assumption for all photons, the Lorentzian photon packets centred around $\omega_1$ have negligible overlap with those centred around $\omega_2$, so that both photonic modes are effectively disjoint. Expression (C2) can be written

$$|\varphi_{f,N}\rangle = \sum_{n=0}^{N} a_n(0)|n, N-n\rangle, \tag{C6}$$

meaning that the amplitudes of the photonic state vectors comprising $|\varphi_{f,N}\rangle$ are equal to the excitation amplitudes of the initial atomic state, such that the amplitude of the photonic state with energy $\hbar(n\omega_1 + (N-n)\omega_2)$ is equal to the amplitude of the atomic state of the same energy with respect to the atomic ground state.

In order to calculate the entanglement negativity, we rewrite (C3) as

$$\hat{\rho}_{f,N} = \sum_{m=0}^{N}\sum_{n=0}^{N} a_m(0) a_n^*(0) |m, N-m\rangle\langle\langle n, N-n|. \tag{C7}$$

and take the partial transpose of $\hat{\rho}_{f,N}$ with respect to either photon mode, and plug it in (C1). We illustrate the entanglement negativity in the main text for various excitations and numbers of indistinguishable V-atoms and demonstrate entanglement between the two photonic modes that are emitted.

We remark that since all atoms are indistinguishable, the symmetrized atomic excitation amplitudes $a_n(0)$ are obtained from

$$a_n(0) = \mathrm{N}\binom{N}{n}\alpha^n \beta^{N-n} \tag{C8}$$

where $\alpha, \beta$ are the excitation amplitudes for a single V-atom, satisfying $|\alpha|^2 + |\beta|^2 = 1$ and the normalization constant is given by $\mathrm{N} = \left(\sum_{j=0}^{N}\left|\binom{N}{j}\alpha^j \beta^{N-j}\right|^2\right)^{-1/2}$.

## APPENDIX D – WIGNER DISTRIBUTIONS

We describe the behavior of our final multipartite photonic states using N-mode Wigner distributions. In this work we analyze systems emitting bi- or tri-partite photonic states, which will be described by the generalization of the Wigner distribution for single-mode Fock states by extending the definition of the parity operator to a joint (multimodal) parity operator [114,115]. The two-mode Wigner distribution for the photonic density matrix $\hat{\rho}$ of the indistinguishable non-degenerate V-atoms ensemble is given by

$$W^{2D}(\alpha,\beta) = \frac{4}{\pi^2}\sum_{k,l}^{\infty}(-1)^{k+l}\langle k,l|\hat{\rho}^{2D}\hat{D}(2\alpha)\hat{D}(2\beta)|k,l\rangle \tag{D1}$$

where $\hat{D}(\xi)$ is the displacement operator. The parameters $\alpha = X_1 + iP_1$, $\beta = X_2 + iP_2$ are respectively the phase-space coordinates for each of the two modes. We use the relation [116]

$$\langle m|\hat{D}(\xi)|n\rangle = \sqrt{\frac{n!}{m!}}\xi^{m-n}e^{-|\xi|^2/2}L_n^{(m-n)}\left(|\xi|^2\right) \tag{D2}$$

with $m \geq n$ where $L_n^{(\zeta)}(x)$ are the generalized Laguerre polynomials. For the case $m < n$ we use an identity of the displacement operator, $\hat{D}(\xi) = \hat{D}^\dagger(-\xi)$. Explicitly substituting the final photonic density matrix $\hat{\rho}^{2D} = \sum_{m,p,n,q}\left[\rho^{2D}\right]_{nq}^{mp}|m,p\rangle\langle n,q|$, we re-write (D1)

$$W^{2D}(\alpha,\beta) = \frac{4}{\pi^2} \sum_{m,p,n,q} \sum_{k,l}^{\infty} (-1)^{k+l} \langle k,l|m,p\rangle\langle n,q|\left[\rho^{2D}\right]_{nq}^{mp} \hat{D}(2\alpha)\hat{D}(2\beta)|k,l\rangle =$$
$$= \frac{4}{\pi^2} \sum_{m,p,n,q} (-1)^{m+p} \left[\rho^{2D}\right]_{nq}^{mp} \langle n|\hat{D}(2\alpha)|m\rangle\langle q|\hat{D}(2\beta)|p\rangle \tag{D3}$$

For visualization purposes we plot six two-dimensional slices of the bimodal Wigner distributions (spanning a 4D phase-space $(\alpha,\beta)=(X_1,P_1,X_2,P_2)$), in each case the four-dimensional phase space is sliced at the origin with respect to the two phase-space coordinates that are not visualized. The results for several combination excitations for four atoms are illustrated in Figure 2(c-e). Multimodal Wigner distributions can be used to indicate intermodal entanglement if they cannot be written as sums of products of single-mode Wigner distributions [117,118]. For example, in the case presented in Figure 2e we see that the two modes are not correlated, since all four cross-mode phase-space coordinate pairs are products of the projections on the coordinate pairs $(X_1,P_1)$ and $(X_2,P_2)$ corresponding single-mode Wigner distribution coordinates. In Figure 8a,b we demonstrate this separability for the bimodal Wigner distribution on the hypersurface $(X_1,0,0,P_2)$; it can be shown that for a deterministic excitation $\alpha=0$ or $\beta=0$ this separability holds for any value of $P_1$ and $X_2$, and generally for any hypersurface, therefore indicating that the two modes are not entangled. In contrast, the case $\alpha=\beta=2^{-1/2}$ is shown in Figure 8c,d on the hypersurface $(X_1,0,X_2,0)$, and it is evident that the bimodal Wigner distribution on that hyperspace is not the product of the single-mode Wigner distributions, and therefore the two modes are entangled. This can be also surmised from the lack of axial symmetries of the cross phase-space coordinates Wigner distribution projections; these are necessary for separability since the single-mode Wigner distributions are symmetric.

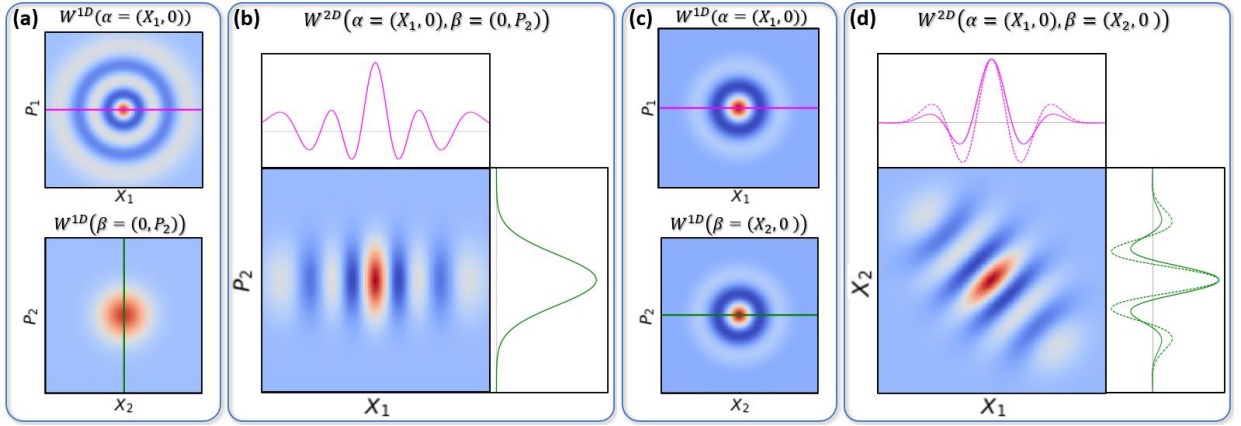

Figure 8 – (a) Single-mode Wigner distributions for four indistinguishable non-degenerate V-atoms with deterministic excitation, shown in Figure 2e. Magenta and green lines indicate the slicing of the Wigner distribution at $P_1=0$ for the first mode and $X_2=0$ for the second mode, respectively. (b) Two-mode Wigner distribution projected on the hypersurface $(X_1,0,0,P_2)$. The magenta and green plots are the marginal quasi-probability distributions along the $X_1$ and $P_2$ phase-space coordinates on the hypersurface. These are identical to the values of the 1D Wigner distributions along the magenta and green lines in subfigure (a), therefore $W^{2D}(X_1,0,0,P_2) = W^{1D}(X_1,0)W^{1D}(0,P_2)$. It can be shown that this holds for any value of $P_1$ and $X_2$, thus indicating separability of the two modes. (c,d) Same as subfigures (a) and (b) but for the superposition excitation $\alpha=\beta=\sqrt{1/2}$ shown in Figure 2c. In subfigure (d) we show the projection on the hyperplane $(X_1,0,X_2,0)$, with the marginal quasi-probability distributions in dashed lines compared to the solid magenta and green lines corresponding to subfigure (c), indicating that $W^{2D}(X_1,0,X_2,0) \neq W^{1D}(X_1,0)W^{1D}(X_2,0)$ so that entanglement between the two modes exists.

The three-mode Wigner distribution relevant for the tripartite photonic state emanated from an ensemble of indistinguishable FLAs will analogously be

$$W^{3D}(\alpha,\beta,\gamma) = \frac{8}{\pi^3} \sum_{k,l,s}^{\infty} (-1)^{k+l+s} \langle k,l,s|\hat{\rho}^{3D}\hat{D}(2\alpha)\hat{D}(2\beta)\hat{D}(2\gamma)|k,l,s\rangle \tag{D4}$$

and for $\hat{\rho}^{3D} = \sum_{m,p,v,n,q,u} \left[\rho^{3D}\right]_{nqu}^{mpv} |m,p,v\rangle\langle n,q,u|$ we have

$$W^{3D}(\alpha,\beta,\gamma) = \frac{8}{\pi^3} \sum_{\substack{m,p,v,\\n,q,u}} (-1)^{m+p+v} \left[\rho^{3D}\right]_{nqu}^{mpv} \langle n|\hat{D}(2\alpha)|m\rangle\langle q|\hat{D}(2\beta)|p\rangle\langle u|\hat{D}(2\gamma)|v\rangle. \tag{D5}$$

The resulting Wigner distribution correspondingly spans a 6D phase-space $(\alpha, \beta, \gamma) = (X_1, P_1, X_2, P_2, X_3, P_3)$. Again, for visualization purposes we show fifteen phase-space hypersurfaces that are each parallel to a coordinate pair, and are located at zero with respect to the other four coordinates, i.e. $(X_1, 0,0, P_2, 0,0)$, $(0, P_1, 0,0,0, P_3)$, etc. Indication of entanglement in the tripartite case is similar to the bipartite case discussed above.

Figure 9 describes the tri-modal Wigner distribution for two indistinguishable FLAs with equal transition rates, as analyzed in APPENDIX H. We see again in Figure 9a that in the case of a deterministic excitation, the Wigner distribution of one of the photonic modes (either $\omega_-$ or $\omega_+$) is that of a vacuum state and its phase-space coordinates are uncorrelated to the phase-space coordinates of the other two modes. This is indicated by the cross-modal Wigner distribution projections that are either a product or a sum of products of the single-mode Wigner distributions. This is expected since for a deterministic excitation the FLA photonic density matrix (H20) is diagonal and can be written as $\hat{\rho}_f^{2-FLA} = \frac{1}{3}\left(|\tilde{\gamma}\rangle\langle\tilde{\gamma}| + |\tilde{\varepsilon}\rangle\langle\tilde{\varepsilon}| + |\tilde{\zeta}\rangle\langle\tilde{\zeta}|\right)$ and therefore the total tri-modal Wigner distribution is equal to the sum of partial tri-modal Wigner distributions of the non-zero density matrix elements,

$$W^{3D} = W^{3D}_{\tilde{\gamma}} + W^{3D}_{\tilde{\varepsilon}} + W^{3D}_{\tilde{\zeta}} \tag{D6}$$

(the dependence on phase-space coordinates is implicit), where e.g. $W^{3D}_{\tilde{\gamma}}$ is obtained from (D5) by substituting $\frac{1}{3}|\tilde{\gamma}\rangle\langle\tilde{\gamma}|$ as the density matrix. In Figure 9b we see as an example that the projection of the tri-modal Wigner distribution on the $(P_0, P_+)$ coordinates is equal to the sum of three projections on $(P_0, P_+)$ of the tri-modal Wigner distributions corresponding to the nonzero density matrix elements $W^{3D}_{\tilde{\gamma}}, W^{3D}_{\tilde{\varepsilon}}$ and $W^{3D}_{\tilde{\zeta}}$; each of these Wigner distributions projections is a product of the corresponding partial single-mode Wigner distributions as can be numerically verified, e.g., $W^{3D}_{\tilde{\gamma}}(0,0,0,P_0,0,P_+) = W^{1D}_{\tilde{\gamma}}(0,P_0)W^{1D}_{\tilde{\gamma}}(0,P_+)$. We therefore conclude that there is no photon entanglement in the deterministic excitation case, as expected from the calculation of the conditional entanglement entropy (see Section III.B).

When the excitation is into a superposition state, cross-modal phase-space coordinate correlations emerge between phase-space coordinates belonging to all three modes, indicating intermodal entanglement, see Figure 7 in the main text.

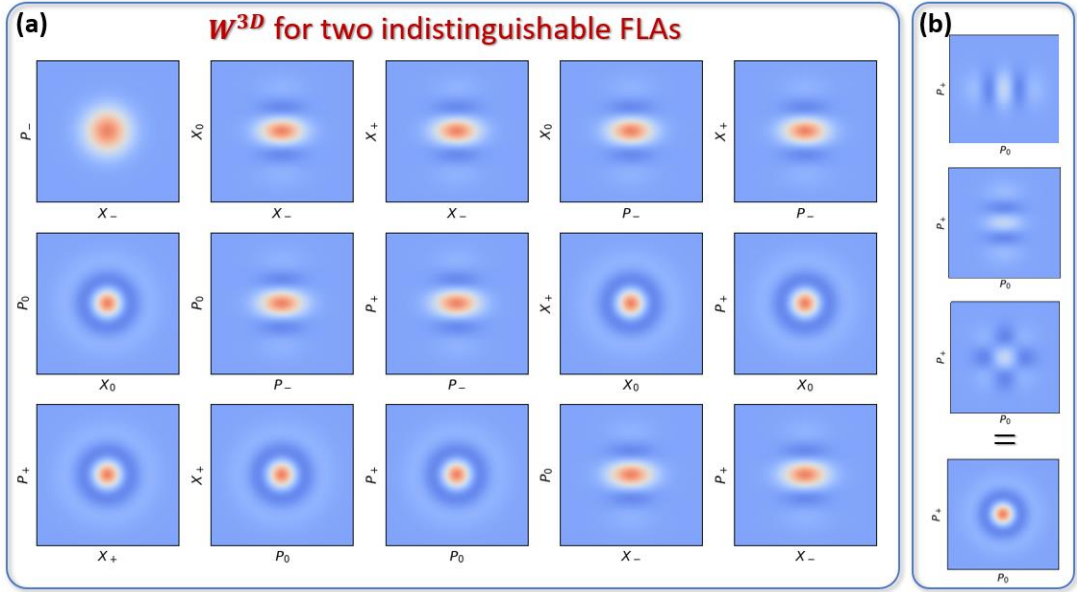

Figure 9 - Three-modal Wigner distribution $W^{3D}(X_-, P_-, X_0, P_0, X_+, P_+)$ for two indistinguishable FLAs with equal radiative decay rates, excited deterministically with $\alpha = 0, \beta = 1$. (a) The fifteen 2D cross-sections, each parallel to a pair of phase-space coordinates and sliced at zero with respect to the others. The left column contains the single-modal Wigner distributions for each mode. The remaining twelve projections exhibit either fringes that are equal to the product of two phase-space coordinates (in a similar fashion as Figure 8a,b), or (b) distributions that are sums of products of two phase-space coordinates for Wigner distributions on parts of the total density matrix.

## APPENDIX E – MASTER EQUATION FOR DEGENERATE V-ATOMS

Calculation of the master equation in the degenerate $N$ indistinguishable V-atom case is straightforward and is based in large on the calculations of APPENDIX A. The difference is that now $\omega_0 \equiv \omega_1 = \omega_2$, so that the photonic modes in (A9) overlap (in fact they are identical) and so we obtain cross-terms in the master equation between the two atomic creation/annihilation operators. To see that, we re-write (A9) for the case of two distinct radiative transitions at angular frequency $\omega_0$,

$$\frac{d}{dt}\hat{\Theta}^I(t) = -Tr_f\left\{\int_0^\infty d\tau\left[\sum_{i=1}^N\sum_{\mathbf{k}}\left\{\hat{a}_{1,\mathbf{k}}^\dagger e^{i(\omega_{\mathbf{k}}-\omega_0)t}\left\{g_{1,\mathbf{k}}^i(\mathbf{r})\hat{\sigma}_{ge_1}^i + g_{2,\mathbf{k}}^i(\mathbf{r})\hat{\sigma}_{ge_2}^i\right\} + \hat{a}_{1,\mathbf{k}}e^{-i(\omega_{\mathbf{k}}-\omega_0)t}\left\{g_{1,\mathbf{k}}^{i*}(\mathbf{r})\hat{\sigma}_{ge_1}^{i\dagger} + g_{2,\mathbf{k}}^{i*}(\mathbf{r})\hat{\sigma}_{ge_2}^{i\dagger}\right\}\right\}\right.\right.$$

$$\left.\left.,\left[\sum_{i=1}^N\sum_{\mathbf{q}}\left\{\hat{a}_{1,\mathbf{q}}^\dagger e^{i(\omega_{\mathbf{q}}-\omega_0)(t-\tau)}\left\{g_{1,\mathbf{q}}^i(\mathbf{r})\hat{\sigma}_{ge_1}^i + g_{2,\mathbf{q}}^i(\mathbf{r})\hat{\sigma}_{ge_2}^i\right\} + \hat{a}_{1,\mathbf{q}}e^{-i(\omega_{\mathbf{q}}-\omega_0)(t-\tau)}\left\{g_{1,\mathbf{q}}^{i*}(\mathbf{r})\hat{\sigma}_{ge_1}^{i\dagger} + g_{2,\mathbf{q}}^{i*}(\mathbf{r})\hat{\sigma}_{ge_2}^{i\dagger}\right\}\right\},\hat{\Theta}^I(t-\tau)\otimes\hat{\Phi}(0)\right]\right\}\quad(E1)$$

Using the property (A10) and substituting the symmetrized atomic operators (A13) we arrive at

$$\frac{d}{dt}\hat{\Theta}^I(t) = -\int_0^\infty d\tau \sum_{l,m=\{1,2\}}\sum_{\mathbf{k}}|g_{j,\mathbf{k}}|^2\left\{e^{-i\omega_{\mathbf{k}}\tau}\left\{e^{i\omega_0\tau}\left[\hat{\Sigma}_{ge_l}^\dagger,\hat{\Sigma}_{ge_m}\hat{\Theta}^I(t-\tau)\right] + e^{-i\omega_0\tau}\left[\hat{\Sigma}_{ge_l},\hat{\Sigma}_{ge_m}^\dagger\hat{\Theta}^I(t-\tau)\right]\right\}\right.$$

$$\left. + e^{i\omega_{\mathbf{k}}\tau}\left\{e^{-i\omega_0\tau}\left[\hat{\Sigma}_{ge_l},\hat{\Sigma}_{ge_m}^\dagger\hat{\Theta}^I(t-\tau)\right] + e^{i\omega_0\tau}\left[\hat{\Sigma}_{ge_l}^\dagger,\hat{\Sigma}_{ge_m}\hat{\Theta}^I(t-\tau)\right]\right\}\right\} \quad (E2)$$

Repeating the procedure of (A15)-(A18) we finally obtain (A18) with the Lindblad operator for the degenerate case

$$\hat{L}_d\{\hat{\Theta}(t)\} = \sum_{l,m=\{1,2\}}\frac{\Gamma_{lm}}{2}\left\{\hat{\Sigma}_{ge_l}^\dagger\hat{\Sigma}_{ge_m}\hat{\Theta}(t) + \hat{\Theta}(t)\hat{\Sigma}_{ge_l}^\dagger\hat{\Sigma}_{ge_m} - 2\hat{\Sigma}_{ge_l}\hat{\Theta}(t)\hat{\Sigma}_{ge_m}^\dagger\right\}. \quad (E3)$$

The decay rates can be obtained in a similar manner,

$$\Gamma_{lm} = \frac{\mathbf{d}_l\cdot\mathbf{d}_m\omega_0^3}{3\pi\varepsilon_0\hbar c^3}. \quad (E4)$$

Despite the similar form of (E3) with comparison to the Lindblad operator defined in (A1), the cross-terms embodied in the "cross" decay rate enrich the superradiance dynamics as discussed in the main text, and the ODE system of the density matrix elements in Schrodinger picture is

$$\frac{\partial}{\partial t}\Theta_{n_{e1},n_{e_2},n_g}^{m_{e1},m_{e_2},m_g} = -i\frac{\omega_0}{2}(m_1+m_2-n_1-n_2)\Theta_{n_{e1},n_{e_2},n_g}^{m_{e1},m_{e_2},m_g}$$

$$-\frac{\Gamma_{11}}{2}\left\{\left[m_1(m_g+1)+n_1(n_g+1)\right]\Theta_{n_{e1},n_{e_2},n_g}^{m_{e1},m_{e_2},m_g} - 2\sqrt{m_g(m_1+1)n_g(n_1+1)}\Theta_{n_{e_1}+1,n_{e_2},n_g-1}^{m_{e_1}+1,m_{e_2},m_g-1}\right\}$$

$$-\frac{\Gamma_{22}}{2}\left\{\left[m_2(m_g+1)+n_2(n_g+1)\right]\Theta_{n_{e1},n_{e_2},n_g}^{m_{e1},m_{e_2},m_g} - 2\sqrt{m_g(m_2+1)n_g(n_2+1)}\Theta_{n_{e_1},n_{e_2}+1,n_g-1}^{m_{e_1},m_{e_2}+1,m_g-1}\right\}$$

$$-\frac{\Gamma_{12}}{2}\left\{(m_g+1)\sqrt{m_2(m_1+1)}\Theta_{n_{e1},n_{e_2},n_g}^{m_{e1}+1,m_{e_2}-1,m_g} + (n_g+1)\sqrt{n_2(n_1+1)}\Theta_{n_{e_1}+1,n_{e_2}-1,n_g}^{m_{e1},m_{e_2},m_g} - 2\sqrt{m_g(m_1+1)n_g(n_2+1)}\Theta_{n_{e_1},n_{e_2}+1,n_g-1}^{m_{e_1}+1,m_{e_2},m_g-1}\right\}$$

$$-\frac{\Gamma_{21}}{2}\left\{(m_g+1)\sqrt{m_1(m_2+1)}\Theta_{n_{e1},n_{e_2},n_g}^{m_{e1}-1,m_{e_2}+1,m_g} + (n_g+1)\sqrt{n_1(n_2+1)}\Theta_{n_{e_1}-1,n_{e_2}+1,n_g}^{m_{e1},m_{e_2},m_g} - 2\sqrt{m_g(m_2+1)n_g(n_1+1)}\Theta_{n_{e_1}+1,n_{e_2},n_g-1}^{m_{e_1},m_{e_2}+1,m_g-1}\right\}\quad(E5)$$

To demonstrate the emergence of a partially dark state in the case of symmetric excitation in the case of an ensemble indistinguishable atoms (rather than distinguishable ones), and vice-versa, consider the ordinary differential equations governing the evolution of the atomic density matrix in the single V-atom case:

$$\frac{d}{dt}\Theta_{100}^{100} = -\Gamma\left(2\Theta_{100}^{100}+\Theta_{010}^{100}+\Theta_{100}^{010}\right), \quad \frac{d}{dt}\Theta_{010}^{010} = -\Gamma\left(2\Theta_{010}^{010}+\Theta_{010}^{100}+\Theta_{100}^{010}\right),$$

$$\frac{d}{dt}\Theta_{100}^{001} = \frac{d}{dt}\Theta_{010}^{001} = -\Gamma\left(\Theta_{100}^{001}+\Theta_{010}^{001}\right), \quad \frac{d}{dt}\Theta_{100}^{010} = -\Gamma\left(2\Theta_{100}^{010}+\Theta_{100}^{100}+\Theta_{010}^{010}\right) \quad (E6)$$

If the atoms are distinguishable, the total density matrix of an $N$ V-atom ensemble is simply the tensor product of $N$ single atom density matrices. This means that an ensemble initially excited to the symmetric or antisymmetric state will fully decay to the ground state or will not evolve at all, respectively. However, when the atoms are indistinguishable, the master equation describing the atomic density matrix is (E5), which is identical to equation (2), with additional six terms on the RHS. To elucidate the difference in the dynamics when the atoms are indistinguishable, consider the simplest case of $N = 2$ indistinguishable V-atoms with the initial antisymmetric excitation

$$\left|\psi_{2-V,i}^{AS}\right\rangle = 2^{-1}\hat{S}\left\{\prod_{j=1}^{2}\left(\left|e_1\right\rangle_j - \left|e_2\right\rangle_j\right)\right\} = 6^{-1/2}\left(\left|2,0,0\right\rangle - 2\left|1,1,0\right\rangle + \left|0,2,0\right\rangle\right) \tag{E7}$$

and transition dipole moments such that $\Gamma_{lm} = \Gamma$ for any $l, m$, the initially excited terms evolve at $t = 0$ as

$$\frac{\partial}{\partial t}\Theta_{200}^{200} = \frac{\partial}{\partial t}\Theta_{020}^{020} = \frac{\partial}{\partial t}\Theta_{200}^{020} = \frac{\partial}{\partial t}\Theta_{020}^{200} = \frac{\Gamma}{3}\left(\sqrt{2}-1\right)$$
$$\frac{\partial}{\partial t}\Theta_{110}^{110} = -\frac{2\Gamma}{3}\left(\sqrt{2}+2\right)$$
$$\frac{\partial}{\partial t}\Theta_{200}^{110} = \frac{\partial}{\partial t}\Theta_{020}^{110} = \frac{\partial}{\partial t}\Theta_{110}^{200} = \frac{\partial}{\partial t}\Theta_{110}^{020} = -\frac{\Gamma}{6}\left(3\sqrt{2}-4\right). \tag{E8}$$

This result stands in contrast to the case of two distinguishable V-atoms mentioned above, in which an antisymmetric ("dark") excitation results in no evolution of the system. The enhancement or suppression of decay for the single degenerate atom case is attributed to constructive or destructive interferences, respectively, between the two superimposed decay paths. When several indistinguishable atoms are considered, the symmetrization operator reduces the dimension of the Hilbert space and some coherence information is inevitably modified. Therefore, we obtain that in the indistinguishable case the evolution paths do not interfere entirely constructively or destructively.

The intra-atomic degeneracy introduces couplings between the transitions $e_1 \rightarrow g$ and $g \rightarrow e_2$ (or the complementary pair) which do not induce photon emission, as all last six terms on the RHS of (E5) do not directly couple population terms with different numbers of excitations – and therefore conserve the energy of the atomic system – unlike the first four real terms on the RHS of (E5).

In an analogous manner to APPENDIX A, the total (coherent) intensity is given by

$$I_t(t) = 2\omega_0\left(\left\langle\Gamma_{11}\hat{\Sigma}_{ge_1}^{\dagger}\hat{\Sigma}_{ge_1} + \Gamma_{22}\hat{\Sigma}_{ge_2}^{\dagger}\hat{\Sigma}_{ge_2}\right\rangle + \left\langle\Gamma_{12}\hat{\Sigma}_{ge_1}^{\dagger}\hat{\Sigma}_{ge_2} + \Gamma_{21}\hat{\Sigma}_{ge_2}^{\dagger}\hat{\Sigma}_{ge_1}\right\rangle\right) \equiv I_1 + I_2 + C_{12} + C_{21}. \tag{E9}$$

The first two terms on the RHS consist of the intensities of photons emitted from the excited $\left|e_1\right\rangle$ or $\left|e_2\right\rangle$ states, whereas the last two terms describe the coherences between the two degenerate transitions, which account for the effect of the virtual transitions. The ensemble undergoes superradiant emission of photons in angular frequency $\omega_0$, however – due to the degeneracy, the total emission intensity $I_t$ consists of contributions from the intensities of the two radiative transitions $I_1$ and $I_2$, and coherence terms (interferences) stemming from the virtual transitions, $C_{12}$ and $C_{21}$, enhancing the emission (on top of the superradiance effect originating from the indistinguishability of the emitters). In Figure 10, we illustrate the total intensity for an ensemble of $N = 8$ indistinguishable degenerate V-atoms excited to the symmetric initial state

$$\left|\psi_{8-V,i}^{S}\right\rangle = N\hat{S}\left\{\prod_{j=1}^{8}\left(\left|e_1\right\rangle_j + \left|e_2\right\rangle_j\right)\right\} \tag{E10}$$

with the normalization constant defined in (C8) and visualize the effect of the coherences on the intensity, for $\Gamma_{11} = \Gamma_{22} = \Gamma$. In Figure 10a we have $\Gamma_{12} = \Gamma_{21} = \Gamma$, and in Figure 10b we have $\Gamma_{12} = \Gamma_{21} = 0$ – meaning no intra-atom virtual transitions take place – which occurs when the two transition dipole moments are orthogonal (see eq. (E4)). From comparing the two figures, we see that the virtual transitions enhance the superradiant emission rate of the ensemble. The insets focus on the emission curves at long times. We see that in the first case, the $I_1$ and $I_2$ intensities attain a positive value, that is negated by the equal and negative values of the coherences $C_{12}$ and $C_{21}$ such that the overall intensity approaches zero as $t \rightarrow \infty$. In this case, the total population of the symmetrized ground state at large times resulting from the modification of the coherence information due to the symmetrizaion operation is numerically found to be ~94.8%, meaning that the remaining ~5.2% of the total excitation populates dark states, in contrast to the distinguishable degenerate V-atom case. In the second case, where no virtual transitions occur, both $I_1$ and $I_2$ approach zero at long times, the coherences are zero, and the entire population occupies the symmetrized ground state. This example illustrates how the correlations stemming from the virtual transitions in the indistinguishable ensemble of degenerate V-atoms are related to the emergence of dark states.

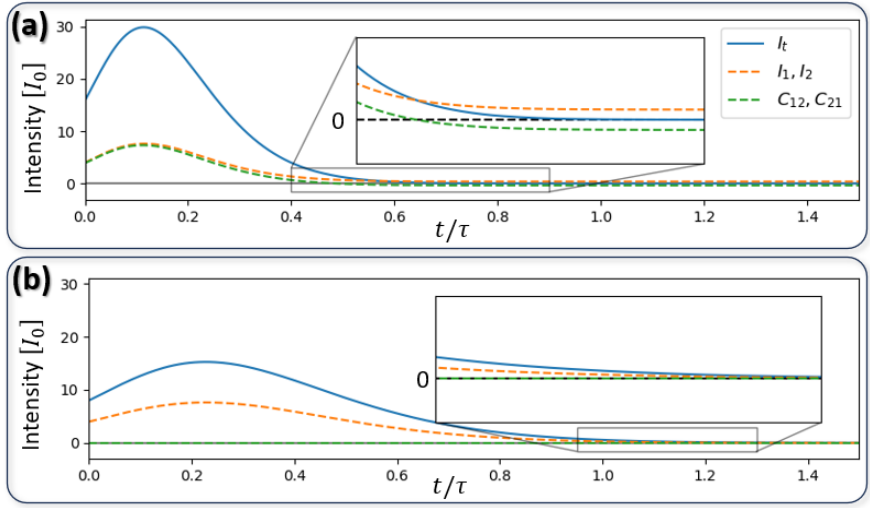

Figure 10 – Intensity vs. time plots for $N = 8$ indistinguishable degenerate V-atoms with $\Gamma_{11} = \Gamma_{22} = \Gamma$. Time is given in units of $\tau = 1/\Gamma$. Intensity is given in units of $I_0$ - intensity of a two-level atom with angular frequency and decay rate $\omega_0$ and $\Gamma$. Solid blue line denotes the total intensity. Dashed orange line denotes the intensities $I_1, I_2$ resulting from the first and second transitions only. Dashed green line denotes the correlations $C_{12}, C_{21}$ between the first and second transitions. (a) virtual transitions occur, with $\Gamma_{12} = \Gamma_{21} = \Gamma$. (b) no virtual transitions occur, i.e. $\Gamma_{12} = \Gamma_{21} = 0$.

## APPENDIX F – MASTER EQUATION FOR PARTIALLY DEGENERATE FOUR-LEVEL ATOMS

The development of the master equation for the indistinguishable FLA ensemble discussed in Section III is straightforward, albeit laborious. The Hamiltonians (A3) and (A6) are modified to include the additional ground level and the four possible radiative transitions supported the FLA,

$$\hat{H}_A^{FLA} = \hbar\omega_0 \sum_{i=1}^N \hat{n}_{e_1}^i + \hbar\omega_+ \sum_{i=1}^N \hat{n}_{e_2}^i + \hbar(\omega_0 - \omega_-)\sum_{i=1}^N \hat{n}_{g_2}^i \tag{F1}$$

$$\hat{H}_f^{FLA} = \hbar\sum_{\mathbf{k}} \omega_\mathbf{k} \left( \hat{a}_{0,\mathbf{k}}^\dagger \hat{a}_{0,\mathbf{k}} + \hat{a}_{+,\mathbf{k}}^\dagger \hat{a}_{+,\mathbf{k}} + \hat{a}_{-,\mathbf{k}}^\dagger \hat{a}_{-,\mathbf{k}} \right) \tag{F2}$$

$$\hat{H}_I^{FLA} = \hbar\sum_{i=1}^N \sum_{\mathbf{k}} \left\{ \hat{a}_{0,\mathbf{k}}^\dagger e^{i(\omega_\mathbf{k}-\omega_0)t} \left[ g_{11,\mathbf{k}}^i(\mathbf{r}_i)\hat{\sigma}_{g_1 e_1}^i + g_{22,\mathbf{k}}^i(\mathbf{r}_i)\hat{\sigma}_{g_2 e_2}^i \right] \right.$$
$$\left. + g_{21,\mathbf{k}}^i(\mathbf{r}_i)\hat{a}_{+,\mathbf{k}}^\dagger \hat{\sigma}_{g_1 e_2}^i e^{i(\omega_\mathbf{k}-\omega_+)t} + g_{12,\mathbf{k}}^i(\mathbf{r}_i)\hat{a}_{-,\mathbf{k}}^\dagger \hat{\sigma}_{g_2 e_1}^i e^{i(\omega_\mathbf{k}-\omega_-)t} + h.c. \right\}. \tag{F3}$$

The coupling coefficients in the interaction Hamiltonian have subscripts commensurate with the atomic levels between which the radiative transition occurs

$$g_{lm,\mathbf{k}}^i(\mathbf{r}_i) \equiv g_{e_l g_m,\mathbf{k}}^i(\mathbf{r}_i) = -i\sqrt{\frac{2\pi ck}{V}} \mathbf{d}_{g_l e_m}^i \cdot \mathbf{e}_\mathbf{k} e^{i\mathbf{k}\cdot\mathbf{r}_i} \tag{F4}$$

for $l, m \in \{1, 2\}$ with the definitions of the parameters as in (A7). The commutation relations (A5) are naturally generalized to the photonic operators $\hat{a}_{0,\mathbf{k}}$, $\hat{a}_{\pm,\mathbf{k}}$, and consequently so are the relations (A10). The integrodifferential operator equation under the Born-Markov approximation in the interaction picture is

$$\frac{d}{dt}\hat{\Theta}^I(t) = -Tr_f\left\{\int_0^\infty d\tau\left[\sum_{i=1}^N\sum_{\mathbf{k}}\left\{\hat{a}_{0,\mathbf{k}}^\dagger e^{i(\omega_\mathbf{k}-\omega_0)t}\left[g_{11,\mathbf{k}}^i(\mathbf{r}_i)\hat{\sigma}_{g_1e_1}^i + g_{22,\mathbf{k}}^i(\mathbf{r}_i)\hat{\sigma}_{g_2e_2}^i\right]\right.\right.\right.$$
$$+ g_{21,\mathbf{k}}^i(\mathbf{r}_i)\hat{a}_{+,\mathbf{k}}^\dagger\hat{\sigma}_{g_1e_2}^i e^{i(\omega_\mathbf{k}-\omega_+)t} + g_{12,\mathbf{k}}^i(\mathbf{r}_i)\hat{a}_{-,\mathbf{k}}^\dagger\hat{\sigma}_{g_2e_1}^i e^{i(\omega_\mathbf{k}-\omega_-)t}$$
$$+ \hat{a}_{0,\mathbf{k}}e^{-i(\omega_\mathbf{k}-\omega_0)t}\left[g_{11,\mathbf{k}}^{i*}(\mathbf{r}_i)\hat{\sigma}_{g_1e_1}^{i\dagger} + g_{22,\mathbf{k}}^{i*}(\mathbf{r}_i)\hat{\sigma}_{g_2e_2}^{i\dagger}\right]$$
$$\left.+ g_{21,\mathbf{k}}^{i*}(\mathbf{r}_i)\hat{a}_{+,\mathbf{k}}\hat{\sigma}_{g_1e_2}^{i\dagger}e^{-i(\omega_\mathbf{k}-\omega_+)t} + g_{12,\mathbf{k}}^{i*}(\mathbf{r}_i)\hat{a}_{-,\mathbf{k}}\hat{\sigma}_{g_2e_1}^{i\dagger}e^{-i(\omega_\mathbf{k}-\omega_-)t}\right\}$$
$$,\left[\sum_{i=1}^N\sum_{\mathbf{q}}\left\{\hat{a}_{0,\mathbf{q}}^\dagger e^{i(\omega_\mathbf{q}-\omega_0)(t-\tau)}\left[g_{11,\mathbf{q}}^i(\mathbf{r}_i)\hat{\sigma}_{g_1e_1}^i + g_{22,\mathbf{q}}^i(\mathbf{r}_i)\hat{\sigma}_{g_2e_2}^i\right]\right.\right.$$
$$+ g_{21,\mathbf{q}}^i(\mathbf{r}_i)\hat{a}_{+,\mathbf{q}}^\dagger\hat{\sigma}_{g_1e_2}^i e^{i(\omega_\mathbf{q}-\omega_+)(t-\tau)} + g_{12,\mathbf{q}}^i(\mathbf{r}_i)\hat{a}_{-,\mathbf{q}}^\dagger\hat{\sigma}_{g_2e_1}^i e^{i(\omega_\mathbf{q}-\omega_-)(t-\tau)}$$
$$+ \hat{a}_{0,\mathbf{q}}e^{-i(\omega_\mathbf{q}-\omega_0)(t-\tau)}\left[g_{11,\mathbf{q}}^{i*}(\mathbf{r}_i)\hat{\sigma}_{g_1e_1}^{i\dagger} + g_{22,\mathbf{q}}^{i*}(\mathbf{r}_i)\hat{\sigma}_{g_2e_2}^i\right]$$
$$\left.\left.\left.+ g_{21,\mathbf{q}}^{i*}(\mathbf{r}_i)\hat{a}_{+,\mathbf{q}}\hat{\sigma}_{g_1e_2}^{i\dagger}e^{-i(\omega_\mathbf{q}-\omega_+)(t-\tau)} + g_{12,\mathbf{q}}^{i*}(\mathbf{r}_i)\hat{a}_{-,\mathbf{q}}\hat{\sigma}_{g_2e_1}^{i\dagger}e^{-i(\omega_\mathbf{q}-\omega_-)(t-\tau)}\right\},\hat{\Theta}^I(t-\tau)\otimes\hat{\Phi}(0)\right]\right]\right\} \quad (F5)$$

As before, we introduce the symmetrized atomic operators accounting for the four possible radiative transitions acting on the atomic levels Fock number states,

$$\hat{\Sigma}_{g_2e_2}|n_{e_2},n_{e_1},n_{g_2},n_{g_1}\rangle = \sqrt{n_{e_2}(n_{g_2}+1)}|n_{e_2}-1,n_{e_1},n_{g_2}+1,n_{g_1}\rangle \quad \hat{\Sigma}_{g_2e_2}^\dagger|n_{e_2},n_{e_1},n_{g_2},n_{g_1}\rangle = \sqrt{n_{g_2}(n_{e_2}+1)}|n_{e_2}+1,n_{e_1},n_{g_2}-1,n_{g_1}\rangle$$
$$\hat{\Sigma}_{g_1e_2}|n_{e_2},n_{e_1},n_{g_2},n_{g_1}\rangle = \sqrt{n_{e_2}(n_{g_1}+1)}|n_{e_2}-1,n_{e_1},n_{g_2},n_{g_1}+1\rangle \quad \hat{\Sigma}_{g_1e_2}^\dagger|n_{e_2},n_{e_1},n_{g_2},n_{g_1}\rangle = \sqrt{n_{g_1}(n_{e_2}+1)}|n_{e_2}+1,n_{e_1},n_{g_2},n_{g_1}-1\rangle$$
$$\hat{\Sigma}_{g_2e_1}|n_{e_2},n_{e_1},n_{g_2},n_{g_1}\rangle = \sqrt{n_{e_1}(n_{g_2}+1)}|n_{e_2},n_{e_1}-1,n_{g_2}+1,n_{g_1}\rangle \quad \hat{\Sigma}_{g_2e_1}^\dagger|n_{e_2},n_{e_1},n_{g_2},n_{g_1}\rangle = \sqrt{n_{g_2}(n_{e_1}+1)}|n_{e_2},n_{e_1}+1,n_{g_2}-1,n_{g_1}\rangle \quad (F6)$$
$$\hat{\Sigma}_{g_1e_1}|n_{e_2},n_{e_1},n_{g_2},n_{g_1}\rangle = \sqrt{n_{e_1}(n_{g_1}+1)}|n_{e_2},n_{e_1}-1,n_{g_2},n_{g_1}+1\rangle \quad \hat{\Sigma}_{g_1e_2}^\dagger|n_{e_2},n_{e_1},n_{g_2},n_{g_1}\rangle = \sqrt{n_{g_1}(n_{e_1}+1)}|n_{e_2},n_{e_1}+1,n_{g_2},n_{g_1}-1\rangle$$

and then calculation of this integral is straightforward, using the relations (A10),

$$\frac{d}{dt}\hat{\Theta}^I(t) = -\int_0^\infty d\tau\sum_{\mathbf{k}}\left\{\sum_{l,m=\{1,2\}}g_{ll,\mathbf{k}}^*g_{mm,\mathbf{k}}\left\{e^{-i\omega_\mathbf{k}\tau}\left\{e^{i\omega_0\tau}\left[\hat{\Sigma}_{g_me_m}^\dagger,\hat{\Sigma}_{g_le_l}\hat{\Theta}^I(t-\tau)\right]+e^{-i\omega_0\tau}\left[\hat{\Sigma}_{g_me_m},\hat{\Sigma}_{g_le_l}^\dagger\hat{\Theta}^I(t-\tau)\right]\right\}\right.\right.$$
$$\left.+ e^{i\omega_\mathbf{k}\tau}\left\{e^{-i\omega_0\tau}\left[\hat{\Sigma}_{g_me_m},\hat{\Sigma}_{g_le_l}^\dagger\hat{\Theta}^I(t-\tau)\right]+e^{i\omega_0\tau}\left[\hat{\Sigma}_{g_me_m}^\dagger,\hat{\Sigma}_{g_le_l}\hat{\Theta}^I(t-\tau)\right]\right\}\right\}$$
$$+|g_{21,\mathbf{k}}|^2\left\{e^{-i\omega_\mathbf{k}\tau}\left\{e^{i\omega_+\tau}\left[\hat{\Sigma}_{g_1e_2}^\dagger,\hat{\Sigma}_{g_1e_2}\hat{\Theta}^I(t-\tau)\right]+e^{-i\omega_+\tau}\left[\hat{\Sigma}_{g_1e_2},\hat{\Sigma}_{g_1e_2}^\dagger\hat{\Theta}^I(t-\tau)\right]\right\}\right.$$
$$\left.+ e^{i\omega_\mathbf{k}\tau}\left\{e^{-i\omega_+\tau}\left[\hat{\Sigma}_{g_1e_2},\hat{\Sigma}_{g_1e_2}^\dagger\hat{\Theta}^I(t-\tau)\right]+e^{i\omega_+\tau}\left[\hat{\Sigma}_{g_1e_2}^\dagger,\hat{\Sigma}_{g_1e_2}\hat{\Theta}^I(t-\tau)\right]\right\}\right\}$$
$$+|g_{12,\mathbf{k}}|^2\left\{e^{-i\omega_\mathbf{k}\tau}\left\{e^{i\omega_-\tau}\left[\hat{\Sigma}_{g_2e_1}^\dagger,\hat{\Sigma}_{g_2e_1}\hat{\Theta}^I(t-\tau)\right]+e^{-i\omega_-\tau}\left[\hat{\Sigma}_{g_2e_1},\hat{\Sigma}_{g_2e_1}^\dagger\hat{\Theta}^I(t-\tau)\right]\right\}\right.$$
$$\left.\left.+ e^{i\omega_\mathbf{k}\tau}\left\{e^{-i\omega_-\tau}\left[\hat{\Sigma}_{g_2e_1},\hat{\Sigma}_{g_2e_1}^\dagger\hat{\Theta}^I(t-\tau)\right]+e^{i\omega_-\tau}\left[\hat{\Sigma}_{g_2e_1}^\dagger,\hat{\Sigma}_{g_2e_1}\hat{\Theta}^I(t-\tau)\right]\right\}\right\}\right\} \quad (F7)$$

The first brackets in the integral describe interference between the two degenerate transitions out of the four radiative transitions, due to the fact that the photonic operators $\hat{a}_{0,\mathbf{k}}$ multiply two distinct atomic operators. The two other brackets describe the remaining two transitions which are non-interfering. After Laplace-transforming (F7) and transforming to Schrodinger picture assuming no dipole-dipole interactions as done in APPENDIX A, the Lindblad operator becomes

$$\hat{L}_{FLA}\{\hat{\Theta}(t)\} = \sum_{l,m=\{1,2\}}\frac{\Gamma_{lm}}{2}\left\{\hat{\Sigma}_{g_me_m}^\dagger\hat{\Sigma}_{g_le_l}\hat{\Theta}(t) + \hat{\Theta}(t)\hat{\Sigma}_{g_me_m}^\dagger\hat{\Sigma}_{g_le_l} - 2\hat{\Sigma}_{g_le_l}\hat{\Theta}(t)\hat{\Sigma}_{g_me_m}^\dagger\right\}$$
$$+\frac{\Gamma_+}{2}\left\{\hat{\Sigma}_{g_1e_2}^\dagger\hat{\Sigma}_{g_1e_2}\hat{\Theta}(t)+\hat{\Theta}(t)\hat{\Sigma}_{g_1e_2}^\dagger\hat{\Sigma}_{g_1e_2}-2\hat{\Sigma}_{g_1e_2}\hat{\Theta}(t)\hat{\Sigma}_{g_1e_2}^\dagger\right\} + \frac{\Gamma_-}{2}\left\{\hat{\Sigma}_{g_2e_1}^\dagger\hat{\Sigma}_{g_2e_1}\hat{\Theta}(t)+\hat{\Theta}(t)\hat{\Sigma}_{g_2e_1}^\dagger\hat{\Sigma}_{g_2e_1}-2\hat{\Sigma}_{g_2e_1}\hat{\Theta}(t)\hat{\Sigma}_{g_2e_1}^\dagger\right\} \quad (F8)$$

and the decay rates are correspondingly

$$\Gamma_{lm} = \frac{\mathbf{d}_{g_l e_l} \cdot \mathbf{d}_{g_m e_m} \omega_0^3}{3\pi\varepsilon_0 \hbar c^3}, \quad \Gamma_+ = \frac{|\mathbf{d}_{g_1 e_2}|^2 \omega_+^3}{3\pi\varepsilon_0 \hbar c^3}, \quad \Gamma_- = \frac{|\mathbf{d}_{g_2 e_1}|^2 \omega_-^3}{3\pi\varepsilon_0 \hbar c^3}. \tag{F9}$$

The first term in (F8) resembles (E4), and includes interference terms between the $e_2 \to g_2$ and $g_1 \to e_1$ transitions, and between the complementary $g_2 \to e_2$ and $e_1 \to g_1$ transitions. The last two terms of (F8) correspond to the $\omega_+$ and $\omega_-$ transitions, respectively, each describing a Dicke-like process with no common atomic states, so that the two processes are seemingly decoupled. However, the new property of (F8) is that the first term also couples between the two last terms, since the transition operators in the first have common excited and ground states with the second and third term. The ODE system of the density matrix elements is

$$\begin{aligned}
\frac{d}{dt}\Theta^{m_{e_2} m_{e_1} m_{g_2} m_{g_1}}_{n_{e_2} n_{e_1} n_{g_2} n_{g_1}} &= -i\left[\omega_+(m_{e_2}-n_{e_2}) + \omega_0(m_{e_1}-n_{e_1}) + \omega_-(m_{g_2}-n_{g_2})\right]\Theta^{m_{e_2} m_{e_1} m_{g_2} m_{g_1}}_{n_{e_2} n_{e_1} n_{g_2} n_{g_1}} \\
&\quad -\frac{\Gamma_+}{2}\left[(m_{e_2}(m_{g_1}+1) + n_{e_2}(n_{g_1}+1))\Theta^{m_{e_2} m_{e_1} m_{g_2} m_{g_1}}_{n_{e_2} n_{e_1} n_{g_2} n_{g_1}} - 2\sqrt{m_{g_1}(m_{e_2}+1)n_{g_1}(n_{e_2}+1)}\Theta^{m_{e_2}+1\, m_{e_1} m_{g_2} m_{g_1}-1}_{n_{e_2}+1\, n_{e_1} n_{g_2} n_{g_1}-1}\right] \\
&\quad -\frac{\Gamma_-}{2}\left[(m_{e_1}(m_{g_2}+1) + n_{e_1}(n_{g_2}+1))\Theta^{m_{e_2} m_{e_1} m_{g_2} m_{g_1}}_{n_{e_2} n_{e_1} n_{g_2} n_{g_1}} - 2\sqrt{m_{g_2}(m_{e_1}+1)n_{g_2}(n_{e_1}+1)}\Theta^{m_{e_2}\, m_{e_1}+1\, m_{g_2}-1\, m_{g_1}}_{n_{e_2}\, n_{e_1}+1\, n_{g_2}-1\, n_{g_1}}\right] \\
&\quad -\frac{\Gamma_{11}}{2}\left[(m_{e_1}(m_{g_1}+1) + n_{e_1}(n_{g_1}+1))\Theta^{m_{e_2} m_{e_1} m_{g_2} m_{g_1}}_{n_{e_2} n_{e_1} n_{g_2} n_{g_1}} - 2\sqrt{m_{g_1}(m_{e_1}+1)n_{g_1}(n_{e_1}+1)}\Theta^{m_{e_2}\, m_{e_1}+1\, m_{g_2}\, m_{g_1}-1}_{n_{e_2}\, n_{e_1}+1\, n_{g_2}\, n_{g_1}-1}\right] \\
&\quad -\frac{\Gamma_{22}}{2}\left[(m_{e_2}(m_{g_2}+1) + n_{e_2}(n_{g_2}+1))\Theta^{m_{e_2} m_{e_1} m_{g_2} m_{g_1}}_{n_{e_2} n_{e_1} n_{g_2} n_{g_1}} - 2\sqrt{m_{g_2}(m_{e_2}+1)n_{g_2}(n_{e_2}+1)}\Theta^{m_{e_2}+1\, m_{e_1}\, m_{g_2}-1\, m_{g_1}}_{n_{e_2}+1\, n_{e_1}\, n_{g_2}-1\, n_{g_1}}\right] \\
&\quad -\frac{\Gamma_{12}}{2}\left[\sqrt{m_{g_2}(m_{e_2}+1)m_{e_1}(m_{g_1}+1)}\Theta^{m_{e_2}+1\, m_{e_1}-1\, m_{g_2}-1\, m_{g_1}+1}_{n_{e_2}\, n_{e_1}\, n_{g_2}\, n_{g_1}} + \sqrt{n_{e_2}(n_{g_2}+1)n_{g_1}(n_{e_1}+1)}\Theta^{m_{e_2}\, m_{e_1}\, m_{g_2}\, m_{g_1}}_{n_{e_2}-1\, n_{e_1}+1\, n_{g_2}+1\, n_{g_1}-1} \right. \\
&\quad \left. - 2\sqrt{m_{g_2}(m_{e_2}+1)n_{g_1}(n_{e_1}+1)}\Theta^{m_{e_2}+1\, m_{e_1}\, m_{g_2}-1\, m_{g_1}}_{n_{e_2}\, n_{e_1}+1\, n_{g_2}\, n_{g_1}-1}\right] \\
&\quad -\frac{\Gamma_{21}}{2}\left[\sqrt{m_{e_2}(m_{g_2}+1)m_{g_1}(m_{e_1}+1)}\Theta^{m_{e_2}-1\, m_{e_1}+1\, m_{g_2}+1\, m_{g_1}-1}_{n_{e_2}\, n_{e_1}\, n_{g_2}\, n_{g_1}} + \sqrt{n_{g_2}(n_{e_2}+1)n_{e_1}(n_{g_1}+1)}\Theta^{m_{e_2}\, m_{e_1}\, m_{g_2}\, m_{g_1}}_{n_{e_2}+1\, n_{e_1}-1\, n_{g_2}-1\, n_{g_1}+1} \right. \\
&\quad \left. - 2\sqrt{m_{g_1}(m_{e_1}+1)n_{g_2}(n_{e_2}+1)}\Theta^{m_{e_2}\, m_{e_1}+1\, m_{g_2}\, m_{g_1}-1}_{n_{e_2}+1\, n_{e_1}\, n_{g_2}-1\, n_{g_1}}\right]
\end{aligned} \tag{F10}$$

The solution of this system of ODEs corresponding to the master equation of the eight-dimensional tensor can then be used to calculate the emission intensities of the three modes in an analogous manner to (A20) and (E9). The results are plotted in Section III.A.

## APPENDIX G – ATOMIC DYNAMICS AND ENTANGLEMENT IN THE FOUR-LEVEL ATOMS

In this appendix we summarize the special atomic dynamics of the FLA ensemble and their entanglement, and show the effect of the atomic degeneracy (that also affects the photon entanglement) on the atoms that is manifested in the master equation (8) due to the coupling between symmetrized atomic operators involving common levels. The atomic entanglement and dynamics are controlled by the radiative transition rates (10), and we show here how the interplay between these parameters affects the evolution of the FLA ensemble.

We start by writing the atomic density matrix of the indistinguishable FLA ensemble in the Schrodinger picture in terms of symmetrized ground states in steady-state, i.e. after the termination of superradiance,

$$\hat{\Theta}(\infty) = \sum_{p,q=0}^{N} \Theta^{00\, p\, N-p}_{00\, q\, N-q} |0,0,p,N-p\rangle\langle 0,0,q,N-q| \equiv \sum_{p,q=0}^{N} \alpha_{p,q} |G_p\rangle\langle G_q|. \tag{G1}$$

Each symmetrized ground state $|G_p\rangle$ describes entanglement between the individual FLAs (this is an extension of the Dicke model). However, the different symmetrized ground states can also be entangled between themselves, thus resulting in a more intricate entanglement scheme. Since in each FLA the states $|g_1\rangle$ and $|g_2\rangle$ have energies that differ by $\hbar\Delta$, the $p$'th term in the sum oscillates with angular frequency $(p-q)\Delta$ with respect to the $q$'th term. Consequently, the coherence terms in the atomic density matrix at $t \to \infty$ oscillate as $\alpha_{k,j} \propto exp[i(p-q)\Delta t]$. This means that even after all photons have been emitted from the ensemble and the atomic populations have reached a steady-state in the form of a superposition of symmetrized ground states, the atomic ensemble will still exhibit beating between the coherence terms. The entries of the density matrix with respect to time are illustrated in Figure 4 for $N=2$. It is clear that after the population terms $\alpha_{pp}$ (marked by dashed lines) reach approximately their steady state values (marked by the vertical red line), the three coherence terms (full lines) oscillate – $\alpha_{01}$ and $\alpha_{12}$ with angular frequency $\Delta$, and $\alpha_{02}$ with angular frequency $2\Delta$.

The entanglement between the different symmetrized ground states $|G_p\rangle$ defined in (G1) originates from the virtual transitions, characterized by the decay rates $\Gamma_{12}, \Gamma_{21}$. To further demonstrate this important notion, as well as the effect of the other atomic intrinsic decay rates on the density matrix if the symmetrized grounds state, we show in Figure 11 the non-zero part of the density matrix of the final state of an FLA ensemble which was initialized by a symmetric excitation for $N=4$ FLAs, namely $\alpha_{p,q} = |\alpha_{p,q}|exp[i(p-q)\Delta t]$. The matrix entries are the absolute values $|\alpha_{p,q}|$, and the phase relative to the population (diagonal) terms is written in white. For the simplified case of identical decay rates for all transitions, $\forall i,j: \Gamma_{ij} = \Gamma_{\pm} = \Gamma$ (Figure 11a), we see that the resulting ground-state density matrix is symmetric and that the $|G_{N/2}\rangle\langle G_{N/2}|$ population term is the most probable.

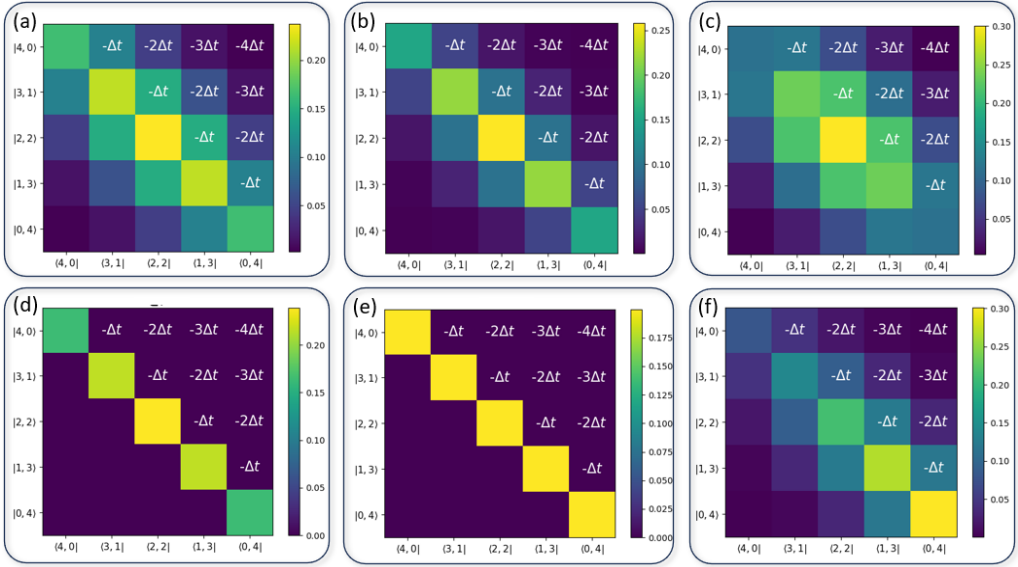

Figure 11 – Density matrix of the symmetrized ground states (eq. (G1)) for $N=4$ FLAs. Each image shows the absolute value of the entry of the density matrix. The phase relative to the population terms (diagonal entries) is denoted in white. For brevity, only the two non-zero occupation numbers $n_{g_1}, n_{g_2}$ are shown, $|0,0,n_{g_2},n_{g_1}\rangle \equiv |n_{g_2}, n_{g_1}\rangle$. (a) $\Gamma_{lm} = \Gamma_{\pm} = \Gamma$ for all $l,m$. (b) $\Gamma_{lm} = \Gamma$ for all $l,m$, $\Gamma_{\pm} = 2.5\Gamma$. (c) $\Gamma_{lm} = \Gamma = 2.5\Gamma_{\pm}$. (d) $\Gamma_{11} = \Gamma_{22} = 2.5\Gamma_{\pm}$, $\Gamma_{12} = \Gamma_{21} = 0$. (e) $\Gamma_{11} = \Gamma_{22} = \Gamma_{\pm}$, $\Gamma_{12} = \Gamma_{21} = 0$. (f) $\Gamma_{lm} = \Gamma_{+}$ for all $l,m$, $\Gamma_{+} = 2.5\Gamma_{-}$.

We also note that the amplitude of the coherence terms becomes smaller as their oscillations become faster (as was also demonstrated in Figure 4 for $N=2$). This means that the farther apart the symmetrized ground states are in energy, the less correlated they become, which can be heuristically attributed to the fact that such states would have smaller overlap of their evolution paths. We can see that when $\forall l,m: \Gamma_{lm} = \Gamma < \Gamma_{\pm}$ (Figure 11b) the coherences between the symmetrized ground states are smaller, and vice versa (Figure 11c). This can be explained again by the fact that the mutual atomic transitions of angular frequency $\omega_0$ introduce overlap between the different evolution paths of the superradiance process, inducing coherences between the symmetrized ground states. When the decay rates $\Gamma_{12} = \Gamma_{21}$ corresponding to the virtual transitions are zero (Figure 11d,e), the coherence terms are zero and the symmetrized ground states become disentangled; this demonstrates that the entanglement between symmetrized ground states is driven by the virtual transitions due to the intra-atomic degeneracy. Lastly, the ratio $\Gamma_{+}/\Gamma_{-}$ is another control knob for the probabilities of the populations and the coherences in the final atomic state (Figure 11f).

We note that at $t \to \infty$ the atomic state is bipartite, because at long times both excitation atomic numbers $n_{e_1}$ and $n_{e_2}$ are completely decayed and the entire information on the atomic system is encapsulated in $n_{g_1}$ and $n_{g_2}$; in other words, measuring $n_{e_1}$ and $n_{e_2}$ at $t \to \infty$ collapses the atomic density matrix into a bipartite density matrix. We show numerically that the two atomic occupation numbers $n_{g_1}$ and $n_{g_2}$ in the final ground state are entangled using the Peres criterion [101], by showing that the partial-transposed density matrix has at least one negative eigenvalue. The entanglement negativity depends on the ratios of the decay rates, as shown in Figure 12. When the decay rate cross-terms $\Gamma_{12}$ and $\Gamma_{21}$ are both zero, describing the case of orthogonal transition dipoles $\mathbf{d}_{11}$ and $\mathbf{d}_{22}$, the oscillations of the coherence terms disappear and the entanglement negativity is zero. Furthermore, the entanglement negativity increases with the $\Gamma_{mm}/\Gamma_{\pm}$ ratio for $m \in \{1,2\}$, demonstrating that the transitions associated with the $\omega_0$ mode introduce the atomic entanglement.

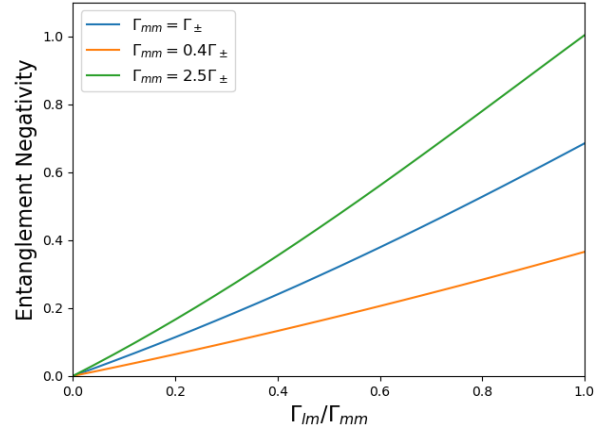

Figure 12 – Entanglement negativity for 4 FLAs as a function of the ratio between the cross-decay rates and the $\omega_0$ transition decay rate, $\Gamma_{lm}/\Gamma_{mm}$ for $l, m \in \{1,2\}$. The blue, orange and green curves denote different ratios between the $\omega_0$ mode decay rates $\Gamma_{11} = \Gamma_{22}$ and the $\omega_+$ and $\omega_-$ modes decay rates $\Gamma_+ = \Gamma_- = \Gamma_{\pm}$.

The non-classical nature of the interatomic interaction can be further demonstrated by the Wigner distributions of the two atomic ground-state occupation numbers at long times. In Figure 13, we plot six two-dimensional atomic Wigner distributions of the two symmetrized ground-state occupation numbers $n_{g_1}$ and $n_{g_2}$,

$$W_A^{2D}(\alpha,\beta) = \frac{4}{\pi^2} \sum_{m,p,n,q} (-1)^{m+p} \Theta_{00nq}^{00mp} \langle n|\hat{D}(2\alpha)|m\rangle\langle q|\hat{D}(2\beta)|p\rangle \qquad (G2)$$

represented by the phase-space coordinates $\alpha$ and $\beta$ respectively, for ensembles of $N = 4$ atoms, where in each plot the four-dimensional phase space is sliced at the origin with respect to the two phase-space coordinates that are not visualized. Details on the derivation of (G2) can be found in APPENDIX D for the photonic states, which can be readily adapted for the atomic degrees of freedom. We see that all six 2D Wigner distributions contain negative parts, attesting to non-classicality of the final atomic state. Figure 13a shows the case of identical intrinsic decay rates, and Figure 13b shows a similar case but with $\Gamma_{lm} = 10\Gamma_{\pm}$. In Figure 13a,b, fringes are evident in the Wigner distribution projections on pairs of phase-space coordinates corresponding to different modes; this interference suggests that the modes are entangled. Note that in Figure 13b more pronounced fringes and more negative values of the Wigner distribution are exhibited than in Figure 13a, due to the stronger $\omega_0$ transition decay rates. Figure 13c shows the case where the $\Gamma_{12}$ and $\Gamma_{21}$ decay rates are zero; in this case we can see that no fringes exist as in the other two subfigures, and the Wigner distributions exhibit four rings, corresponding to a superposition of Fock states with excitation numbers of up to four. In correspondence with the oscillatory behavior of the ensemble at steady state (as demonstrated for two FLAs in Figure 4), these distributions rotate periodically with time with a period $T = 1/\Delta$.

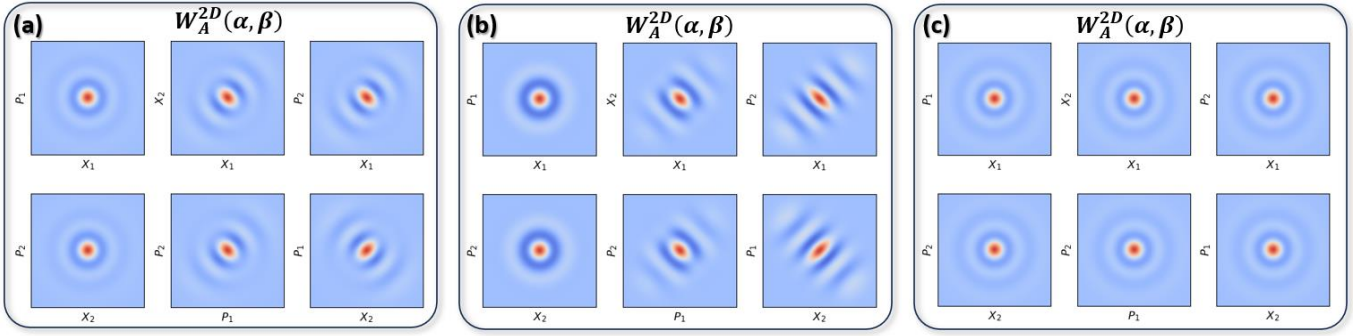

Figure 13 – 2D Wigner distributions of the two symmetrized ground-state occupation numbers $n_{g_1}$ and $n_{g_2}$, represented by the phase-space coordinates $\alpha$ and $\beta$ respectively, for ensembles of $N = 4$ atoms. (a) $\Gamma_{lm} = \Gamma_{\pm} = \Gamma$. (b) $\Gamma_{lm} = \Gamma = 10\Gamma_{\pm}$. (c) $\Gamma_{11} = \Gamma_{22} = \Gamma_{\pm}$, $\Gamma_{12} = \Gamma_{21} = 0$.

## APPENDIX H – ENTANGLEMENT MEASURE BETWEEN PHOTONIC FIELDS IN FLA

Here we follow the same procedure as in APPENDIX B in order to describe the photonic state emitted from the indistinguishable FLA ensemble. We then proceed to calculate a measure of entanglement between the photonic modal Fock numbers. The single FLA state vector is the pure state

$$|\psi_{FLA}\rangle(t) = a_2(t)|1,0,0,0\rangle|0,0,0\rangle + a_1(t)|0,1,0,0\rangle|0,0,0\rangle$$
$$+ \sum_{\mathbf{k}} |0,0,1,0\rangle \left( b_{22,\mathbf{k}}(t)|0,1_{\mathbf{k}},0\rangle + b_{21,\mathbf{k}}(t)|1_{\mathbf{k}},0,0\rangle \right) + |0,0,0,1\rangle \left( b_{12,\mathbf{k}}(t)|0,0,1_{\mathbf{k}}\rangle + b_{11,\mathbf{k}}(t)|0,1_{\mathbf{k}},0\rangle \right) \quad \text{(H1)}$$

At $t \to \infty$, $a_1(\infty) = a_2(\infty) = 0$. The other constants are obtained from Schrödinger's equations with the Weisskopf-Wigner ansatz,

$$\frac{d}{dt}a_1(t) = -\frac{\Gamma_{11} + \Gamma_-}{2}a_1(t), \quad \frac{d}{dt}b_{11,\mathbf{k}}(t) = -ig_{11,\mathbf{k}}a_1(t)e^{i(\omega_{\mathbf{k}}-\omega_0)t}, \quad \frac{d}{dt}b_{12,\mathbf{k}}(t) = -ig_{12,\mathbf{k}}a_2(t)e^{i(\omega_{\mathbf{k}}-\omega_+)t},$$
$$\frac{d}{dt}a_2(t) = -\frac{\Gamma_{22} + \Gamma_+}{2}a_2(t), \quad \frac{d}{dt}b_{21,\mathbf{k}}(t) = -ig_{21,\mathbf{k}}a_1(t)e^{i(\omega_{\mathbf{k}}-\omega_-)t}, \quad \frac{d}{dt}b_{22,\mathbf{k}}(t) = -ig_{22,\mathbf{k}}a_2(t)e^{i(\omega_{\mathbf{k}}-\omega_0)t}$$
(H2)

Applying the same process as for the V-atoms in APPENDIX B we obtain at infinite times

$$b_{11,\mathbf{k}}(\infty) = \frac{g_{11,\mathbf{k}}a_1(0)}{\omega_{\mathbf{k}} - \omega_0 + i(\Gamma_{11} + \Gamma_-)/2}, \quad b_{12,\mathbf{k}}(\infty) = \frac{g_{12,\mathbf{k}}a_2(0)}{\omega_{\mathbf{k}} - \omega_+ + i(\Gamma_{22} + \Gamma_+)/2},$$
$$b_{21,\mathbf{k}}(\infty) = \frac{g_{21,\mathbf{k}}a_1(0)}{\omega_{\mathbf{k}} - \omega_- + i(\Gamma_{11} + \Gamma_-)/2}, \quad b_{22,\mathbf{k}}(\infty) = \frac{g_{22,\mathbf{k}}a_2(0)}{\omega_{\mathbf{k}} - \omega_0 + i(\Gamma_{22} + \Gamma_+)/2}$$
(H3)

for the decay rates (F9). The reduced density matrix of the photonic degrees of freedom is

$$\rho_f^{FLA}(\infty) = Tr_A\{\rho_{FLA}(\infty)\} = \langle 0,0,1,0|\psi_{FLA}\rangle\langle\psi_{FLA}|0,0,1,0\rangle(\infty) + \langle 0,0,0,1|\psi_{FLA}\rangle\langle\psi_{FLA}|0,0,0,1\rangle(\infty)$$
$$= \sum_{\mathbf{k},\mathbf{k}'} \left[ \left( b_{22,\mathbf{k}}(\infty)|0,1_{\mathbf{k}},0\rangle + b_{21,\mathbf{k}}(\infty)|1_{\mathbf{k}},0,0\rangle \right) \left( b_{22,\mathbf{k}'}^*(\infty)\langle 0,1_{\mathbf{k}'},0| + b_{21,\mathbf{k}'}^*(\infty)\langle 1_{\mathbf{k}'},0,0| \right) \right.$$
$$\left. + \left( b_{12,\mathbf{k}}(\infty)|0,0,1_{\mathbf{k}}\rangle + b_{11,\mathbf{k}}(\infty)|0,1_{\mathbf{k}},0\rangle \right) \left( b_{12,\mathbf{k}'}^*(\infty)\langle 0,0,1_{\mathbf{k}'}| + b_{11,\mathbf{k}'}^*(\infty)\langle 0,1_{\mathbf{k}'},0| \right) \right]$$
(H4)

For brevity will henceforth drop the notation for $t \to \infty$. We aim to calculate the conditional entanglement entropy which requires us to calculate exactly the entries of the photonic density matrix. To simplify the calculations, which otherwise need to be performed numerically, we assume that all photon packets have identical spectral shapes. In this case, it is straightforward to see from (H3) that this assumption requires that $\Gamma_\pm = \Gamma_{mm} = \Gamma$ for $m \in \{1,2\}$. From (F9), we see that this amounts to $\omega_0^3 d_{g_1 e_1}^2 = \omega_0^3 d_{g_2 e_2}^2 = \omega_+^3 d_{g_1 e_2}^2 = \omega_-^3 d_{g_2 e_1}^2$, and for the atomic excitations $a_1(0) = \alpha$ and $a_2(0) = \beta$ the we can write the coefficients as $\tilde{b}_k \alpha \equiv b_{11,k} = b_{21,k}$ and $\tilde{b}_k \beta \equiv b_{22,k} = b_{12,k}$. We define the orthonormal basis

$$|010\rangle\rangle = \sqrt{2} \sum_k \tilde{b}_k |0,1_k,0\rangle, \quad |100\rangle\rangle = \sqrt{2} \sum_k \tilde{b}_k |1_k,0,0\rangle, \quad |001\rangle\rangle = \sqrt{2} \sum_k \tilde{b}_k |0,0,1_k\rangle \quad \text{(H5)}$$

and plugging this into (H4) yields

$$\rho_f^{FLA}(\infty) = \frac{1}{2}\left(\beta|010\rangle\rangle + \alpha|100\rangle\rangle\right)\left(\beta^* \langle\langle 010| + \alpha^* \langle\langle 100|\right) + \frac{1}{2}\left(\alpha|010\rangle\rangle + \beta|001\rangle\rangle\right)\left(\alpha^* \langle\langle 010| + \beta^* \langle\langle 001|\right) \quad \text{(H6)}$$

appearing in matrix form in (14) in the main text. The $\sqrt{2}$ factor in (H5) is necessary for the orthonormality condition of this basis, since from (H3) in the continuum limit we obtain

$$\sum_k |\tilde{b}_k|^2 \to \frac{1}{2}. \quad \text{(H7)}$$

In order to quantify entanglement between the three Fock-numbers describing the three photonic fields, we will use the measure of conditional entropy of entanglement. Given a bipartite state $\hat{\rho}^{\phi\theta}$, the conditional entropy of entanglement [104] of the partition $\phi$ conditioned on $\theta$ is given by

$$S(\phi|\theta) = S(\hat{\rho}^{\phi\theta}) - S(\theta) \quad \text{(H8)}$$

where $S(...)$ is the von-Neumann entropy and $S(\theta) = S(Tr_\phi \hat{\rho}^{\phi\theta})$ is the reduced entropy of subsystem $\theta$. In our case, we wish to calculate the conditional entanglement entropy $S(\mathbf{2,3}|\mathbf{1}) = S_f - S_{f1}$, where $\mathbf{1}$ is either one of the three photonic modes $\{-,0,+\}$ constituting one partition and $\mathbf{2,3}$ are the two remaining modes constituting the other partition. The expression $S_f = S(\hat{\rho}_{f,N})$ denotes the von-Neumann entropy of the photonic fields and $S_{f1} = S(Tr_1 \hat{\rho}_{f,N})$ the reduced entropy of the photonic subsystem of the $\omega_1$ photons modal occupation number. Calculation of the von-Neumann entropy therefore amounts to calculating

$$S_f \equiv S\left(\rho_f^{FLA}\right) = Tr\left\{\rho_f^{FLA} \log \rho_f^{FLA}\right\}. \quad \text{(H9)}$$

The logarithm of $\rho_f^{FLA}$ can be numerically calculated using singular value decomposition. Since $\rho_f^{FLA}$ is positive-semidefinite and Hermitian we can write it as

$$\rho_f^{FLA} = VDV^{-1}, \quad \text{(H10)}$$

for a unitary matrix $V$ and a diagonal matrix with non-negative entries $D$. From the unitarity of $V$ it follows that

$$\log \rho_f^{FLA} = V(\log D) V^{-1} \quad \text{(H11)}$$

where $\log D$ is a diagonal matrix of the logarithm values of the elements of $D$. This can be immediately verified by exponentiation of (H11), using the identity $\exp(A) = \sum_{n=0}^{\infty} \frac{A^n}{n!}$ for a square matrix $A$.

To find the conditional entropies for the various modes, we calculate the partial traces of $\rho_f^{FLA}$ with respect to its three modes. Tracing-out the first photonic degree of freedom we obtain

$$\rho_{f,-}^{FLA} = Tr_-\left\{\rho_f^{FLA}\right\} = \frac{1}{2}\left(|10_-\rangle\rangle\langle\langle 10_-| + |\alpha|^2 |00_-\rangle\rangle\langle\langle 00_-| + |\beta|^2 |01_-\rangle\rangle\langle\langle 01_-| + \alpha\beta^* |10_-\rangle\rangle\langle\langle 01_-| + \alpha^*\beta |01_-\rangle\rangle\langle\langle 10_-|\right) \quad \text{(H12)}$$

with the definition

$$|10_-\rangle\rangle = \sqrt{2}\sum_{\mathbf{k}}\tilde{b}_{\mathbf{k}}|1_{\mathbf{k}},0\rangle_-, \quad |00_-\rangle\rangle = \sqrt{2}\sum_{\mathbf{k}}\tilde{b}_{\mathbf{k}}|0,0\rangle_-, \quad |01_-\rangle\rangle = \sqrt{2}\sum_{\mathbf{k}}\tilde{b}_{\mathbf{k}}|0,1_{\mathbf{k}}\rangle_-. \tag{H13}$$

Here the photonic vector kets with a subscript "−" denote photonic states describing only the $\omega_0$ and $\omega_+$ modes, encoded by the vector $|n_{\omega_0},n_{\omega_+}\rangle_-$ (since the $\omega_-$ mode is traced-out). From (H12) we can obtain the conditional entropy $S_f^{FLA}(\mathbf{0},+|-) = S_f^{FLA} - S_{f,-}^{FLA}$.

The conditional entropy $S_f^{FLA}(-,\mathbf{0}|+) = S_f^{FLA} - S_{f,+}^{FLA}$ can be calculated numerically in a similar manner due to the similarity of the expressions $S_{f,-}^{FLA}$ and $S_{f,+}^{FLA}$. We denote

$$|10_+\rangle\rangle = \sqrt{2}\sum_{\mathbf{k}}\tilde{b}_{\mathbf{k}}|1_{\mathbf{k}},0\rangle_+, \quad |00_+\rangle\rangle = \sqrt{2}\sum_{\mathbf{k}}\tilde{b}_{\mathbf{k}}|0,0\rangle_+, \quad |01_+\rangle\rangle = \sqrt{2}\sum_{\mathbf{k}}\tilde{b}_{\mathbf{k}}|0,1_{\mathbf{k}}\rangle_+$$
$$|10_0\rangle\rangle = \sqrt{2}\sum_{\mathbf{k}}\tilde{b}_{\mathbf{k}}|1_{\mathbf{k}},0\rangle_0, \quad |00_0\rangle\rangle = \sqrt{2}\sum_{\mathbf{k}}\tilde{b}_{\mathbf{k}}|0,0\rangle_0, \quad |01_0\rangle\rangle = \sqrt{2}\sum_{\mathbf{k}}\tilde{b}_{\mathbf{k}}|0,1_{\mathbf{k}}\rangle_0 \tag{H14}$$

The subscript "+" for the photonic state vector denotes a state composed of only the $\omega_-$ and $\omega_0$ modes encoded $|n_{\omega_-},n_{\omega_0}\rangle_+$, in an analogous manner to (H13), and the "0" subscript is analogously encoded $|n_{\omega_-},n_{\omega_+}\rangle_0$.

Then, one obtains by tracing out the $\omega_+$ mode

$$\rho_{f,+}^{FLA} = \frac{1}{2}\left(|01_+\rangle\rangle\langle\langle 01_+| + |\alpha|^2|10_+\rangle\rangle\langle\langle 10_+| + |\beta|^2|00_+\rangle\rangle\langle\langle 00_+| + \alpha^*\beta|01_+\rangle\rangle\langle\langle 10_+| + \alpha\beta^*|10_+\rangle\rangle\langle\langle 01_+|\right) \tag{H15}$$

and tracing out the $\omega_0$ mode from (H4) similarly yields

$$\rho_{f,\mathbf{0}}^{FLA} = \frac{1}{2}\left(|00_0\rangle\rangle\langle\langle 00_0| + |\alpha|^2|10_0\rangle\rangle\langle\langle 10_0| + |\beta|^2|01_0\rangle\rangle\langle\langle 01_0|\right) \tag{H16}$$

This is a diagonal matrix, therefore it describes no correlations between the $\omega_-$ and $\omega_+$ photonic modes, as explained in the main text.

**Two indistinguishable FLAs**

For two indistinguishable FLAs, we may write the final state as

$$|\psi_{2-FLA}\rangle = |0,0,2,0\rangle\sum_{\mathbf{k},\mathbf{q}}^{N}\left(c_{1,\mathbf{kq}}|2_{\mathbf{qk}},0,0\rangle + c_{2,\mathbf{kq}}|1_{\mathbf{k}},1_{\mathbf{q}},0\rangle + c_{3,\mathbf{kq}}|0,2_{\mathbf{qk}},0\rangle\right)$$
$$+|0,0,1,1\rangle\sum_{\mathbf{k},\mathbf{q}}^{N}\left(c_{4,\mathbf{kq}}|1_{\mathbf{k}},1_{\mathbf{q}},0\rangle + c_{5,\mathbf{kq}}|0,2_{\mathbf{qk}},0\rangle + c_{6,\mathbf{kq}}|1_{\mathbf{k}},0,1_{\mathbf{q}}\rangle + c_{7,\mathbf{kq}}|0,1_{\mathbf{k}},1_{\mathbf{q}}\rangle\right)$$
$$+|0,0,0,2\rangle\sum_{\mathbf{k},\mathbf{q}}^{N}\left(c_{8,\mathbf{kq}}|0,2_{\mathbf{qk}},0\rangle + c_{9,\mathbf{kq}}|0,1_{\mathbf{k}},1_{\mathbf{q}}\rangle + c_{10,\mathbf{kq}}|0,0,2_{\mathbf{qk}}\rangle\right). \tag{H17}$$

The coefficients $c_{n,\mathbf{kq}}$ can be obtained by extending (H2) to symmetrized states, in an analogous manner to the coefficients for the V-atom case in APPENDIX B. After tracing out the atomic degrees of freedom from the density matrix of two indistinguishable FLAs at long times, $\rho^{2-FLA} = |\psi_{2-FLA}\rangle\langle\psi_{2-FLA}|$, we obtain the photonic density matrix

$$\rho_f^{2-FLA} = \sum_{\mathbf{k},\mathbf{k}'}\sum_{\mathbf{q},\mathbf{q}'}\Big[\left(c_{1,\mathbf{kq}}|2_{\mathbf{qk}},0,0\rangle + c_{2,\mathbf{kq}}|1_{\mathbf{k}},1_{\mathbf{q}},0\rangle + c_{3,\mathbf{kq}}|0,2_{\mathbf{qk}},0\rangle\right)\left(c_{1,\mathbf{k'q'}}^*\langle 2_{\mathbf{q'k'}},0,0| + c_{2,\mathbf{k'q'}}^*\langle 1_{\mathbf{k'}},1_{\mathbf{q'}},0| + c_{3,\mathbf{k'q'}}^*\langle 0,2_{\mathbf{q'k'}},0|\right)$$
$$+\left(c_{4,\mathbf{kq}}|1_{\mathbf{k}},1_{\mathbf{q}},0\rangle + c_{5,\mathbf{kq}}|0,2_{\mathbf{qk}},0\rangle + c_{6,\mathbf{kq}}|1_{\mathbf{k}},0,1_{\mathbf{q}}\rangle + c_{7,\mathbf{kq}}|0,1_{\mathbf{k}},1_{\mathbf{q}}\rangle\right)\times$$

$$\left(c_{4,\mathbf{k'q'}}^* \langle 1_{\mathbf{k'}}, 1_{\mathbf{q'}}, 0| + c_{5,\mathbf{k'q'}}^* \langle 0, 2_{\mathbf{q'k'}}, 0| + c_{6,\mathbf{k'q'}}^* \langle 1_{\mathbf{k'}}, 0, 1_{\mathbf{q'}}| + c_{7,\mathbf{k'q'}}^* \langle 0, 1_{\mathbf{k'}}, 1_{\mathbf{q'}}|\right)$$
$$+\left(c_{8,\mathbf{kq}} |0, 2_{\mathbf{qk}}, 0\rangle + c_{9,\mathbf{kq}} |0, 1_{\mathbf{k}}, 1_{\mathbf{q}}\rangle + c_{10,\mathbf{kq}} |0, 0, 2_{\mathbf{qk}}\rangle\right)\left(c_{8,\mathbf{k'q'}}^* \langle 0, 2_{\mathbf{q'k'}}, 0| + c_{9,\mathbf{k'q'}}^* \langle 0, 1_{\mathbf{k'}}, 1_{\mathbf{q'}}| + c_{10,\mathbf{k'q'}}^* \langle 0, 0, 2_{\mathbf{q'k'}}|\right)\Big] \quad \text{(H18)}$$

We will again assume that all decay rates are identical, so that the frequency-domain profiles of the photon packets containing the same photonic composition are identical Lorentzian distributions, up to a factor consisting of the initial excitation amplitude and a path-dependent combinatoric factor stemming from the indistinguishability of the two atoms. We denote the constituent mode of the 2D Lorentzian distribution with wavevectors $\mathbf{k}$ and $\mathbf{q}$ as $c_{\mathbf{kq}}$, and define the orthonormal vectors

$$|200\rangle\rangle = \sqrt{3}\sum_{\mathbf{kq}} c_{\mathbf{kq}} |2_{\mathbf{kq}}, 0, 0\rangle, \quad |110\rangle\rangle = \sqrt{3}\sum_{\mathbf{kq}} c_{\mathbf{kq}} |1_{\mathbf{k}}, 1_{\mathbf{q}}, 0\rangle, \quad |020\rangle\rangle = \sqrt{3}\sum_{\mathbf{kq}} c_{\mathbf{kq}} |0, 2_{\mathbf{kq}}, 0\rangle,$$
$$|101\rangle\rangle = \sqrt{6}\sum_{\mathbf{kq}} c_{\mathbf{kq}} |1_{\mathbf{k}}, 0, 1_{\mathbf{q}}\rangle, \quad |011\rangle\rangle = \sqrt{3}\sum_{\mathbf{kq}} c_{\mathbf{kq}} |0, 1_{\mathbf{k}}, 1_{\mathbf{q}}\rangle, \quad |002\rangle\rangle = \sqrt{3}\sum_{\mathbf{kq}} c_{\mathbf{kq}} |0, 0, 2_{\mathbf{kq}}\rangle \quad \text{(H19)}$$

and using (H19) we rewrite (H18) as a mixture of pure states

$$\rho_f^{2-FLA} = \frac{1}{3}|\varphi_{f,(0)}^{2-FLA}\rangle\langle\varphi_{f,(0)}^{2-FLA}| + \frac{1}{3}|\varphi_{f,(1)}^{2-FLA}\rangle\langle\varphi_{f,(1)}^{2-FLA}| + \frac{1}{3}|\varphi_{f,(2)}^{2-FLA}\rangle\langle\varphi_{f,(2)}^{2-FLA}| \quad \text{(H20)}$$

with

$$|\varphi_{f,(0)}^{2-FLA}\rangle = a_0 |200\rangle\rangle + a_1 |110\rangle\rangle + a_2 |020\rangle\rangle$$
$$|\varphi_{f,(1)}^{2-FLA}\rangle = a_0 |110\rangle\rangle + 2^{-1/2} a_1 (|020\rangle\rangle + |101\rangle\rangle) + a_2 |011\rangle\rangle$$
$$|\varphi_{f,(2)}^{2-FLA}\rangle = a_0 |020\rangle\rangle + a_1 |011\rangle\rangle + a_2 |002\rangle\rangle \quad \text{(H21)}$$

where $a_0 = N\alpha^2$, $a_1 = 2N\alpha\beta$ and $a_2 = N\beta^2$ with the normalization constant $N = \left(|\alpha|^4 + 4|\alpha\beta|^2 + |\beta|^4\right)^{-1/2}$ are the excitation amplitudes of the symmetrized atomic state. Note that in the case of a deterministic excitation, namely $a_0, a_1 = 0$ or $a_1, a_2 = 0$ the density matrix (H20) is diagonal, as in the equivalent case of $\Lambda$ atoms. The von-Neumann entropy of (H20) can now be numerically calculated. We trace out the first photonic degree of freedom from (H18) and define the orthogonal vectors

$$|00_-\rangle\rangle = \sqrt{3}\sum_{\mathbf{kq}} c_{\mathbf{kq}} |0,0\rangle_-, \quad |10_-\rangle\rangle_- = \sqrt{3}\sum_{\mathbf{kq}} c_{\mathbf{kq}} |1_{\mathbf{k}}, 0\rangle, \quad |20_-\rangle\rangle = \sqrt{3}\sum_{\mathbf{kq}} c_{\mathbf{kq}} |2_{\mathbf{kq}}, 0\rangle_-$$
$$|01_-\rangle\rangle = \sqrt{6}\sum_{\mathbf{kq}} c_{\mathbf{kq}} |0, 1_{\mathbf{q}}\rangle_-, \quad |11_-\rangle\rangle = \sqrt{3}\sum_{\mathbf{kq}} c_{\mathbf{kq}} |1_{\mathbf{k}}, 1_{\mathbf{q}}\rangle_-, \quad |02_-\rangle\rangle = \sqrt{3}\sum_{\mathbf{kq}} c_{\mathbf{kq}} |0, 2_{\mathbf{qk}}\rangle_- \quad \text{(H22)}$$

Here the photonic vector kets with a subscript "-" denote photonic states describing only the $\omega_0$ and $\omega_+$ modes, encoded $|n_{\omega_0}, n_{\omega_+}\rangle_-$ as in the single FLA case. Then, one obtains

$$\rho_{f,-}^{2-FLA} = \Big[|a_0|^2 |00_-\rangle\rangle\langle\langle 00_-| + |a_1|^2 |10_-\rangle\rangle\langle\langle 10_-| + |a_2|^2 |20_-\rangle\rangle\langle\langle 20_-|$$
$$+\left(a_0 |10_-\rangle\rangle + 2^{-1/2} a_1 |01_-\rangle\rangle\right)\left(a_0^* \langle\langle 10_-| + 2^{-1/2} a_1^* \langle\langle 01_-|\right) + \left(2^{-1/2} a_1 |20_-\rangle\rangle + a_2 |11_-\rangle\rangle\right)\left(2^{-1/2} a_1^* \langle\langle 20_-| + a_2^* \langle\langle 11_-|\right)$$
$$+\left(a_0 |20_-\rangle\rangle + a_1 |11_-\rangle\rangle + a_2 |02_-\rangle\rangle\right)\left(a_0^* \langle\langle 20_-| + a_1^* \langle\langle 11_-| + a_2^* \langle\langle 02_-|\right)\Big] \quad \text{(H23)}$$

and we may calculate the von-Neumann entropy $S_{f,-}$ and finally obtain $S_f^{2-FLA}(\mathbf{0},+|-)$. Likewise, from

$$\rho_{f,0}^{2-FLA} = \Big[|a_0|^2 |20_0\rangle\rangle\langle\langle 20_0| + |a_1|^2 |10_0\rangle\rangle\langle\langle 10_0| + 2^{-1} |a_1|^2 |11_0\rangle\rangle\langle\langle 11_0| + |a_1|^2 |01_0\rangle\rangle\langle\langle 01_0| + |a_2|^2 |02_0\rangle\rangle\langle\langle 02_0|$$
$$+\left(|a_0|^2 + 2^{-1} |a_1|^2 + |a_2|^2\right)|00_0\rangle\rangle\langle\langle 00_0| + \left(a_0 |10_0\rangle\rangle + a_2 |01_0\rangle\rangle\right)\left(a_0^* \langle\langle 10_0| + a_2^* \langle\langle 01_0|\right)\Big] \quad \text{(H24)}$$

and

$$\begin{aligned}\rho_{f,+}^{2-FLA} = &\Big[|a_2|^2 |20_+\rangle\rangle\langle\langle 20_+| + |a_1|^2 |11_+\rangle\rangle\langle\langle 11_+| + |a_0|^2 |02_+\rangle\rangle\langle\langle 02_+|\\ &+\big(a_2|11_+\rangle\rangle + 2^{-1/2}a_1|10_+\rangle\rangle\big)\big(a_2^*\langle\langle 11_+| + 2^{-1/2}a_1^*\langle\langle 10_+|\big) + \big(2^{-1/2}a_1|02_+\rangle\rangle + a_0|01_+\rangle\rangle\big)\big(2^{-1/2}a_1^*\langle\langle 02_+| + a_0^*\langle\langle 01_+|\big)\\ &+\big(a_2|02_+\rangle\rangle + a_1|01_+\rangle\rangle + a_0|00_+\rangle\rangle\big)\big(a_2^*\langle\langle 02_+| + a_1^*\langle\langle 01_+| + a_0^*\langle\langle 00_+|\big)\Big]\end{aligned} \quad (H25)$$

we may numerically perform the calculations of $S_f^{2-FLA}(-,+|\mathbf{0})$ and $S_f^{2-FLA}(-,\mathbf{0}|+)$. The vectors with subscripts "$\mathbf{0}$" and "+" are defined in an analogous manner to the definition (H22), with the kets describing the $\omega_-,\omega_+$ and $\omega_-,\omega_0$ respectively, encoded $|n_{\omega_-},n_{\omega_+}\rangle_\mathbf{0}$ and $|n_{\omega_-},n_{\omega_0}\rangle_+$ respectively,

$$\begin{aligned}|20_0\rangle\rangle &= \sqrt{3}\sum_{\mathbf{kq}}c_{\mathbf{kq}}|2_{\mathbf{qk}},0\rangle_0, & |10_0\rangle\rangle &= \sqrt{3}\sum_{\mathbf{kq}}c_{\mathbf{kq}}|1_\mathbf{k},0\rangle_0, & |00_0\rangle\rangle &= \sqrt{3}\sum_{\mathbf{kq}}c_{\mathbf{kq}}|0,0\rangle_0,\\ |11_0\rangle\rangle &= \sqrt{6}\sum_{\mathbf{kq}}c_{\mathbf{kq}}|1_\mathbf{k},1_\mathbf{q}\rangle_0, & |01_0\rangle\rangle &= \sqrt{3}\sum_{\mathbf{kq}}c_{\mathbf{kq}}|0,1_\mathbf{k}\rangle_0, & |02_0\rangle\rangle &= \sqrt{3}\sum_{\mathbf{kq}}c_{\mathbf{kq}}|0,2_{\mathbf{qk}}\rangle_0\end{aligned} \quad (H26)$$

and

$$\begin{aligned}|20_+\rangle\rangle &= \sqrt{3}\sum_{\mathbf{kq}}c_{\mathbf{kq}}|2_{\mathbf{kq}},0\rangle_+, & |11_+\rangle\rangle &= \sqrt{3}\sum_{\mathbf{kq}}c_{\mathbf{kq}}|1_\mathbf{k},1_\mathbf{q}\rangle_+, & |02_+\rangle\rangle &= \sqrt{3}\sum_{\mathbf{kq}}c_{\mathbf{kq}}|0,2_{\mathbf{kq}}\rangle_+,\\ |10_+\rangle\rangle &= \sqrt{6}\sum_{\mathbf{kq}}c_{\mathbf{kq}}|1_\mathbf{k},0\rangle_+, & |01_+\rangle\rangle &= \sqrt{3}\sum_{\mathbf{kq}}c_{\mathbf{kq}}|0,1_\mathbf{k}\rangle_+, & |00_+\rangle\rangle &= \sqrt{3}\sum_{\mathbf{kq}}c_{\mathbf{kq}}|0,0\rangle_+\end{aligned}. \quad (H27)$$

The three conditional entanglement entropies are illustrated in Figure 6 in the main text.